\documentclass[fleqn,usenatbib]{mnras}

\usepackage{newtxtext,newtxmath}
\usepackage{float}% If comment this, figure moves to Page 2
\usepackage{lipsum}
\usepackage[T1]{fontenc}
\usepackage[utf8]{inputenc}\usepackage[T1]{fontenc}
\usepackage{pdflscape}
\usepackage{afterpage}
\usepackage{capt-of}% or use the larger `caption` package

\DeclareRobustCommand{\VAN}[3]{#2}
\let\VANthebibliography\thebibliography
\def\thebibliography{\DeclareRobustCommand{\VAN}[3]{##3}\VANthebibliography}

\usepackage{color,soul}

\usepackage{graphicx}	% Including figure files
\usepackage{amsmath}	% Advanced maths commands
\usepackage{multicol}
\usepackage{hyperref}
\usepackage[capitalize]{cleveref}
\crefname{table}{table}{tables}
\Crefname{table}{Table}{Tables}
\usepackage{pdflscape}
\usepackage{tabularx}

\usepackage{caption}
\usepackage{subcaption}
\title[ML Classification of Repeating FRBs]{Machine Learning Classification of Repeating FRBs from FRB121102} 

\author[B.J. Raquel et al.]{
Bjorn Jasper R. Raquel$^{1,2,3\thanks{E-mail: bjornjasper1@gmail.com}}$,
Tetsuya Hashimoto$^{2}$,
Tomotsugu Goto$^{4}$,
Bo Han Chen$^{4,5,6}$,\newauthor
Yuri Uno$^{2}$,
Tiger Yu-Yang Hsiao$^{4,7}$,
Seong Jin Kim$^{4}$, and
Simon C.-C. Ho$^{4,8}$
\\
$^{1}$Department of Earth and Space Sciences, Rizal Technological University, Boni Avenue, Mandaluyong City, 1550 Metro Manila, Philippines\\
$^{2}$Department of Physics, National Chung Hsing University, No. 145, Xingda Rd., South Dist., Taichung, 40227, Taiwan (R.O.C.)\\
$^{3}$National Institute of Physics, College of Science, University of the Philippines, Diliman, Quezon City, 1101 Metro Manila, Philippines\\
$^{4}$Institute of Astronomy, National Tsing Hua University, 101, Section 2. Kuang-Fu Road, Hsinchu, 30013, Taiwan (R.O.C.)\\
$^{5}$Department of Physics, National Tsing Hua University, 101, Section 2. Kuang-Fu Road, Hsinchu, 30013, Taiwan (R.O.C.)\\
$^{6}$Graduate School of Data Science, Seoul National University,  1, Gwanak-ro, Gwanak-gu, Seoul 08826, Korea\\
$^{7}$Department of Physics and Astronomy, Johns Hopkins University, Baltimore, MD 21218, USA\\
$^{8}$Research School of Astronomy and Astrophysics, The Australian National University, Canberra, ACT 2611, Australia\\
}

\date{Accepted 2023 June 15. Received 2023 June 12; in original form 2022 October 9}

\pubyear{2023}

\begin{document}
\label{firstpage}
\pagerange{\pageref{firstpage}--\pageref{lastpage}}
\maketitle

\begin{abstract}
Fast Radio Bursts (FRBs) are mysterious bursts in the millisecond timescale at radio wavelengths. Currently, there is little understanding about the classification of repeating FRBs, based on difference in physics, which is of great importance in understanding their origin. Recent works from the literature focus on using specific parameters to classify FRBs to draw inferences on the possible physical mechanisms or properties of these FRB subtypes. In this study, we use publicly available 1652 repeating FRBs from FRB121102 detected with the Five-hundred-meter Aperture Spherical Telescope (FAST), and studied them with an unsupervised machine learning model. By fine-tuning the hyperparameters of the model, we found that there is an indication for four clusters from the bursts of FRB121102 instead of the two clusters ("Classical" and "Atypical") suggested in the literature. Wherein, the “Atypical” cluster can be further classified into three sub-clusters with distinct characteristics. Our findings show that the clustering result we obtained is more comprehensive not only because our study produced results which are consistent with those in the literature but also because our work uses more physical parameters to create these clusters. Overall, our methods and analyses produced a more holistic approach in clustering the repeating FRBs of FRB121102.
\end{abstract}

\begin{keywords}
(transients:) fast radio bursts -- stars: magnetars -- stars: neutron -- methods: data analysis
\end{keywords}

%%%%%%%%%%%%%%%%%%%%%%%%%%%%%%%%%%%%%%%%%%%%%%%%%%

%%%%%%%%%%%%%%%%% BODY OF PAPER %%%%%%%%%%%%%%%%%%

\section{INTRODUCTION}\label{sec1}

    \; \; Fast Radio Bursts (FRBs) are bright millisecond-duration radio flashes of extragalactic origin (\citealt{lorimer2007bright,thornton2013population,petroff2016frbcat}). They are characterized by their anomalously high dispersion measure (DM) and millisecond duration, indicating high brightness temperature and isotropic energy release (\citealt{ravi2015fast,tendulkar2017host,zhang2018fast,bannister2019single,ravi2019fast,li2021long,bochenek2020fast}). FRBs are usually classified as either ‘repeating’ or ‘non-repeating.’  Repeating FRBs have multiple bursts, while non-repeating FRBs have one-off bursts (\citealt{cordes2019}). Currently, there are $>$ 600 FRBs that are reported as of April 2022 (\citealt{petroff2016frbcat,li2021long,CHIME/FRB2021}).

	FRB121102, first discovered in 2014 (\citealt{spitler2014fast}) and identified as a repeater in 2016 (\citealt{spitler2016repeating}), is the most extensively studied FRB across a broad range of radio frequencies from 600 MHz up to 8 GHz (\citealt{josephy2019chime,gajjar2018highest}).The repetition allowed for localization with a high precision of 100 mas, leading to the first unambiguous identification of an FRB host galaxy at $\sim$1 Gpc ($z = 0.193$) and its association with a persistent radio source \citep{chatterjee2017direct,bassa2017frb,marcote2017repeating,tendulkar2017host,kokubo2017}). Many theoretical models have been developed to explain the physical nature of FRB121102 (see \citealt{platts2019living} for review). In particular, it has been suggested that FRB121102 might have originated from a young magnetar \citep[][]{kashiyama2017testing,metzger2017millisecond,beloborodov2017,margalit2018unveiling}. Performing follow-up observations using the Arecibo Telescope, \citealt{spitler2016repeating} found ten additional bursts for FRB121102. Shortly after, \citealt{scholz2016repeating} found six bursts from two different telescopes. Five from the Green Bank Telescope (GBT) at 2 GHz, and one from the Arecibo Telescope at 1.4 GHz.
    \citealt{michilli2018extreme} detected 16 bursts from FRB121102 using the William E. Gordon Telescope at the Arecibo Observatory at 4.1-4.9 GHz. Most recently, \citealt{2021Natur.598..267L} found 1652 bursts using the Five-hundred-meter Aperture Spherical radio Telescope (FAST) at 1.05-1.45 GHz. In addition to these, \citealt{rajwade2020possible} discovered a tentative period of 157 d with a duty cycle of 56 percent, and \citealt{2019Hessels} showed that FRB121102 exhibits a complex time-frequency structure. %This characteristic is not observed for radio-emitting magnetars or radio pulsars.

    Machine Learning (ML) has been proven helpful in astronomy and its related fields. In the field of FRB research, ML has found its applications in the works of \citealt{zhangetal2018fast}, wherein they used a combination of neural network detection with dedispersion verification to work on pulse detection and periodicity of FRB121102; \citealt{wagstaff2016machine} in the development of automated methods in identifying events of interest; \citealt{connor2018applying} in applying deep learning to single-pulse classification and developing a hierarchical framework for ranking events by their probability of being astrophysical transients; and most recently, \citealt{chen2022uncloaking} where an unsupervised machine learning algorithm, namely Uniform Manifold Approximation and Projection or UMAP \citep{2018arXiv180203426M}, was used to understand, classify, and identify possible FRB repeaters from a sample of 501 non-repeating and 93 repeating FRBs. %\textcolor{red}{Following the similar method, the works of \citealt{2022MNRAS.514.5987K} and \citealt{2022MNRAS.511.1961H} goes to show that using unsupervised machine learning is a powerful tool in studying FRBs}.

	Despite these developments, there is still little understanding about the nature of the repeating FRBs (e.g., \cite{2022MNRAS.514.5987K,2022MNRAS.511.1961H}). Thus, the main purpose of this research is to shed light on the underlying physical mechanisms of repeating FRBs by studying FRB121102. Specifically, this study focuses on determining and characterizing burst subtypes of FRB121102 in order to unveil latent features or properties of repeating FRBs. Also, we would limit the focus of this paper to classifying FRBs leaving the discussion of the possible mechanisms to future theoretical studies. 

    This paper is structured as follows: Section \ref{sec2} (Data Preprocessing) discusses the selection of the samples from the archival data shown in the Supplementary Table 1 of the \citet[dataset]{2021Natur.598..267L} paper. Section \ref{sec3} (Unsupervised Machine Learning) is divided into two subsections. Section \ref{sec3.1} (Uniform Manifold Approximation and Projection (UMAP)) focuses on finding the low-dimensional representation of the data using UMAP and Section \ref{sec3.2} (Hierarchical Density-Based Spatial Clustering of Applications with Noise (HDBSCAN)) discusses how HDBSCAN was used to cluster the data. In Section \ref{sec4} (Results) we show the parameter coloring of the UMAP embedding results to see any trends and investigate the properties of each cluster. In Section \ref{sec5} (Discussion) we discuss the implication of the results, change in cluster membership (Section~\ref{clusterchange}, Cluster membership change), and compare it to the results found in the literature (Section~\ref{comparison}, Comparison with other results). Lastly, Section \ref{sec6} (Conclusions) summarizes the findings and conclusions of this study. An Appendix~\ref{appendix} has also been included to show other important results that is used in the analysis but not central to the goal of the paper.

\section{DATA PREPROCESSING}\label{sec2}

        \; \;In this paper, we used the archival data from FAST as presented in the Supplementary Table 1 of \citet[dataset]{2021Natur.598..267L}. Wherein, they reported 1652 independent bursts in a total of $59.5$ hours all throughout the continuous monitoring campaign of FRB121102 from August 29, 2019, up until October 29, 2019, using the FAST. The archival data from the Supplementary Table 1 of \citet[dataset]{2021Natur.598..267L} have the following parameters:
            \begin{itemize}
                \item Burst Arrival Time (MJD)
                \item Dispersion Measure (pc $\cdot$ cm$^{-3}$)
                \item Time Width (ms)
                \item Bandwidth (GHz)
                \item Peak Flux (mJy) 
                \item Fluence (Jy $\cdot$ ms)
                \item Energy (erg)
            \end{itemize}
            
        We want to include as many parameters as possible to ensure the veracity of the results. Thus, we included waiting time, which is defined to be the arrival-time difference between two subsequent bursts.  
        
        For the parameters that we used in the unsupervised machine learning we excluded the Burst Time (MJD) parameter because the observational periods of the monitoring campaign are not uniform based on Figure 1.a of \citet[dataset]{2021Natur.598..267L}. The Dispersion Measure (pc $\cdot$ cm$^{-3}$) is also excluded for the reason that it is not intrinsic to the FRB source, and it is mainly related to the distance of the source. Some of these parameters are known to have correlation with each other, but we still included them in the analysis because their inclusion does not introduce bias as what \cite{lindner2020misconceptions} found in their analysis and exploration on the treatment of collinearity in quantitative empirical research. Additionally, it does not hurt to include as many parameters as possible. Thus, the parameters that are used for the unsupervised machine learning are Time Width (ms), Bandwidth (GHz), Peak Flux (mJy), Fluence (Jy $\cdot$ ms), Energy (erg), and Waiting Time (s).

\begin{figure}
            \centering
            \includegraphics[width=\columnwidth]{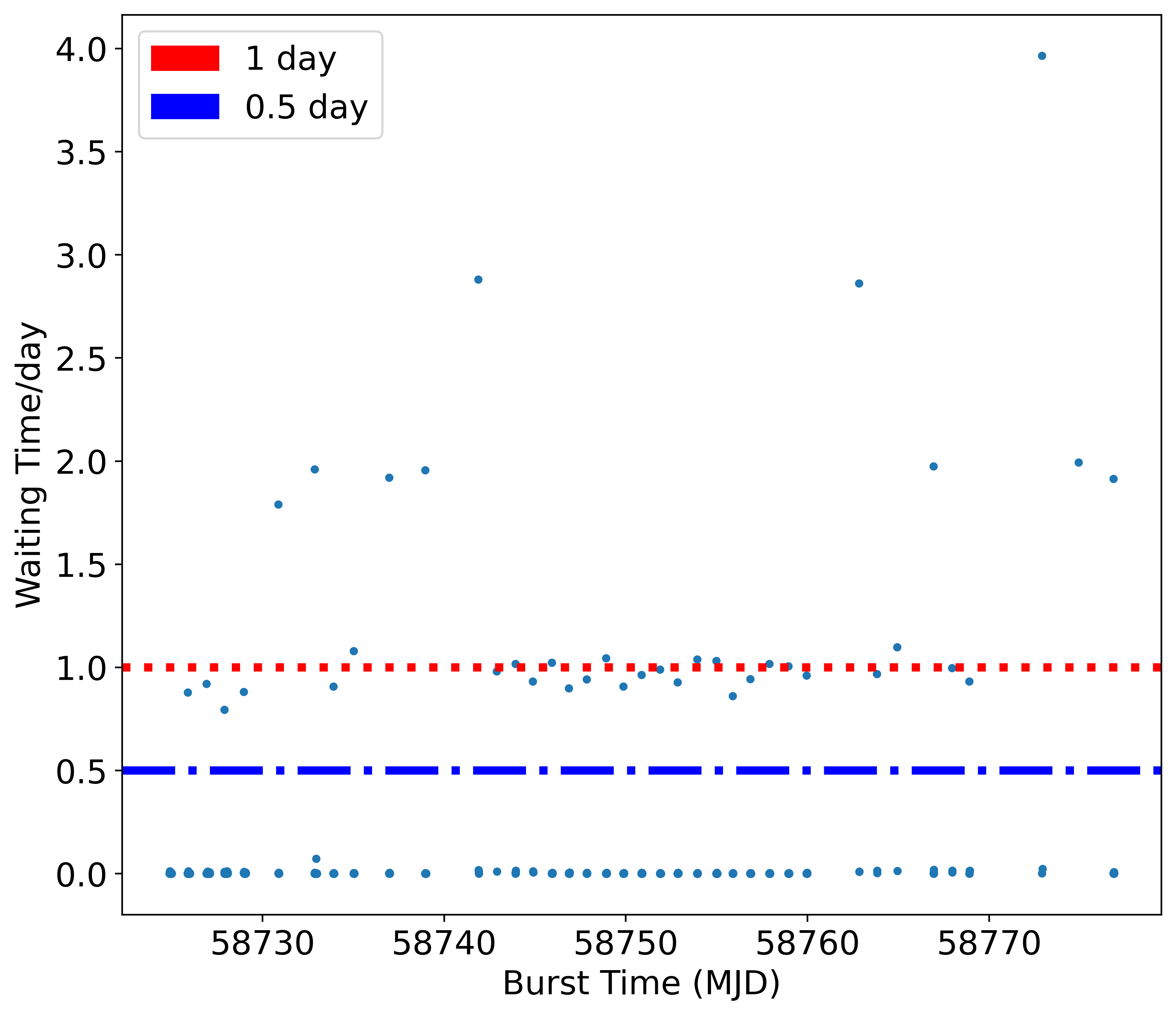}
            \caption{ \large   Waiting time values of each burst throughout the monitoring campaign done by \citet[dataset]{2021Natur.598..267L}. }
            \label{WT}
        \end{figure}

        It is also important to realize that since the observation period is not uniform, we need to exclude the data points that have a waiting time of longer than or equal to one day. As this is just an artifact of the monitoring campaign and has no use for our analysis in this paper. As we can see from Figure \ref{WT}, the red dotted line represents a waiting time value of one day and the blue dash-dotted line represents a waiting time of half a day. We exclude the data points beyond the blue dash dotted line because of the observational cadence of FAST.
        %Fig 1
        
\begin{figure}
            \centering
            \includegraphics[width=\columnwidth]{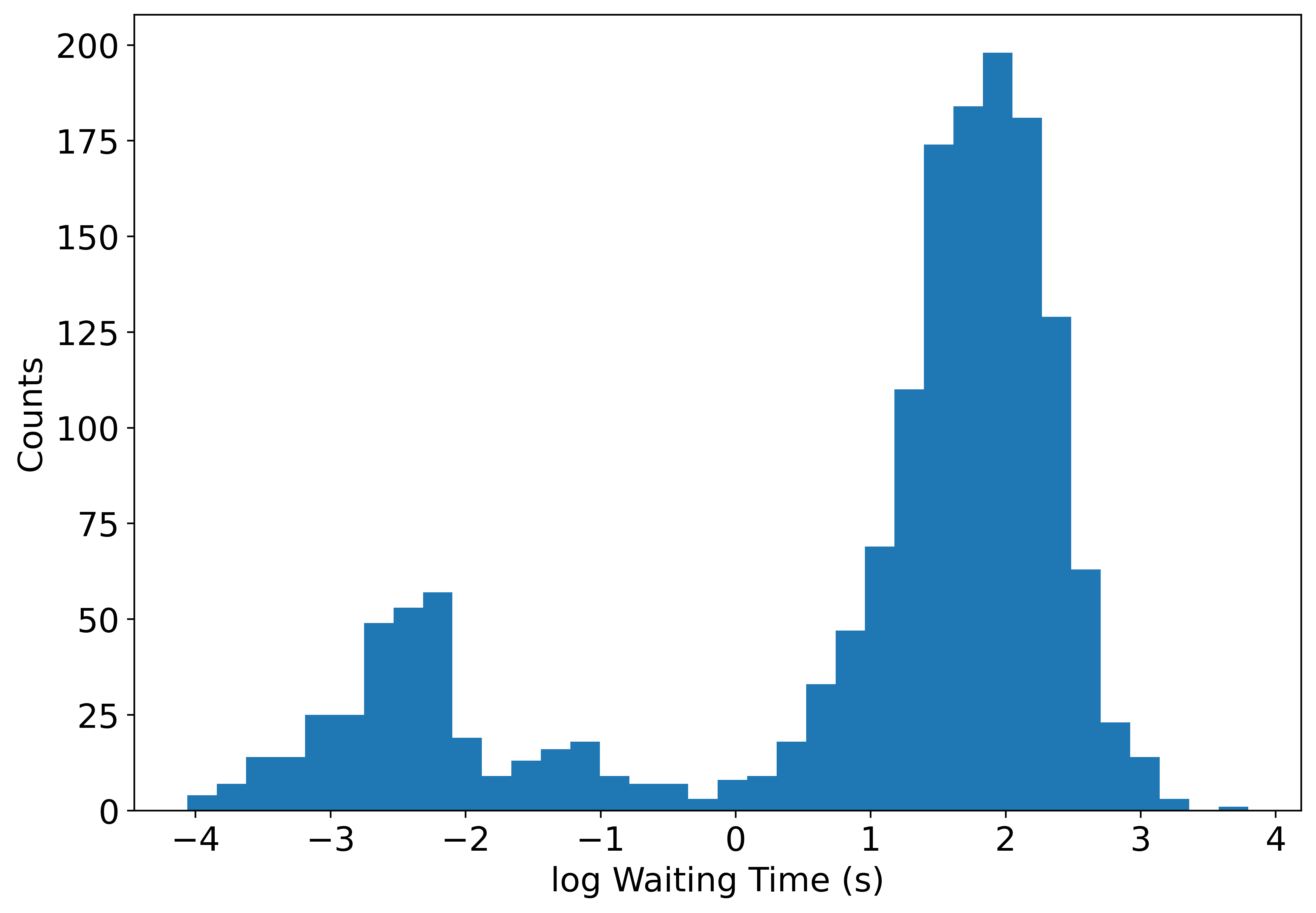}
            \caption{ \large   Bimodal waiting time distribution of FRB121102.}
            \label{WThisto}
        \end{figure}

        From the 1652 independent bursts reported by \citet[dataset]{2021Natur.598..267L}, after following the data selection method explained above, the number of independent burst samples we will use for the unsupervised machine learning is 1613. It is known that FRB121102 have a bimodal waiting time distribution \citep[e.g.,][]{2021Natur.598..267L} and as Figure \ref{WThisto} shows; the exclusion of 39 data points did not affect this property.
        %Fig 2
        
\section{UNSUPERVISED MACHINE LEARNING}\label{sec3}
    \subsection{Uniform Manifold Approximation and Projection (UMAP)}\label{sec3.1}
    
        \; \; Our data have 1613 rows and 7 columns after the preprocessing. Now, we employ a dimension reduction algorithm to visualize our data and conduct our unsupervised learning. This can be done by using Uniform Manifold Approximation and Projection (UMAP) (\citealt{2018arXiv180203426M}). Based on the ideas from topological data analysis and manifold learning techniques, UMAP finds a low dimensional representation of a given data by using basic Riemannian geometry to bring the data much closer to the underlying assumptions of the topological data analysis algorithm.
    
        UMAP has four basic hyperparameters which significantly affect the resulting embedding of the data. These hyperparameters are {\ttfamily min\_dist}, {\ttfamily metric}, {\ttfamily n\_components}, and {\ttfamily n\_neighbors}. It is important to realize that in this work we would like to uncover if there are underlying physical mechanisms or properties that make an FRB a repeater. Thus, we tune these parameters in a way that we will be able to notice a structure in the embedding.
    
        {\ttfamily min\_dist} restricts the clumping of the points in the resulting embedding. Providing the minimum distance apart that the points are allowed to be in the low dimensional representation. Meaning the closer the value of {\ttfamily min\_dist} is to zero the clumpier the embedding of the locally connected points. Since we would like to, as much as possible, see clustering in the embedding, we set {\ttfamily min\_dist = 0}.
    
        {\ttfamily metric} defines the way the distance between two points is measured. For our purpose of extracting intuitive realizations, we set {\ttfamily metric = euclidean}.
    
        {\ttfamily n\_components} is just the dimension of the resulting embedding. This hyperparameter helps us to visualize the data in the reduced dimension space of our own choosing and since we want to visualize our result in the two-dimensional (2D) plane, we set {\ttfamily n\_components = 2}.
    
        {\ttfamily n\_neighbors} constrains the size of the local neighborhood UMAP considers when estimating the manifold structure of the data. This hyperparameter focuses much more on the local structure when it has low values and on the global structure when it has a higher value. In our analysis, we considered a range of values for {\ttfamily n\_neighbors}. Namely, {\ttfamily n\_neighbors = 5,6,7,8,} and {\ttfamily 9} which is a reasonable range of values as these provide us with distinct clusters. However, for our interests, we will be only focusing on the clustering result of {\ttfamily n\_neighbors = 9}.
        \begin{figure}
            \centering
            \includegraphics[width=\columnwidth]{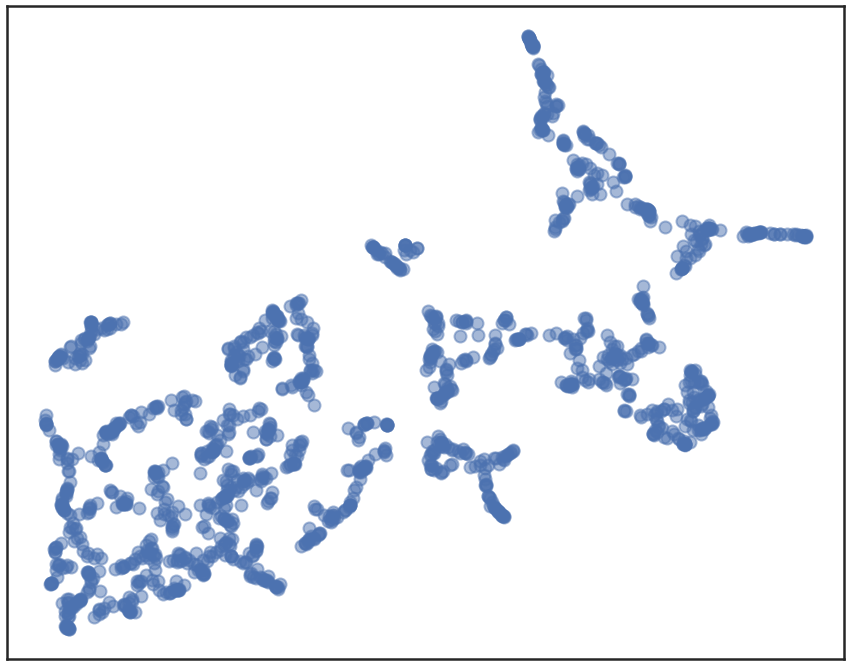}
            \caption{2-dimensional UMAP embedding for {\ttfamily n\_neighbors = 9}}
            \label{umap_nn9}
        \end{figure}
        
        As shown in Figure \ref{umap_nn9}, the UMAP embedding shows the lower left of the plot has a higher density of data points compared to the upper right of the plot. It is also evident that as the value of {\ttfamily n\_neighbors} increases more focus on the overall structure of the data is highlighted (See \ref{apnn5-9}). This is why the researchers only considered these values for the {\ttfamily n\_neighbors} because if we include higher values we will see embeddings that do not have a clear division or separation between data points and this will not prove useful to us in investigating the underlying mechanisms of the FRB.
    
    \subsection{Hierarchical Density-Based Spatial Clustering of Applications with Noise (HDBSCAN)}\label{sec3.2}
        
        \; \; In this paper, we use the clustering algorithm  developed by  Campello, Moulavi, and Sander \citep{campello2013}, namely Hierarchical Density-Based Spatial Clustering of Applications with Noise (HDBSCAN), to cluster the UMAP results we got from Section~\ref{sec3.1}. In HDBSCAN there are only four major parameters that can be tuned. These parameters are {\ttfamily min\_cluster\_size}, {\ttfamily min\_samples}, {\ttfamily cluster\_selection\_epsilon}, and {\ttfamily alpha}. Each of these have significant effects on the clustering result of the data points. 
        
        {\ttfamily min\_cluster\_size} affects the size of grouping that can be considered a cluster. The bigger the value of this parameter the lesser the number of the resulting clusters. For our purposes, we set {\ttfamily min\_cluster\_size = 200}.
        
        {\ttfamily min\_samples} controls the number of points that will be declared as noise. Considering points that are far from dense areas to be noise. The larger the value of this parameter the larger number of points that will be considered noise. In this study, we have {\ttfamily min\_samples = 10}.
        
        {\ttfamily cluster\_selection\_epsilon} when tuned allows micro-clusters in  high concentration region to be merged. Preventing clusters to be split up any further than the given threshold. Since we obtained our desired clustering we allowed this parameter to retain its default value, {\ttfamily cluster\_selection\_epsilon = 0}.
        
        Lastly {\ttfamily alpha}, this parameter is usually not changed or modified. However, if the clustering result from the changes made in {\ttfamily min\_samples} and {\ttfamily cluster\_selection\_epsilon} is unwieldy, one can then adjust {\ttfamily alpha} and make the clustering more conservative. Meaning more points will be considered as noise. Similar to {\ttfamily cluster\_selection\_epsilon}, the researchers allowed this parameter to retain its value, {\ttfamily alpha = 1}.
    
\begin{figure}
            \centering
            \includegraphics[width = \columnwidth]{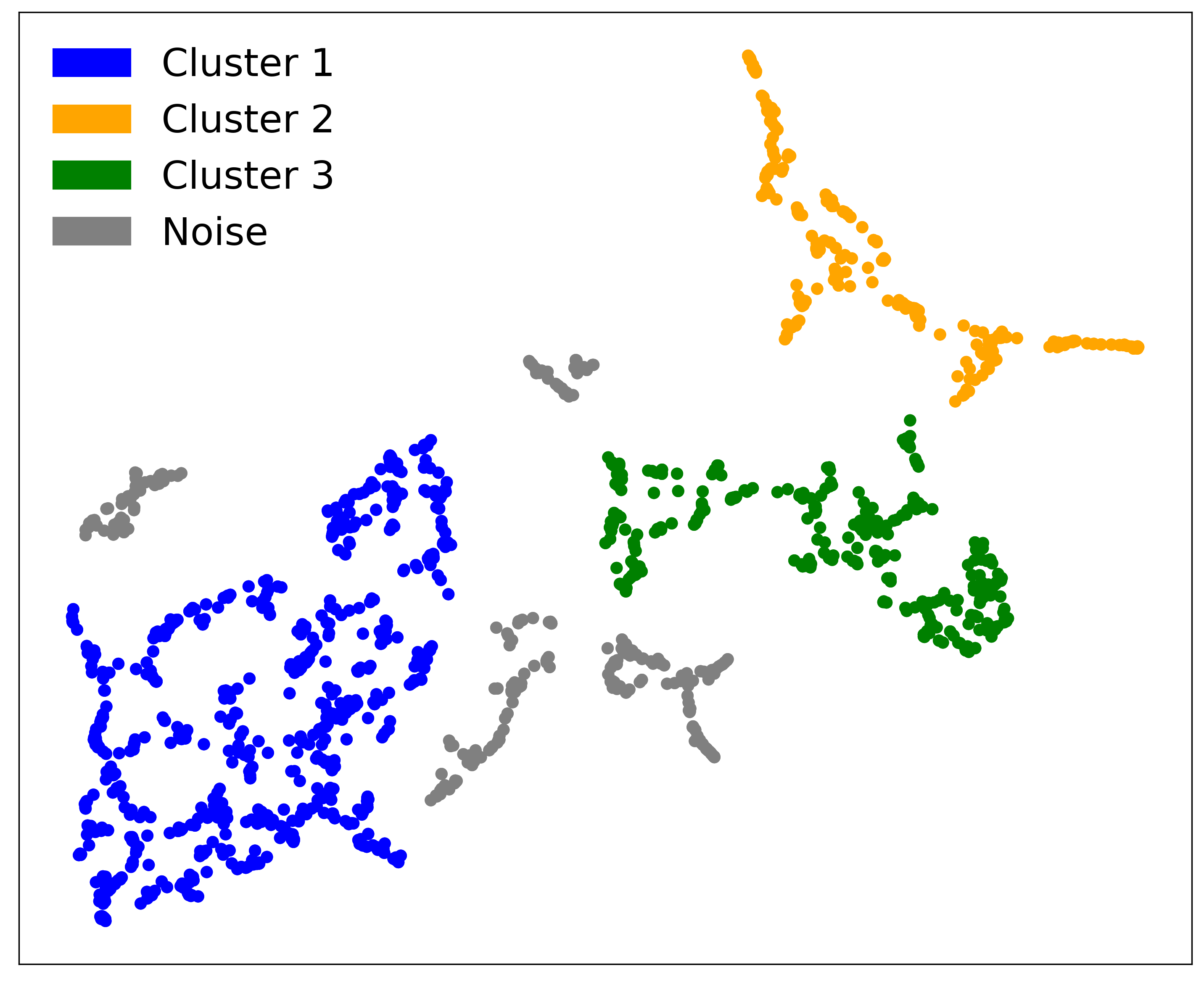}
            \caption{HDBSCAN Clustering result for {\ttfamily n\_neighbors = 9}}
            \label{fig4}
        \end{figure}

        After using UMAP to find a low dimensional representation of our data, we now use HDBSCAN to cluster this embedding.Looking at Fig. \ref{fig4} and the figures of \ref{aphdbscan}, it is evident that the HDBSCAN Clustering results for {\ttfamily n\_neighbors = 7}, {\ttfamily n\_neighbors = 8}, and {\ttfamily n\_neighbors = 9} have three clusters with noise while {\ttfamily n\_neighbors = 5} have three clusters without noise, and {\ttfamily n\_neighbors = 6}  have two clusters without noise. Since similar cluster number of different {\ttfamily n\_neighbors} value are not identical to each other we introduce the following nomenclature: nn[{\ttfamily n\_neighbors}value].c[clusternumber]. As an example, nn5.c1 refers to the  cluster 1 of {\ttfamily n\_neighbors = 5}. For the Noise clusters we will use a "N" in place of the [clusternumber], e.g., nn5.cN.

\section{RESULTS}\label{sec4}
\begin{figure*}
                \centering
                \begin{multicols}{2}
                    \subcaptionbox{\label{bw9}}{\includegraphics[width=\linewidth, height=0.8\columnwidth]{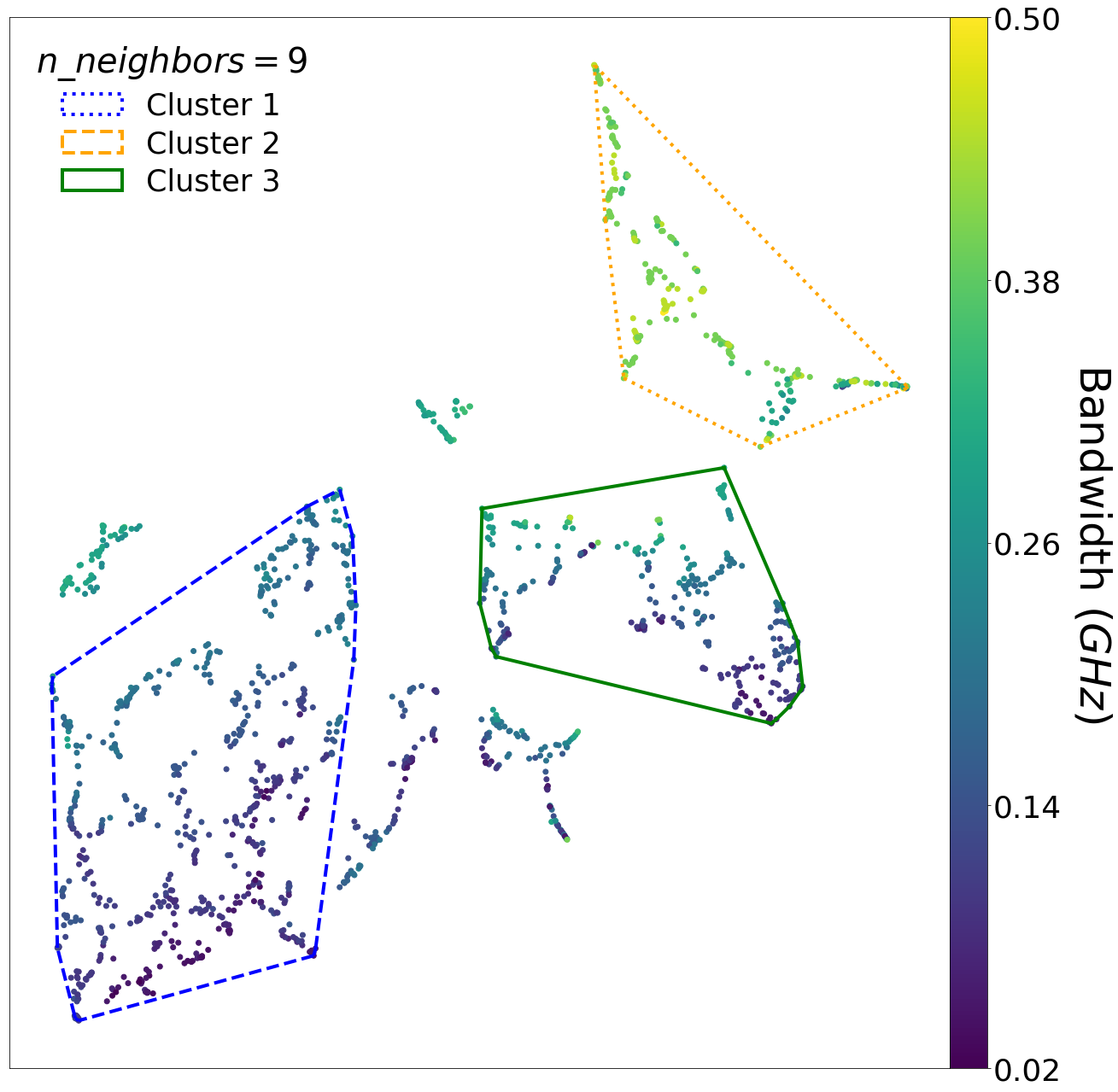}}\par 
                    \subcaptionbox{\label{e9}}{\includegraphics[width=\linewidth, height=0.8\columnwidth]{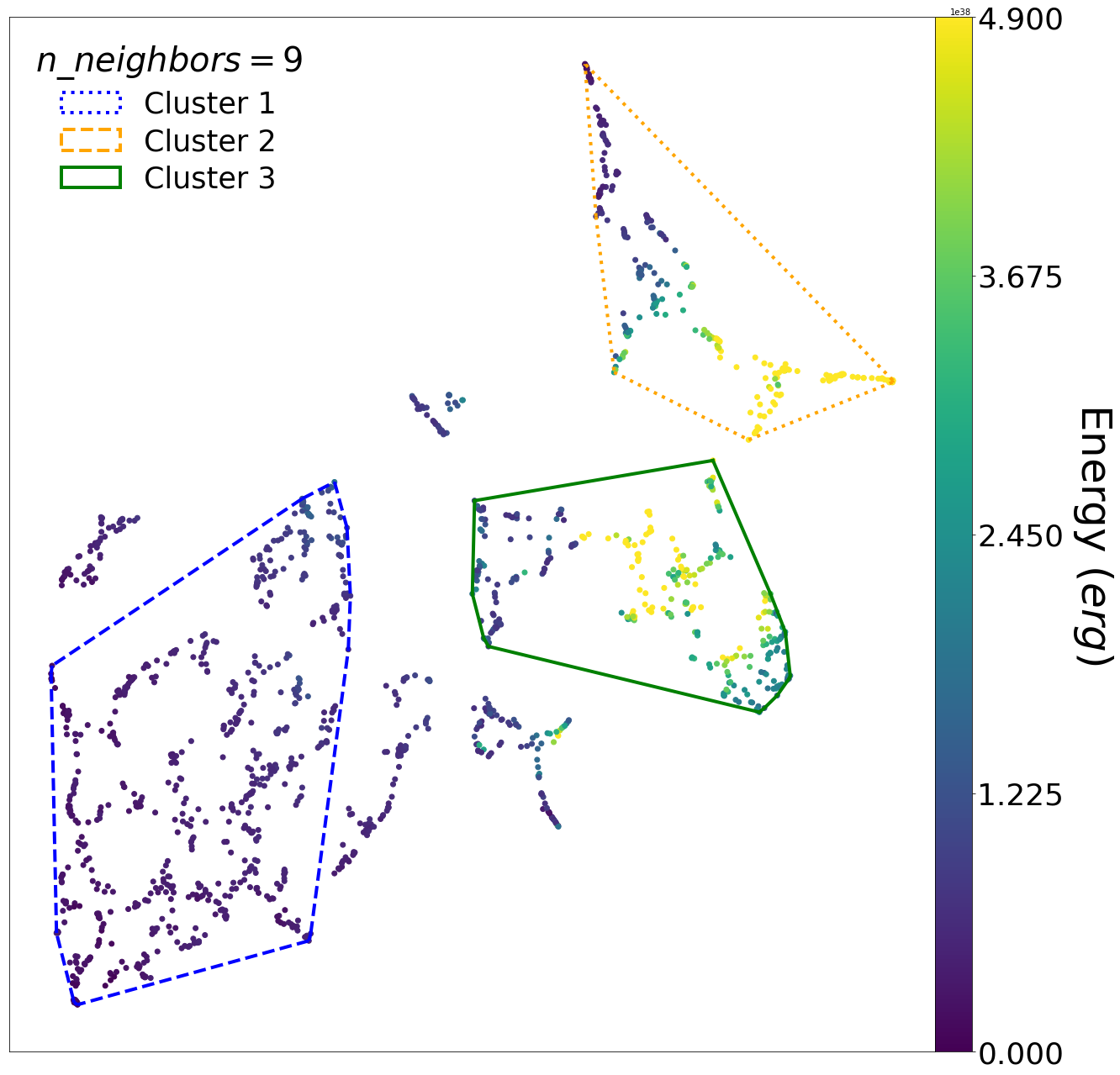}}\par 
                \end{multicols}
                \begin{multicols}{2}
                    \subcaptionbox{\label{fl9}}{\includegraphics[width=\linewidth, height=0.8\columnwidth]{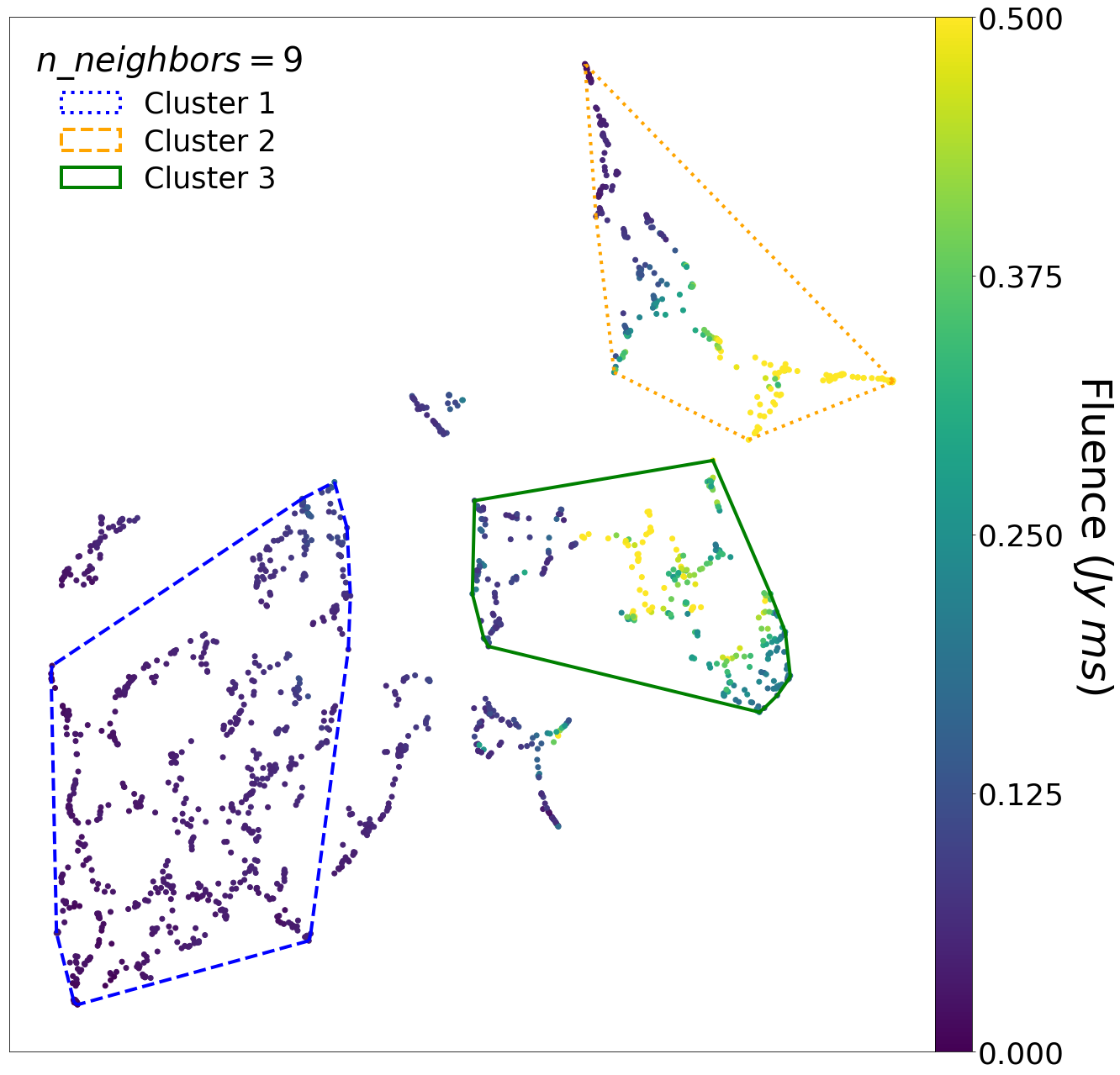}}\par
                    \subcaptionbox{\label{pf9}}{\includegraphics[width=\linewidth, height=0.8\columnwidth]{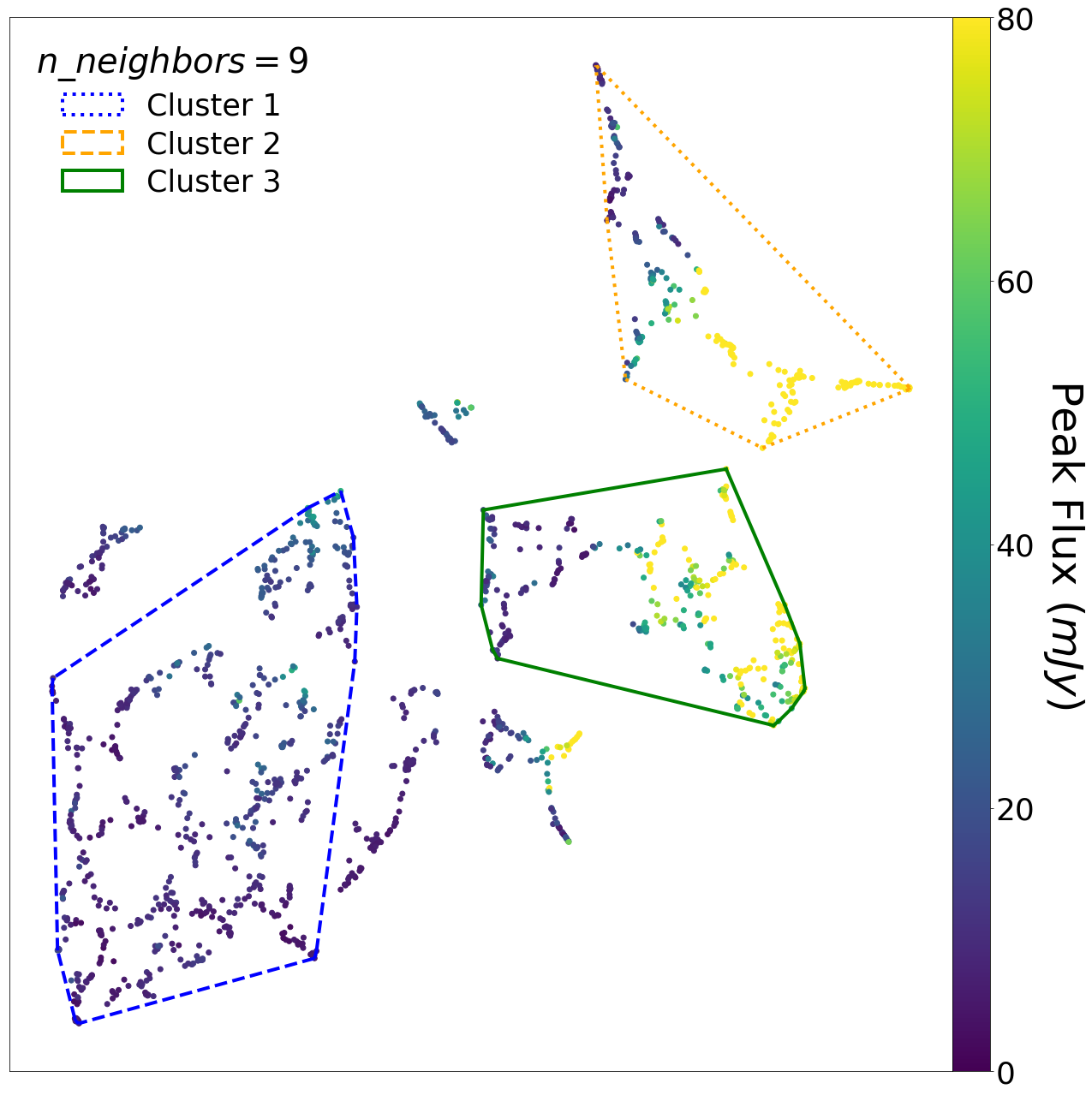}}\par
                \end{multicols}
                \begin{multicols}{2}
                    \subcaptionbox{\label{tw9}}{\includegraphics[width=\linewidth, height=0.8\columnwidth]{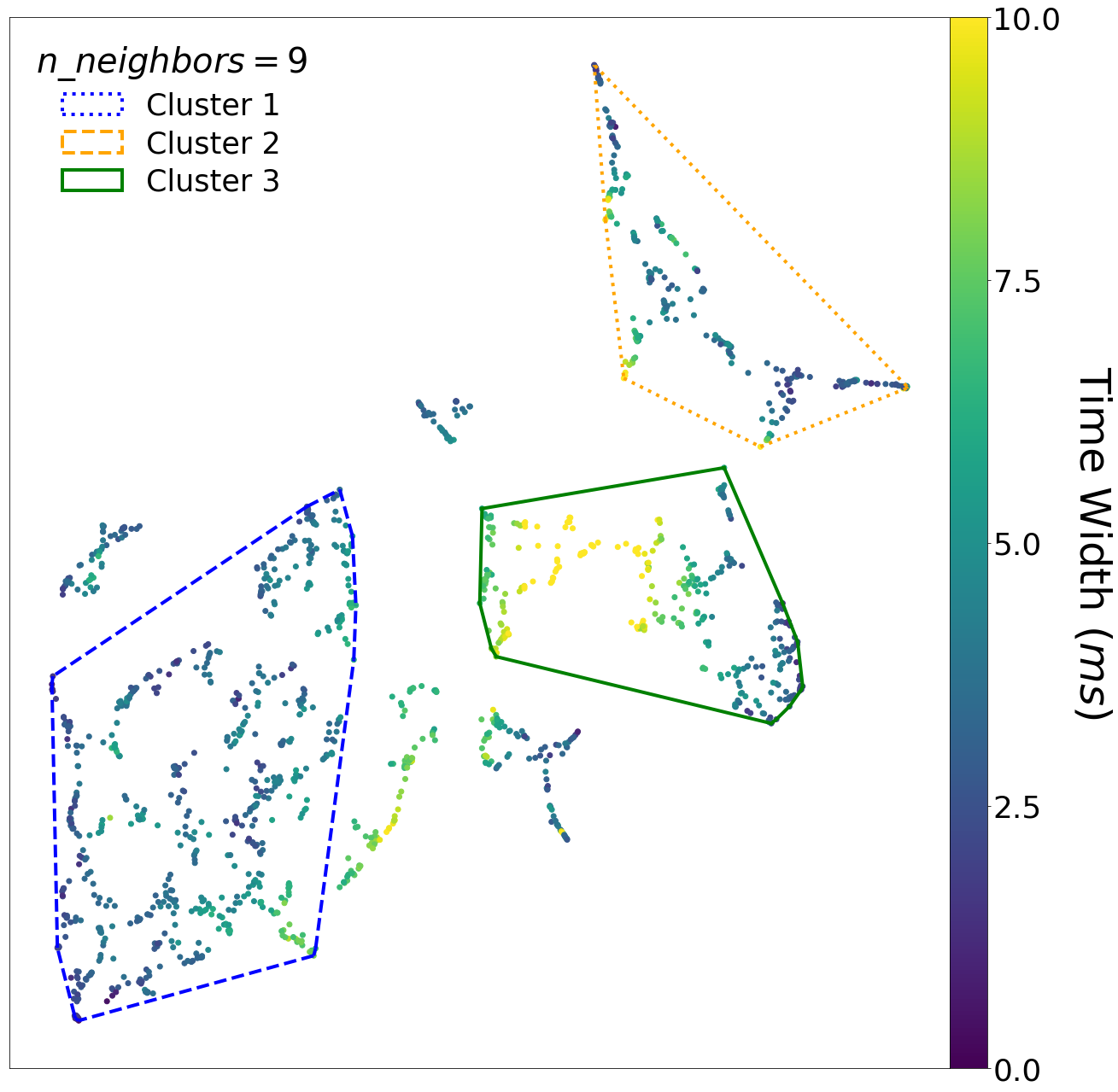}}\par
                    \subcaptionbox{\label{wt9}}{\includegraphics[width=\linewidth, height=0.8\columnwidth]{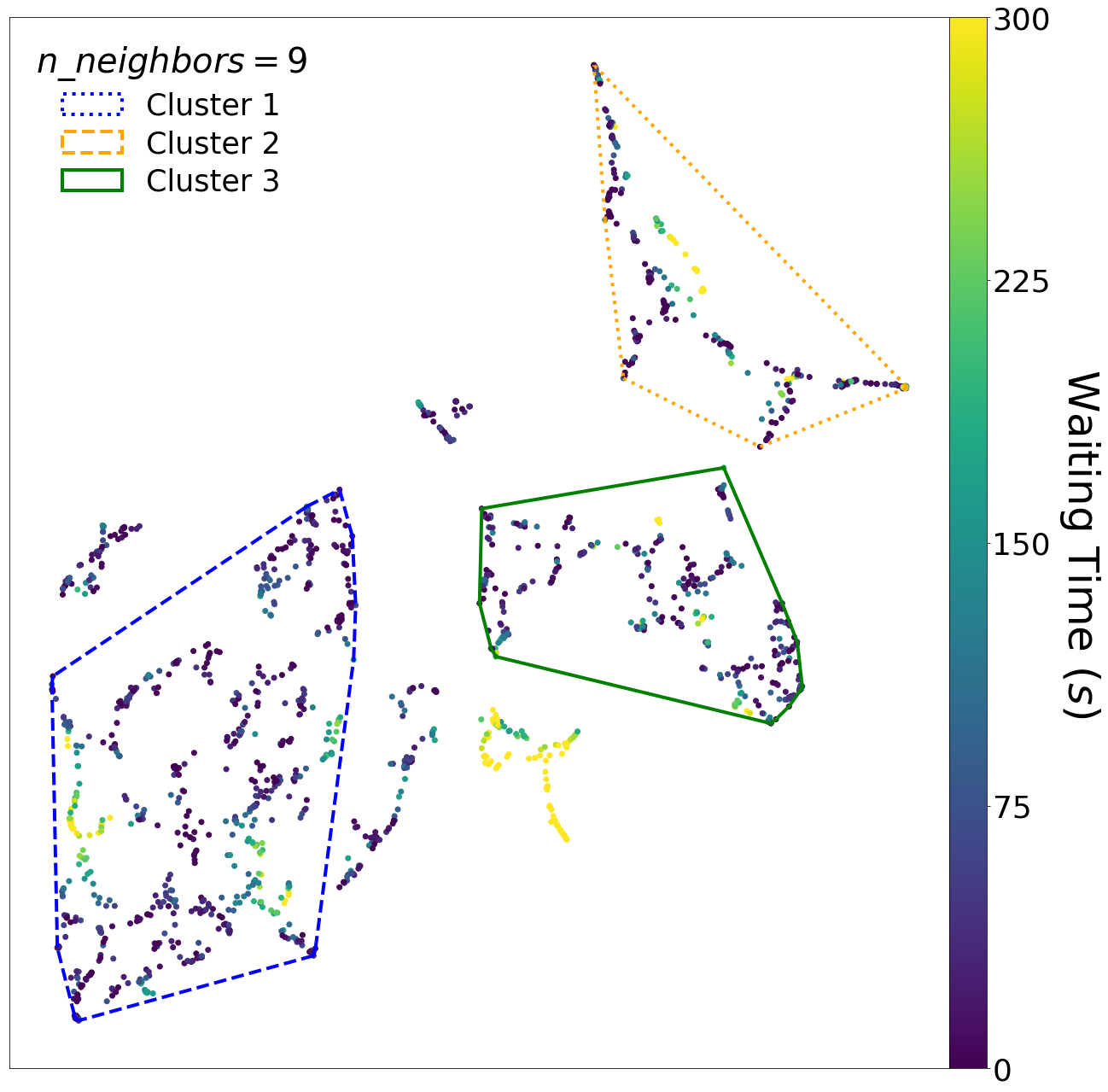}}\par
                \end{multicols}
                \caption{ \large Parameter colouring of the clustering result for {\ttfamily n\_neighbors = 9}. The data which are not surrounded by lines correspond to Noise clusters.}
            \label{nn9parcol}
            \end{figure*}

    \; \; Since this paper is focused on determining what physical mechanisms or properties underlie repeating FRBs. It is important to see how the parameters of our data behave or manifest in our clustering. This can be achieved by colouring the clustering result with the parameters. Doing so will allow us to recognize trends, patterns, and properties among the clusters that can be helpful in investigating the repeating FRBs. Since we have six parameters, we have six plots with colouring based on a single parameter. In Figure \ref{nn9parcol}, we have the Bandwidth colouring (GHz) \ref{bw9}, the Energy colouring (erg) \ref{e9}, the Fluence colouring (Jy ms) \ref{fl9}, the Peak Flux colouring (mJy) \ref{pf9}, the Time Width colouring (ms) \ref{tw9}, and the Waiting Time colouring (s) \ref{wt9}. The other parameter colouring results that we have considered in our analyses can be found at Figures \ref{apbwparcol} -- \ref{apwtparcol}.
    
    From these, we can see that parameter coloring enables us to easily characterize each cluster. From Figure \ref{bw9} (Bandwidth), nn9.c1, nn9.c3, and nn9.cN are narrowband and nn9.c2 is broadband. Figures \ref{e9} -- \ref{pf9} (Energy, Fluence, and Peak Flux) all exhibit a similar trend for their clusters. nn9.c1 and nn9.cN both exhibits low energy, low fluence, and low peak flux. Majority of nn9.c3 have high energy, high fluence, and high peak flux while nn9.c2 have diverse energy, diverse fluence, and diverse peak flux.  Figure \ref{tw9} (Time Width) exhibits an interesting coloring among the clustering results. Points with longer duration tend to be located near the center of the clustering, and points with shorter duration tend to be located away from the center of the embedding. Lastly, Figure \ref{wt9} (Waiting Time) tend to cluster points with long waiting times into small clusters and scattered among the bigger clusters of short waiting time points. Showing that regardless of cluster, FRB 121102 can be described to have a very short waiting time. In addition, all of these clusters do have a bimodal distribution for the waiting time. It is also important to realize that the significance of these properties were also supported by the histograms which we included in Appendix~\ref{histogrammers}. Inspecting the histograms, especially the Bandwidth \ref{apbwparcolhisto}, Energy \ref{apeparcolhisto}, Fluence \ref{apflparcolhisto}, and Peak Flux \ref{appfparcolhisto} histograms, we see that the clusters were distinct from each other, showing different distributions for each cluster. One of the most notable is the bandwidth distributions, where all clusters show different distributions. Thus, supporting the idea that the resulting clusters are significantly different.
    
    We can then summarize the characterization of each cluster regardless of {\ttfamily n\_neighbors} value as shown in Table~\ref{tab1}. It is important to keep in mind that the qualitative description of the clusters is based relative to the range of values for each parameter on a given cluster. As shown in Table~\ref{tab1}, each cluster contains a unique set of properties that remains constant throughout the change of {\ttfamily n\_neighbors} value which highlights the difference in physics among clusters. We may refer to these properties as "invariant" cluster properties. Identifying these invariant cluster properties is essential in describing the underlying physical mechanisms that we might discover based on the number of clusters that we found. Thus, supporting the idea that the resulting clusters are significant. 

        \begin{table}
            \centering
            \begin{tabular}{|c|c|c|c|c|}
                    \hline
                 Invariant &  &  & &   \\
                 Cluster & Cluster 1 & Cluster 2 & Cluster 3 & Noise  \\
                 Properties &  &  &  &   \\
                 \hline
                 Bandwidth & Narrowband & Broadband & Narrowband & Narrowband  \\
                 \hline
                 Peak Flux & Low & Diverse & Diverse & Diverse  \\
                 \hline
                 Fluence & Diverse & Diverse & Diverse & Low  \\
                 \hline
                 Energy & Low & Diverse & High & Low  \\
                 \hline
                 Time width & Short & Short & Short & Short  \\
                 \hline 
                 Waiting time & Very short & Very short & Very short & Very short \\ 
                 \hline
            \end{tabular}
            \caption{ \large Cluster properties that remain constant regardless of {\ttfamily n\_neighbors} value. The qualitative description of the clusters is based relative to the range of values for each parameter on a given cluster}
            \label{tab1}
        \end{table}

\section{DISCUSSION}\label{sec5}
    
    \; \;The model used in this study is similar to what was employed in the work of \citealt{chen2022uncloaking}. UMAP was used to find the low dimensional representation of the data and then HDBSCAN was used to cluster the data points. Several key differences can be pointed out between our work and \citealt{chen2022uncloaking}. First, \citealt{chen2022uncloaking} used both non-repeater and repeater sources in their study while we used a single repeating FRB which is FRB121102. Second, one of the main goals of their work is to evaluate the assumption that non-repeating FRBs are contaminated by repeating FRBs while this work focuses on characterizing or identifying the underlying properties of FRB121102 to further understand repeating FRBs. Lastly, the work of \citealt{chen2022uncloaking} provided a new way to classify repeating and non-repeating FRBs while this study aims to provide  a classification of repeating FRBs from FRB121102. Nevertheless, this study also introduced additional analysis which helped us to qualitatively characterize FRB121102, such as parameter colouring, identification of invariant cluster properties, and the cluster membership change of the data points/FRBs that will be discussed in the succeeding subsection.
    
    \subsection{Cluster membership change} \label{clusterchange}
    
    \; \; In Section~\ref{sec4}, we found the invariant cluster properties (see Table~\ref{tab1}) regardless of the {\ttfamily n\_neighbors} values. However, it is easy to see that the number of clusters as the {\ttfamily n\_neighbors} value change did not remain the same. Four clusters (including the Noise cluster) were found for {\ttfamily n\_neighbors = 7,8,9}, three clusters were found for {\ttfamily n\_neighbors = 5}, and two clusters were found for {\ttfamily n\_neighbors = 6}. This suggests that the FRB clusters of {\ttfamily n\_neighbors = 5,6} might be shared in more than one cluster of {\ttfamily n\_neighbors = 7,8,9}. Therefore, we investigate the change in the cluster membership of the FRBs as the {\ttfamily n\_neighbors} values varies to see whether this is true. This can be done using an alluvial diagram, as shown below in Figure~\ref{allu}.
    
    \begin{figure*}
        \centering
        \includegraphics[width=\textwidth]{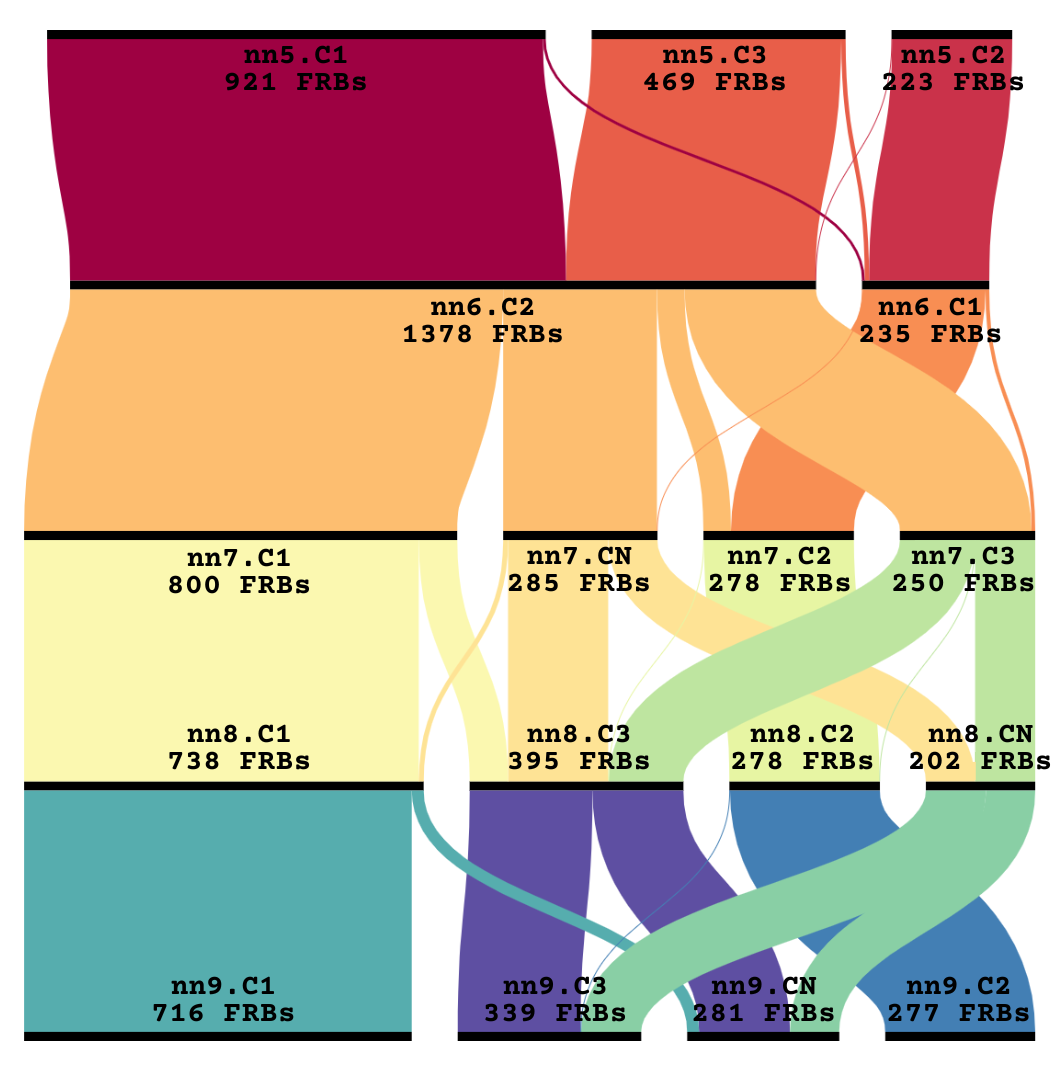}
        \caption{ \large Cluster membership change of each FRB as {\ttfamily n\_neighbors} is varied.}
        \label{allu}
    \end{figure*}
    
    Looking at Figure~\ref{allu}, the axes (oriented vertically) represents the {\ttfamily n\_neighbors} values {\ttfamily 5,6,7,8,} and {\ttfamily 9}, respectively. The stratum contained in each axis represents the clusters we found using HDBSCAN which are Cluster 1, Cluster 2, Cluster 3, and Noise. The flow between axes connecting two strata represent the change of the cluster membership of those data points across an {\ttfamily n\_neighbors} value change. An alluvium in our diagram consists of four flows connecting different strata or clusters on different axes or clustering. From the diagram we find the following observations; (i) Majority of the data points tend to retain their cluster membership over {\ttfamily n\_neighbors} value change, (ii) The number of clusters increases as the {\ttfamily n\_neighbors} value increases, and (iii) If we only consider "thick flows" (i.e., flows that contains majority of the data points) to be significant, we can divide alluvial diagram into two alluviums. One of which is the alluvium connecting the clusters nn5.c2, nn6.c1, nn7.c2, nn8.c2, and nn9.c2.
    
    The result (i) we have, based on the diagram, shows that the clustering we found on the data is because of the data itself and not an effect of the clustering algorithm. This then lends to a more physical interpretation and characterization of the clusters.  Result (ii) compared to (i) can also be attributed to how the clustering algorithm works. Particularly, the hyperparameters {\ttfamily n\_neighbors} from UMAP, {\ttfamily min\_cluster\_size}, and {\ttfamily min\_samples} from HDBSCAN affect the resulting embedding and clustering of the data points. But to this effect, looking at the alluvium from nn5.c2 to nn9.c2 we can see that nn5.c2 evidently bears greater and greater significance as we increase the {\ttfamily n\_neighbors} value. Showing that there are data points that are once grouped into nn5.c2 are now considered a major part of nn9.c2. This observation is also true for alluviums with significant flows.  Lastly, result (iii) implies that the two significant clusters (nn6.c2 and nn6.c1) are really made up of four clusters (nn9.c1, nn9.c2, nn9.c3, nn9.cN) and one of these two major clusters, nn6.c2, can be split up further into three clusters, namely nn9.c1, nn9.cN, and nn9.cN.

    Since the data have noise, it is to be expected that our unsupervised learning might produce non-physical clusters. Regarding this matter, we can use the alluvial diagram which keeps track of the cluster membership change of each FRB as an additional cross-check. Enabling us to look past the cluster membership assigned to each FRB by considering thick flows to be significant and see how each cluster evolves throughout {\ttfamily n\_neighbors} change. This eliminates complete dependency of our final results on a certain set of hyperparameter values, showing that certain groups of FRB/data points remain as a group all throughout {\ttfamily n\_neighbors} change. Also, this supports the idea of certain cluster properties being carried over from one cluster with different/similar {\ttfamily n\_neighbors} value into another cluster with different/similar {\ttfamily n\_neighbors} value. This entails that the clustering result is based on the difference in physics which is also supported by the one-dimensional histograms (see Section \ref{histogrammers}).
   
    \subsection{Comparison with other results} \label{comparison}
    \; \; In relation to the number of clusters we have from the HDBSCAN clustering, there are similar results from literature that found the same number of clusters as our results for the {\ttfamily n\_neighbors = 6} clustering. Using the same dataset from the Supplementary Table 1 of \citealt{2021Natur.598..267L}, \citealt{Xiao&Dai2022} found two clusters by assigning a  critical brightness temperature ($T_{B,cri}$) of $10^{33}$ K. In their work, they used the brightness temperature of the FRBs as a criterion to cluster the bursts of FRB121102 because it directly relates to the radiation mechanism of FRBs. These clusters contain bursts depending on whether the bursts have a $T_B$ value greater than or equal or less than the $T_{B,cri}$. "Classical" bursts are bursts that have $T_B \geq T_{B,cri}$ while the "Atypical" bursts are bursts that have $T_B < T_{B,cri}$. \citealt{Xiao&Dai2022} also found that the 76 "Classical" bursts have a tight  width - fluence (T - $\mathcal{F}_\nu$) relation described by $\log$(T) $= 0.306\cdot\log(\mathcal{F}_\nu) + 0.399$ with a correlation coefficient of $r = 0.936$. Given that this relation does not hold true for the case of "Atypical" bursts and the total bursts of FRB121102, it leads \citealt{Xiao&Dai2022} to suggest that these "Atypical" bursts may be further grouped into several subtypes consisting of different radio transient types. 
    
\begin{figure*}
                \centering
                \begin{multicols}{2}
                    \subcaptionbox{\label{xd5}}{\includegraphics[width=0.95\linewidth, height=0.8\columnwidth]{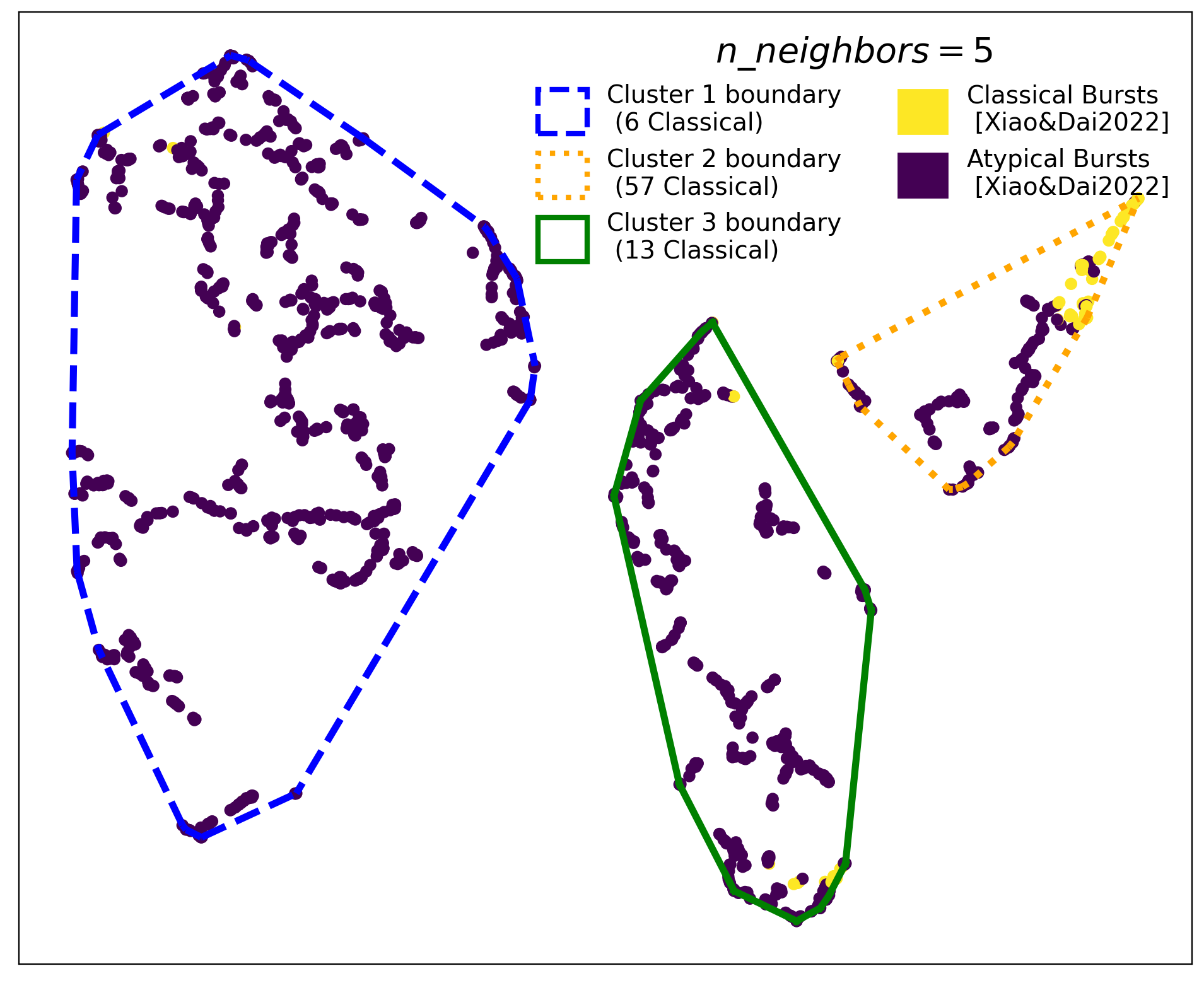}}\par 
                    \subcaptionbox{\label{xd6}}{\includegraphics[width=0.95\linewidth, height=0.8\columnwidth]{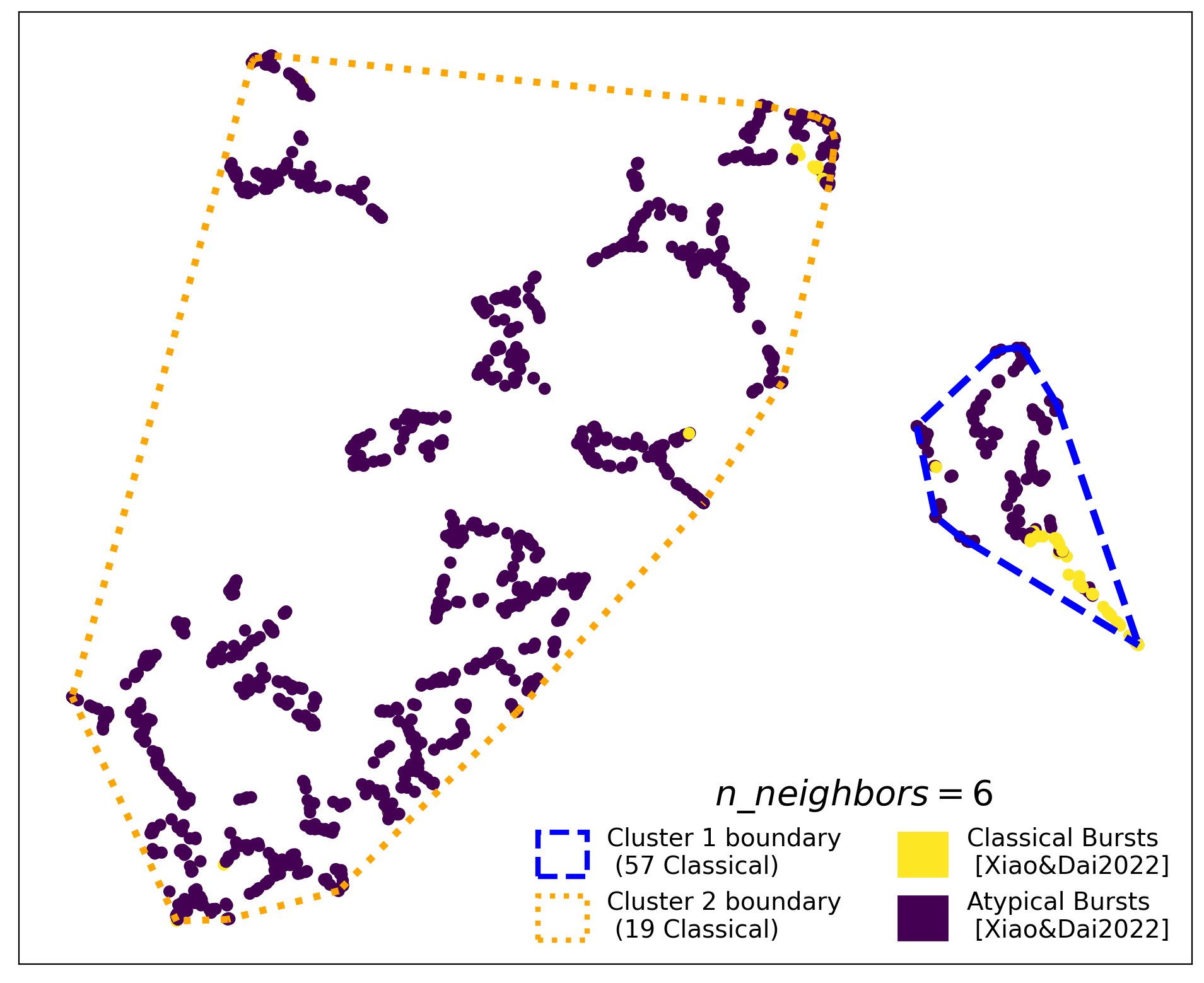}}\par 
                \end{multicols}
                \begin{multicols}{2}
                    \subcaptionbox{\label{xd7}}{\includegraphics[width=0.95\linewidth, height=0.8\columnwidth]{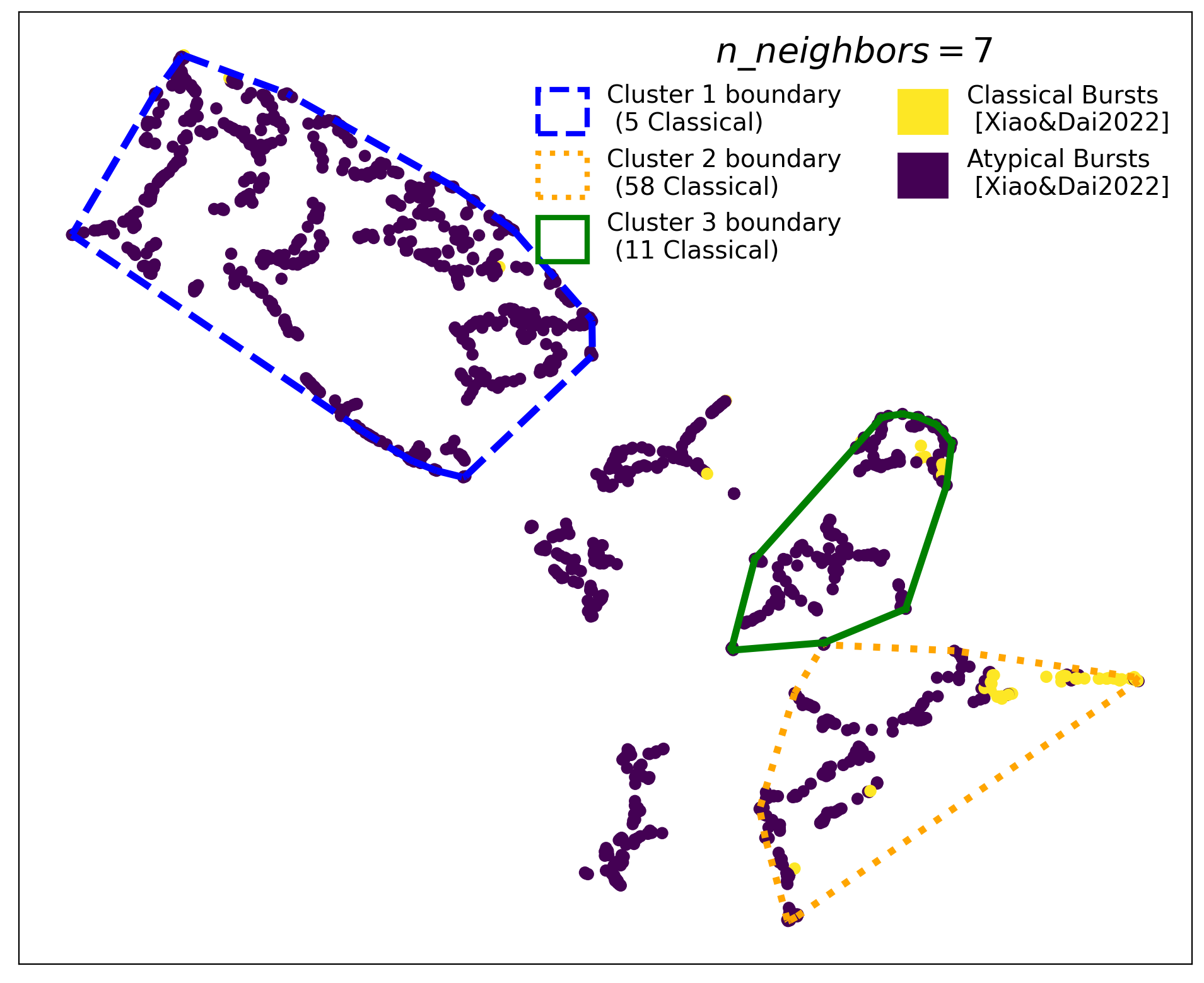}}\par
                    \subcaptionbox{\label{xd8}}{\includegraphics[width=0.95\linewidth, height=0.8\columnwidth]{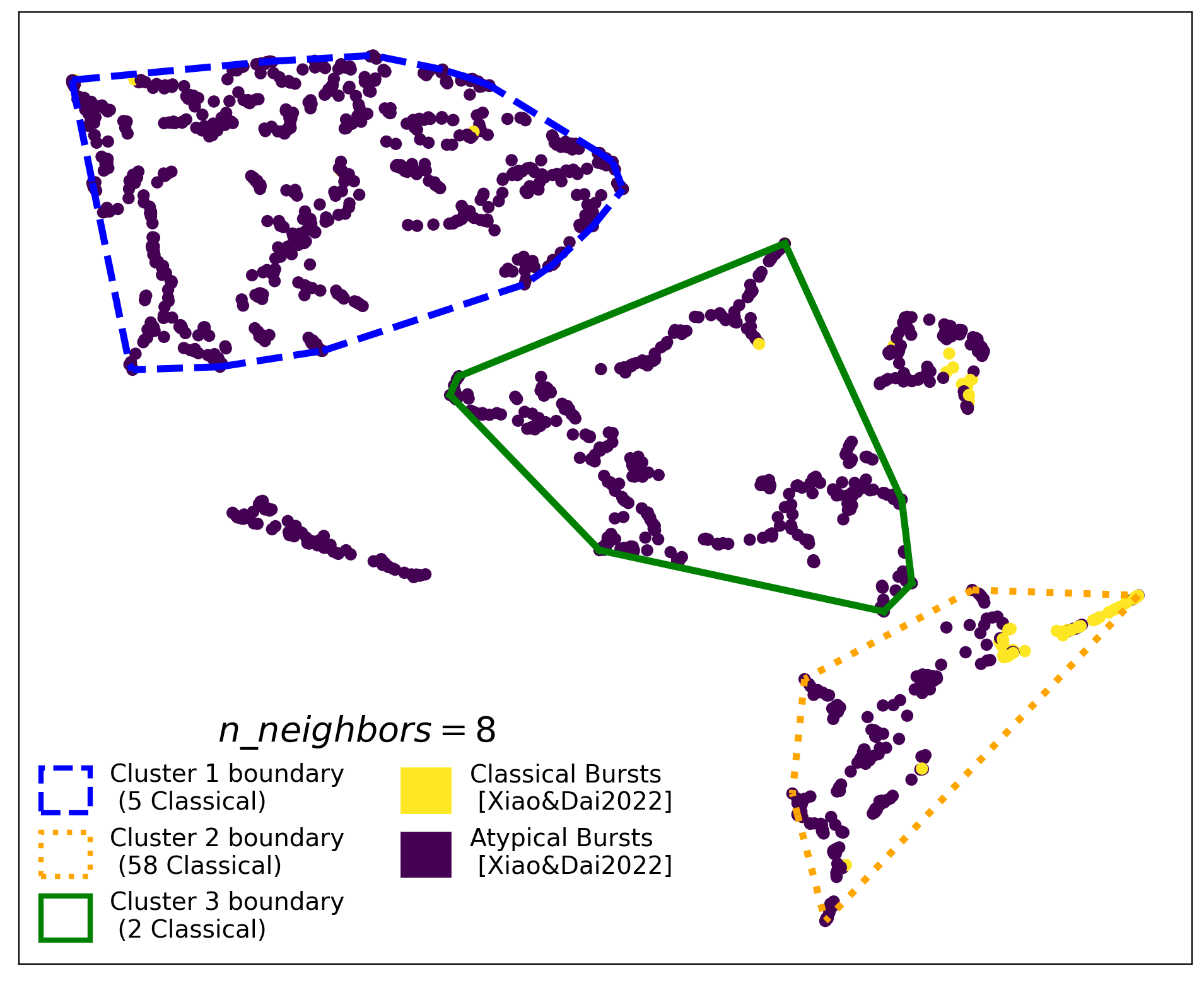}}\par
                \end{multicols}
                \begin{multicols}{2}
                    \subcaptionbox{\label{xd9}}{\includegraphics[width=0.95\linewidth, height=0.8\columnwidth]{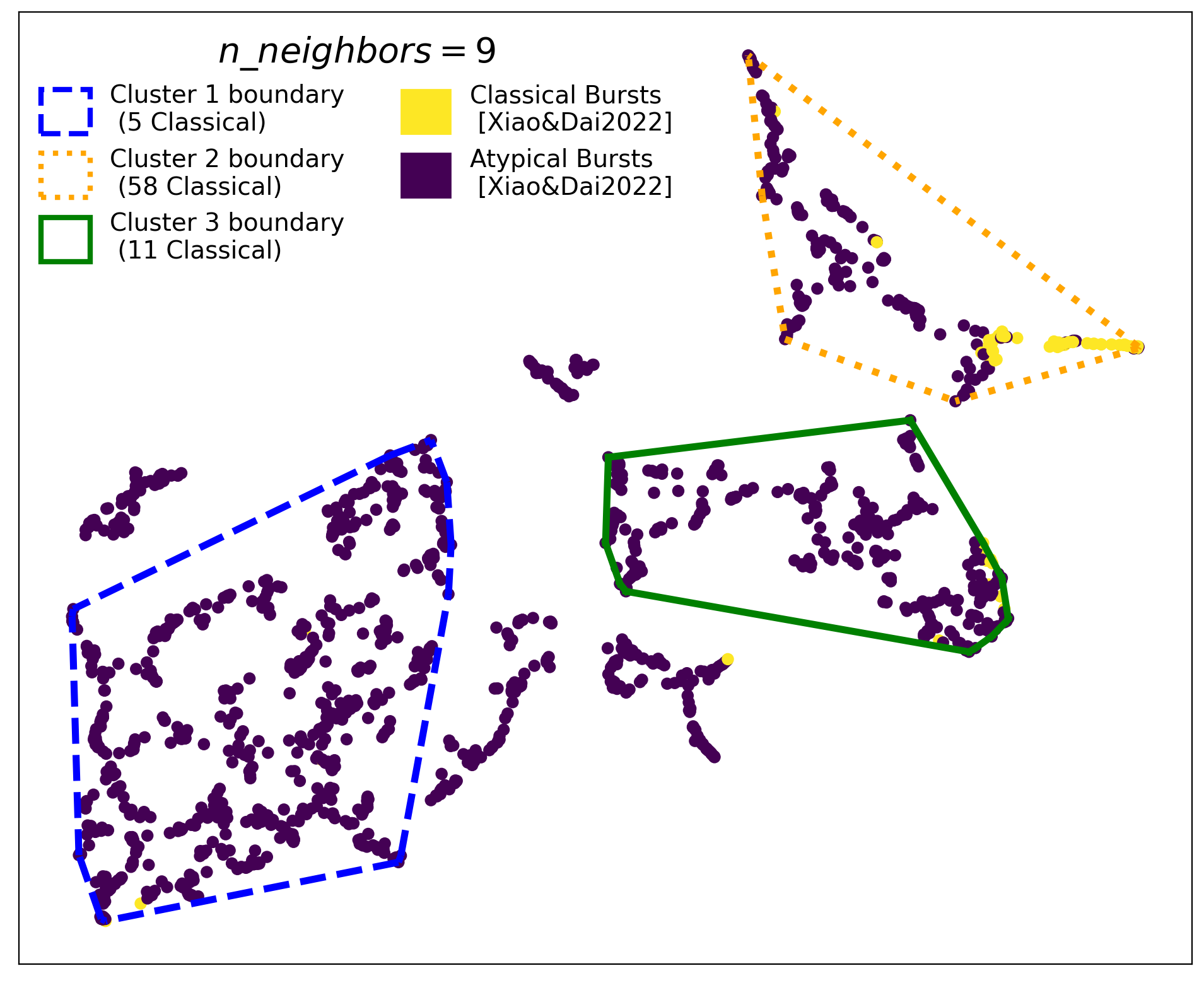}}\par
                \end{multicols}
                \caption{\large "Classical" and "Atypical" bursts in UMAP results for (\ref{xd5}) {\ttfamily n\_neighbors = 5}, (\ref{xd6}) {\ttfamily n\_neighbors = 6}, (\ref{xd7}) {\ttfamily n\_neighbors = 7}, (\ref{xd8}) {\ttfamily n\_neighbors = 8}, and (\ref{xd9}) {\ttfamily n\_neighbors = 9}.}
            \label{xdclustering}
    \end{figure*}

    Marking the "Classical" and "Atypical" bursts of \citealt{Xiao&Dai2022} in our UMAP results and checking their cluster membership based on the clusters we identified, we have Figure~\ref{xdclustering}. In the figure, the clustering based on HDBSCAN is represented by a convex hull boundary and the number of "Classical" bursts within a cluster is indicated within the parenthesis. Since there are $76$ "Classical" bursts and not all "Classical" bursts are members of clusters 1, 2, and 3. Then, these other "Classical" bursts are members of the Noise cluster. Tracking the change of "Classical" bursts membership using Figure~\ref{allu}. We found that the change in cluster membership of the majority ($\geq 75\% = 57$) of "Classical" bursts follow the alluvium connecting nn5.c2 and nn9.c2. Suggesting that nn6.c1 corresponds to the "Classical" bursts while nn6.c2 corresponds to the "Atypical" bursts. 
    
    However, compared to the work of \citealt{Xiao&Dai2022} which only used a single parameter to group or cluster the FRBs of FRB121102. This study used seven of the parameters from the Supplementary Table 1 of \citealt{2021Natur.598..267L} to cluster the FRBs giving a more robust clustering result. Nevertheless, the agreement between the clustering results implies that there is an existing structure in the FRBs of FRB121102 regardless of clustering method that is used. 
    
    The work of \citealt{Chaikova+2022} also agrees with the findings of \citealt{Xiao&Dai2022} and this study in terms of the number of clusters. Using the version of the CHIME/FRB catalog data \citep{amiri+2018} which contains $536$ events (repeaters and non-repeaters). \citealt{Chaikova+2022}, also found two clusters with significant differences in their morphology by using the {\ttfamily frbmclust} software \citep{Chaikova2022frbmclust}. The first cluster is described to have broad widths, low flux, several peaks per event (13.4\% of events have >1 peak), mean boxcar width = 24.79 ms, median flux = 0.56 Jy, and has 28 repeaters. The second cluster is described to have narrow widths, high flux, single peaks per event (6.3\% of events have >1 peak), mean boxcar width = 4.12 ms, median flux = 1.08 Jy, and has 33 repeaters. From these descriptions of each cluster, \citealt{Chaikova+2022} concluded that what they identified as second cluster corresponds to the "Classical" bursts of \citealt{Xiao&Dai2022} and what they identified as first cluster resembles the findings of \citealt{Xiao&Dai2022} for the broad population of sources. This result of \citealt{Chaikova+2022} then suggests that the clustering of FRB121102 into two according to \citealt{Xiao&Dai2022} must also be the same to the $536$ events (repeaters and non-repeaters) they studied.
    
    In addition to these, \citealt{2021Natur.598..267L} also suggested that the bimodality of the energy distribution points to more than one emission mechanism or emission site or beam shape. Whereas \citealt{Xiao&Dai2022} pointed out that the bimodal burst energy distribution found by \citealt{2021Natur.598..267L} already hints (if not indicate) that there are two subtypes of FRBs and the subsequent work of \citealt{Chaikova+2022} supports this result and along with ours. Thus, we find that the number of significant clusters we found for {\ttfamily n\_neighbors = 6} corresponds to clusters found by \citealt{Xiao&Dai2022}, \citealt{Chaikova+2022}, and \citealt{2021Natur.598..267L}.
    
    Since we have established that the clusters of {\ttfamily n\_neighbors = 6} is consistent with the results of \citealt{Xiao&Dai2022}, \citealt{Chaikova+2022}, and \citealt{2021Natur.598..267L} and that the nn6.c2 based on Figure~\ref{allu} is really composed of three clusters. It then follows that what we found as nn9.c1, nn9.c3, and nn9.cN must correspond to the "Atypical" bursts described in the work of \citealt{Xiao&Dai2022}. Showing that these "Atypical" bursts can be further split into three clusters with distinct properties (see Table~\ref{tab1}). Therefore, we can describe the properties of these "Atypical" bursts based on the properties of nn9.c1, nn9.c3, and nn9.cN. However, since the primary focus of our work revolves around classifying repeating FRBs, discussions of the physical mechanisms of each cluster will be left for future theoretical works. 

    \subsection{Clustering performance} \label{clusteringperformance}
    \; \; In evaluating the agreement or similarity of the clustering results presented in this paper. A clustering performance metric is employed namely the Rand Index \citep{Hubert1985} and its corrected-for-chance version Adjusted Rand index \citep{Steinley2004}. This metric and its adjusted form will give us as a Rand score and Adjusted Rand score for each pair of clustering result we compare. Wherein a high score suggests that the two clustering result are in very good agreement. In addition, we also considered the Rand score and Adjusted score for the case where the Cluster 3 and Noise of each clustering result is merged. The reason for this is rooted in the results presented in Figures~\ref{apbwparcolhisto} -- \ref{apwtparcolhisto} where the existence of Cluster 3 is not a “firm detection” but an “indicator” of another cluster; one that is an intermediate one between Cluster 1 and Cluster 2.

    \begin{figure}
        \centering
        \begin{subfigure}[b]{0.55\textwidth}
        \includegraphics[width=0.9\linewidth]{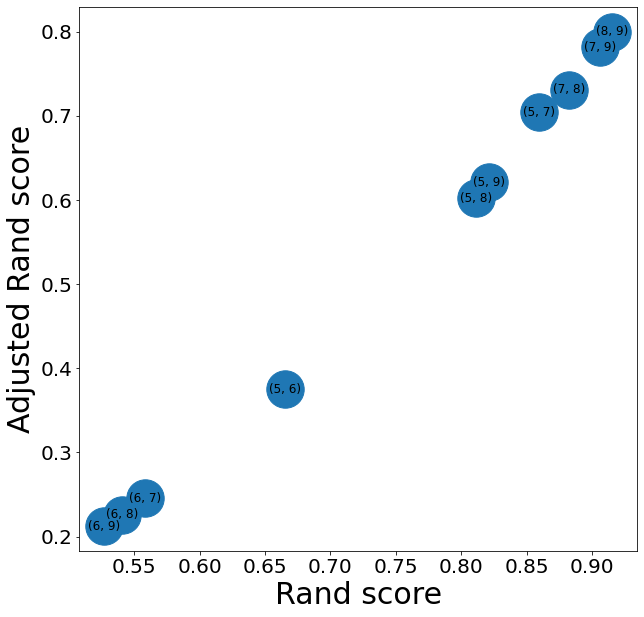}
        \caption{}
        \label{adjrandvrandplot-wnoise} 
        \end{subfigure}

        \begin{subfigure}[b]{0.55\textwidth}
        \includegraphics[width=0.9\linewidth]{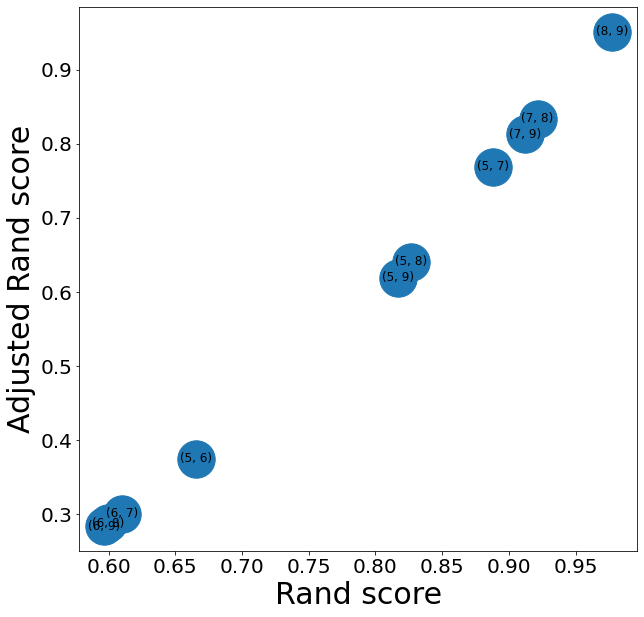}
        \caption{}
        \label{adjrandvrandplot}
        \end{subfigure}
    \caption{
    Adjusted Rand score and Rand score scatter plot of \ref{adjrandvrandplot-wnoise}) Clustering results with Cluster 3 separated from Noise and \ref{adjrandvrandplot}) Clustering results with Cluster 3 merged with Noise.
    Each point is annotated with the pair of {\ttfamily n\_neighbors} values that the scores are calculated. 
    Thus, all the data points are pairs of clustering results with their corresponding Rand and Adjusted Rand scores as their coordinates.
    }
    \label{adjrandvrand}
    \end{figure}

    As shown in Figure \ref{adjrandvrand}, we present the Rand and Adjusted Rand scores using a scatter plot of their values. Each point is annotated with the pair of {\ttfamily n\_neighbors} values that the scores are calculated. Thus, all the data points are pairs of clustering results with their corresponding Rand and Adjusted Rand scores as their coordinates. Figure \ref{adjrandvrandplot-wnoise} shows the calculated scores for each pair of clustering result where Cluster 3 and Noise is not merged while Figure \ref{adjrandvrandplot} shows the calculated scores for each pair of clustering result where Cluster 3 and Noise is merged. In both scatter plots, the clustering results for {\ttfamily n\_neighbors = 5,7,8,} and {\ttfamily 9} exhibited Rand scores greater than $0.80$ and Adjusted Rand scores of at least $0.60$. This indicates the clustering results were in very good agreement with one another with the exclusion of {\ttfamily n\_neighbors = 6}. Now, comparing the scores for the clustering results for the two cases one where the Noise is not merged with Cluster 3 (Figure \ref{adjrandvrandplot-wnoise}) and one where it is merged (Figure \ref{adjrandvrandplot}). We can see that there is no significant difference or significant improvement in the scores of the clustering result pairs after merging the Noise and Cluster 3. This implies that there is no significant merit in merging Cluster 3 and Noise. Lastly, both figures show that either the clustering result for {\ttfamily n\_neighbors = 8} and {\ttfamily 9} is a good representative of the clustering for our dataset. In this paper, we adopt the case of Noise separate from Cluster 3.

\section{CONCLUSIONS}\label{sec6}
    \; \; With the above underpinnings, this paper concludes the following: 
    \begin{itemize}
        \item Using parameter colouring, we have identified the invariant cluster properties of each cluster regardless of {\ttfamily n\_neighbors} value. Showing that describing the FRB subtypes without any dependence on the set value of {\ttfamily n\_neighbors} permits comparison with other works that aims at classifying FRBs. Invariant cluster properties also aids in determining possible physical mechanisms that corresponds to the characterization of each cluster which can be further discussed in future theoretical works. 
        \item Investigating and plotting the change in cluster membership of the FRBs proved to be useful in pointing out connections between clusters of different {\ttfamily n\_neighbors} value. This analysis led us to understand more about the underlying structure of the FRBs of FRB121102 by showing that certain clusters may have complex composition and consist of smaller distinct clusters. As shown, we have found that the "Atypical" cluster of FRB121102 can be further split up into three smaller clusters. 
        \item Compared to the existing results in the literature, our clustering result does not only depend on a single parameter to create a grouping of FRBs. Using pertinent physical parameters, if not all parameters, the model we have used created a more robust classification of the repeating FRBs from FRB121102.  Nevertheless, a certain degree of agreement with other results (e.g., being able to recover the FRB classification used by \citealt{Xiao&Dai2022}) exhibits consistency and foundation on physical parameters of the clusters.
    \end{itemize}

\section*{ACKNOWLEDGMENTS}
    We thank the anonymous referee for many insightful comments, which improved the paper significantly.
    TG acknowledges the support of the National Science and Technology Council of Taiwan through grants 108-2628-M-007-004-MY3, 111-2112-M-007-021, and 111-2123-M-001-008-.
    TH acknowledges the support of the National Science and Technology Council of Taiwan through grants 110-2112-M-005-013-MY3, 110-2112-M-007-034-, and 111-2123-M-001-008-. 
    The authors would also like to extend their utmost gratitude to Professor Wang Pei and his colleagues for sharing and making the data openly available in Science Data Bank. The authors would also like to thank Dr. Shotaro Yamasaki for his valuable insights and suggestions.

%%%%%%%%%%%%%%%%%%%%%%%%%%%%%%%%%%%%%%%%%%%%%%%%%%
    \section*{DATA AVAILABILITY}
%The data underlying this article will be shared upon reasonable request to the corresponding author.
    The data underlying this article is available in the work of \citet[dataset]{2021Natur.598..267L}. The dataset were derived from Science Data Bank, at \url{http://doi.org/10.11922/sciencedb.01092. DOI:10.11922/sciencedb.01092}. 


\bibliographystyle{mnras}
\bibliography{output} % if your bibtex file is called example.bib

%\begin{thebibliography}{99}
%\end{thebibliography}

%%%%%%%%%%%%%%%%%%%%%%%%%%%%%%%%%%%%%%%%%%%%%%%%%%

%%%%%%%%%%%%%%%%% APPENDICES %%%%%%%%%%%%%%%%%%%%%

\hspace{150pt}

\appendix

\section{APPENDIX}
\label{appendix}
Additional results that are important to the analyses in our paper.
\subsection{UMAP results for {\ttfamily n\_neighbors = 5,6,7, and 8}}
  
            \begin{figure*}
                \centering
                \begin{multicols}{2}
                    \subcaptionbox{\label{umap5}}{\includegraphics[width=1\linewidth]{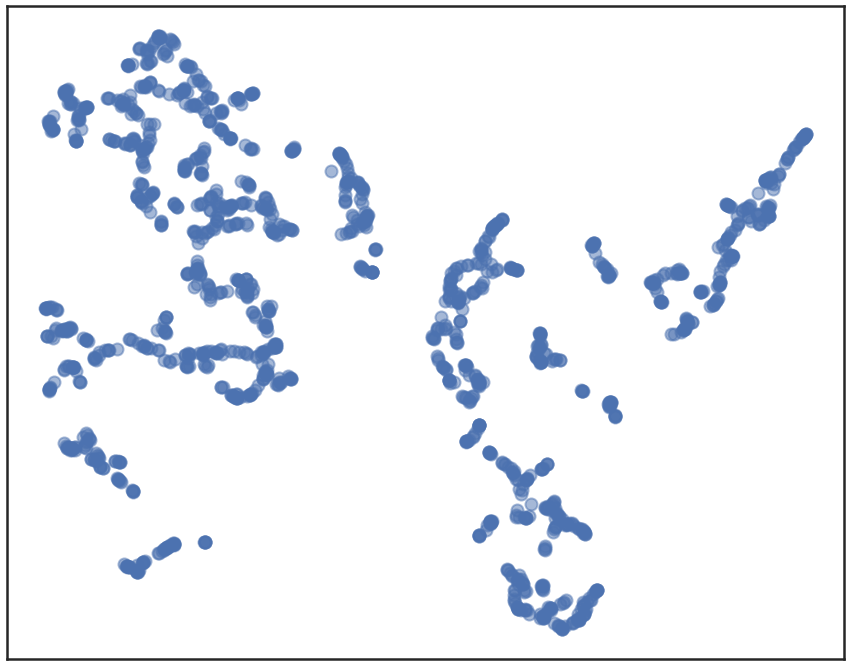}}\par 
                    \subcaptionbox{\label{umap6}}{\includegraphics[width=1\linewidth]{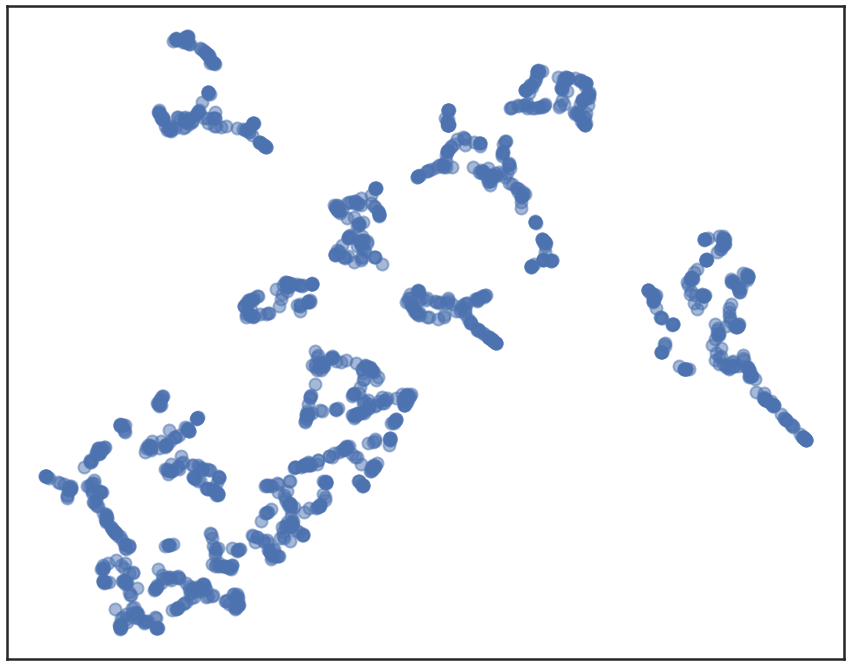}}\par 
                \end{multicols}
                \begin{multicols}{2}
                    \subcaptionbox{\label{umap7}}{\includegraphics[width=1\linewidth]{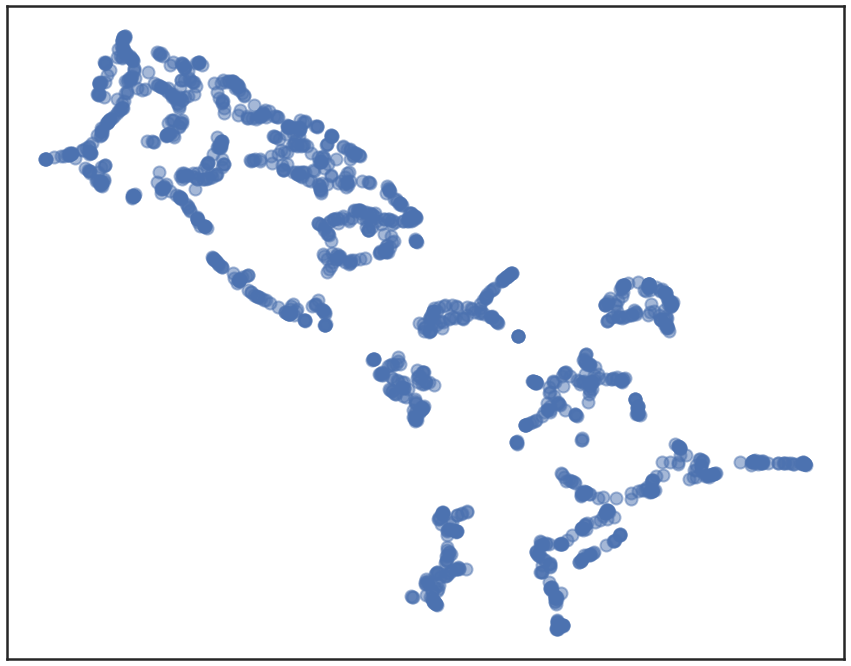}}\par
                    \subcaptionbox{\label{umap8}}{\includegraphics[width=1\linewidth]{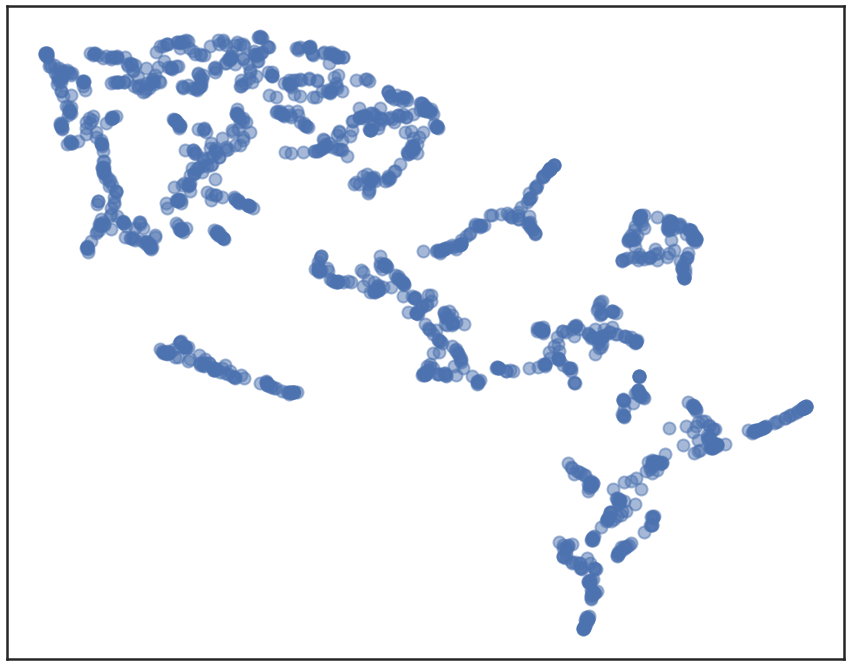}}\par
                \end{multicols}
                \caption{\large UMAP embedding results for (\ref{umap5}) {\ttfamily n\_neighbors = 5}, (\ref{umap6}) {\ttfamily n\_neighbors = 6}, (\ref{umap7}) {\ttfamily n\_neighbors = 7}, and (\ref{umap8}) {\ttfamily n\_neighbors = 8}.}
            \label{apnn5-9}
            \end{figure*}
%\clearpage
\subsection{HDBSCAN results for {\ttfamily n\_neighbors = 5,6,7, and 8}}
  
%%%%%%%%%% COMBINE FIGURES 4-8 INTO ONE FIGURE %%%%%%%%%%%%%%%
        
        \begin{figure*} 
                \centering
                \begin{multicols}{2}
                    \subcaptionbox{\label{hdbscan5}}{\includegraphics[width=1\linewidth]{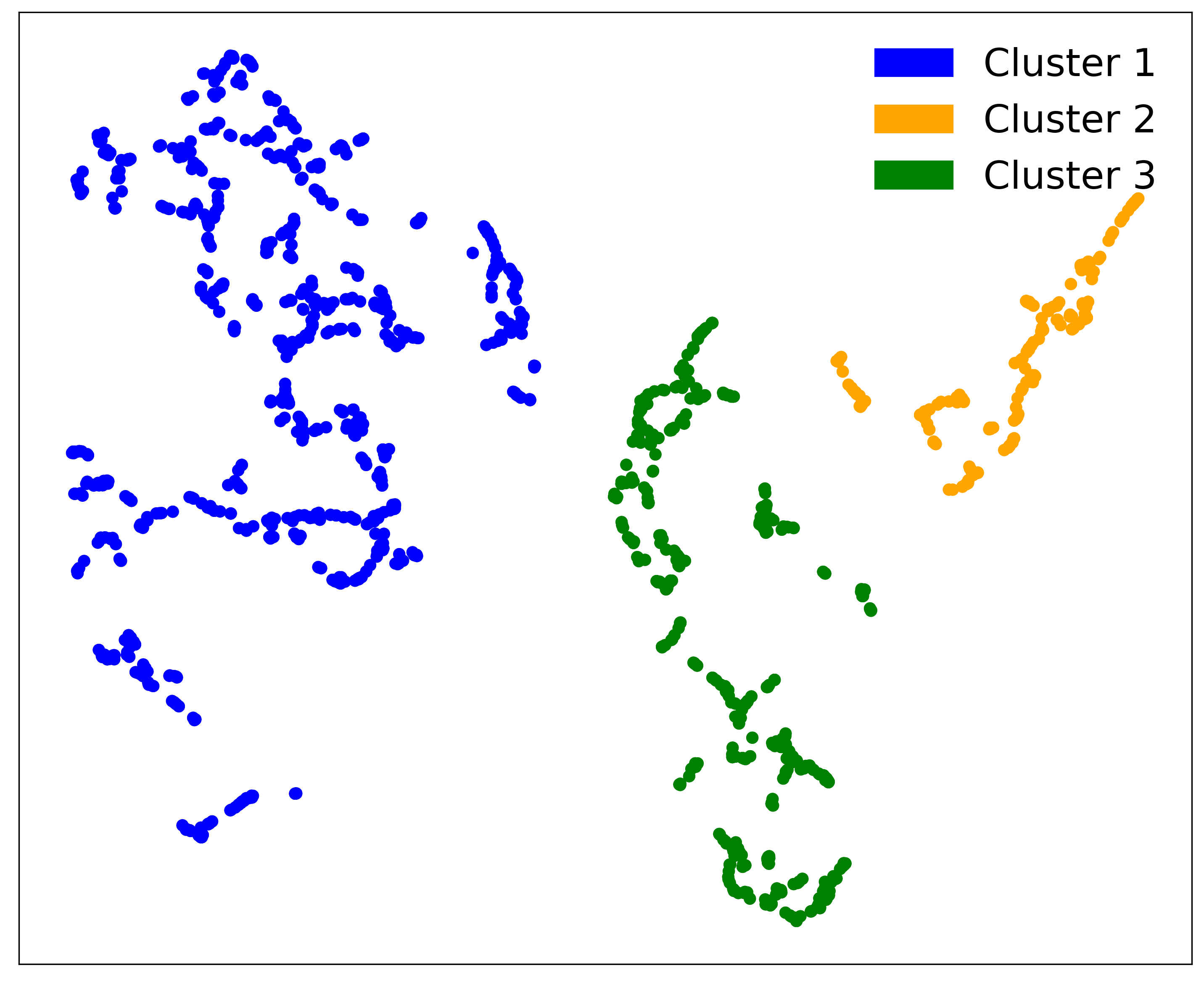}}\par 
                    \subcaptionbox{\label{hdbscan6}}{\includegraphics[width=1\linewidth]{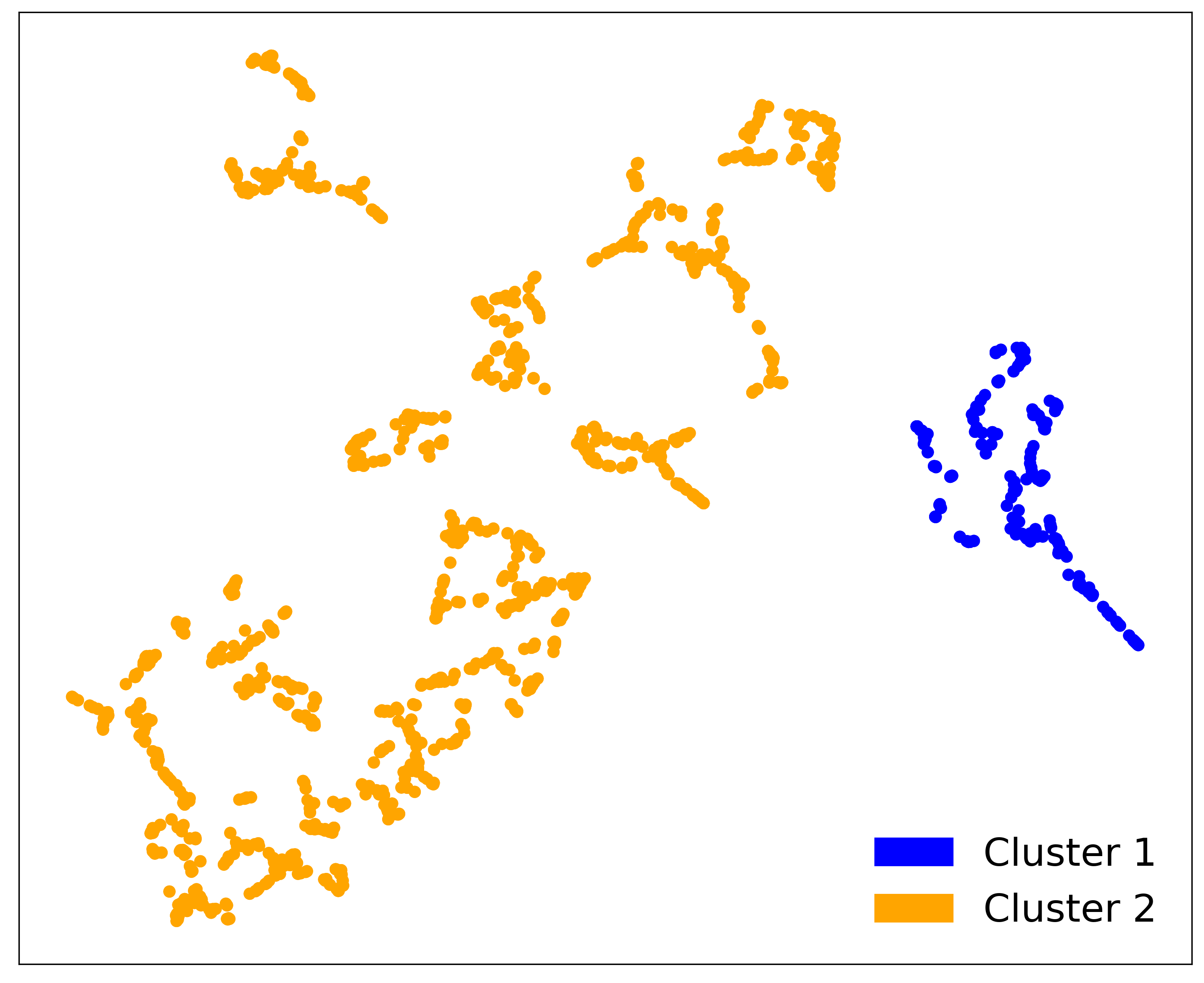}}\par 
                \end{multicols}
                \begin{multicols}{2}
                    \subcaptionbox{\label{hdbscan7}}{\includegraphics[width=1\linewidth]{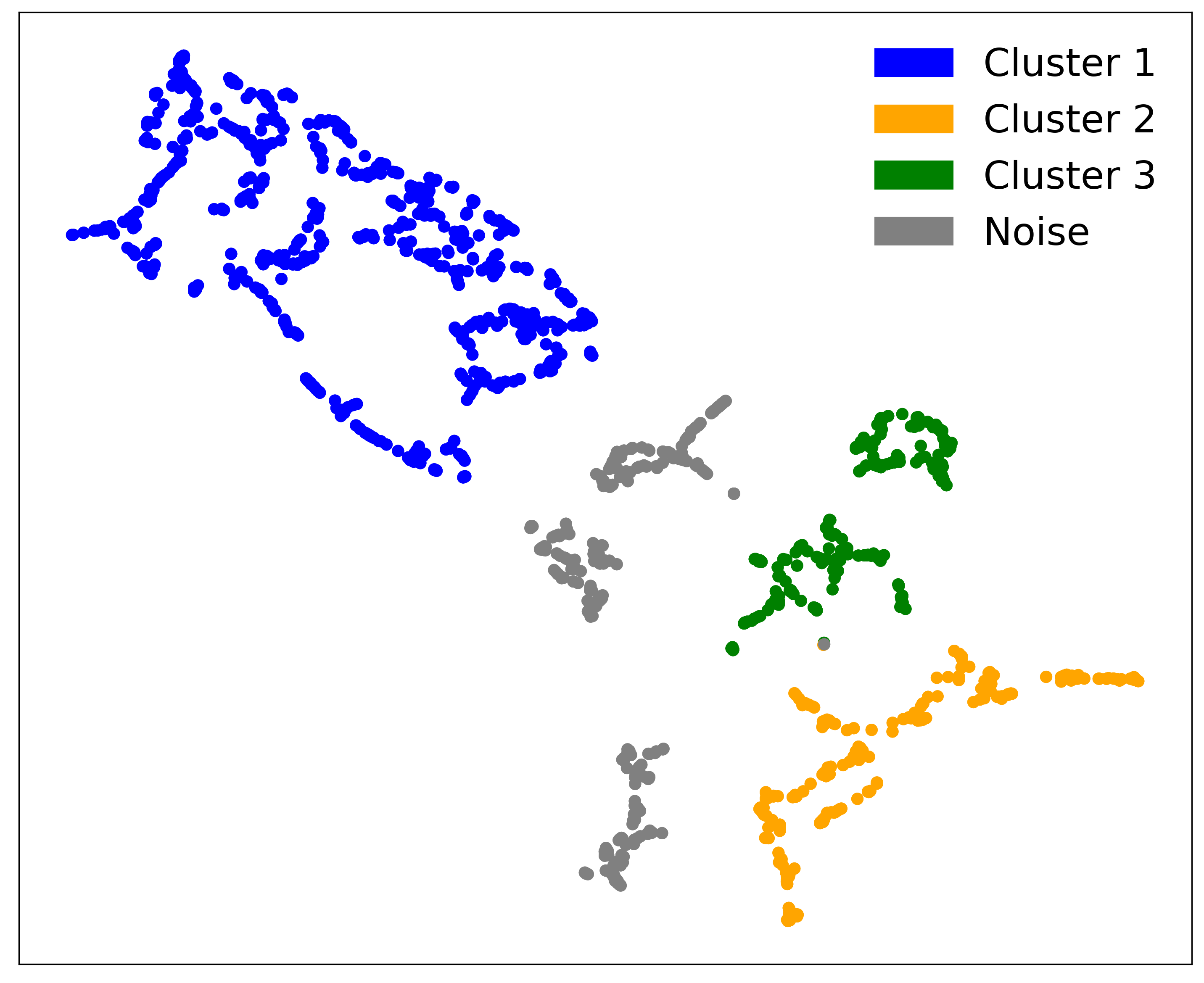}}\par
                    \subcaptionbox{\label{hdbscan8}}{\includegraphics[width=1\linewidth]{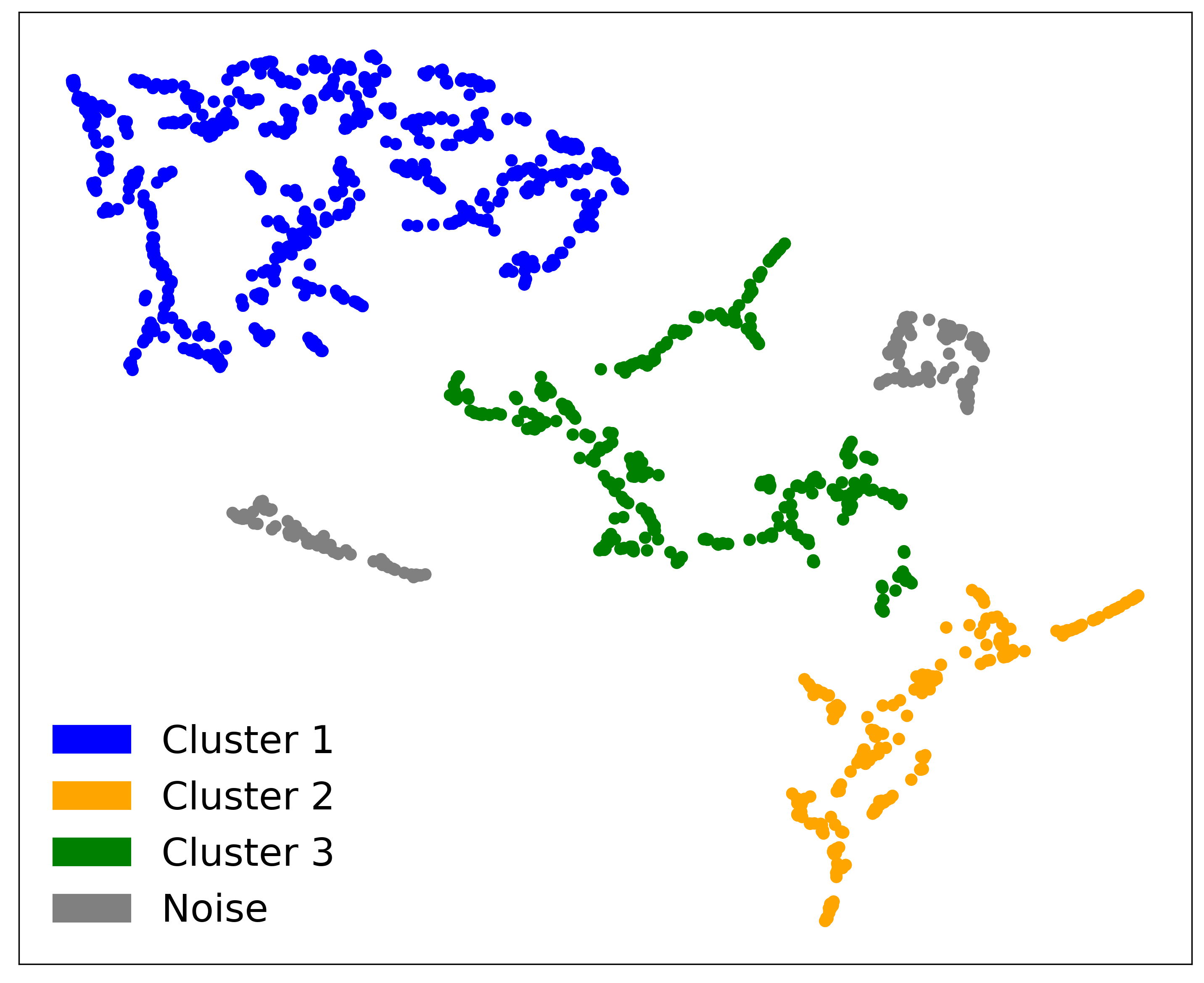}}\par
                \end{multicols}
                \caption{  \large HDBSCAN Clustering result for  (\ref{hdbscan5}) {\ttfamily n\_neighbors = 5}, (\ref{hdbscan6}) {\ttfamily n\_neighbors = 6}, (\ref{hdbscan7}) {\ttfamily n\_neighbors = 7}, and (\ref{hdbscan8}) {\ttfamily n\_neighbors = 8}.}
            \label{aphdbscan}
            \end{figure*}

\subsection{Parameter Colouring results for {\ttfamily n\_neighbors = 5,6,7, and 8}}
  
                \begin{figure*} 
                \centering
                \begin{multicols}{2}
                    \subcaptionbox{\label{apcbw5}}{\includegraphics[width=\linewidth, height=0.8\columnwidth]{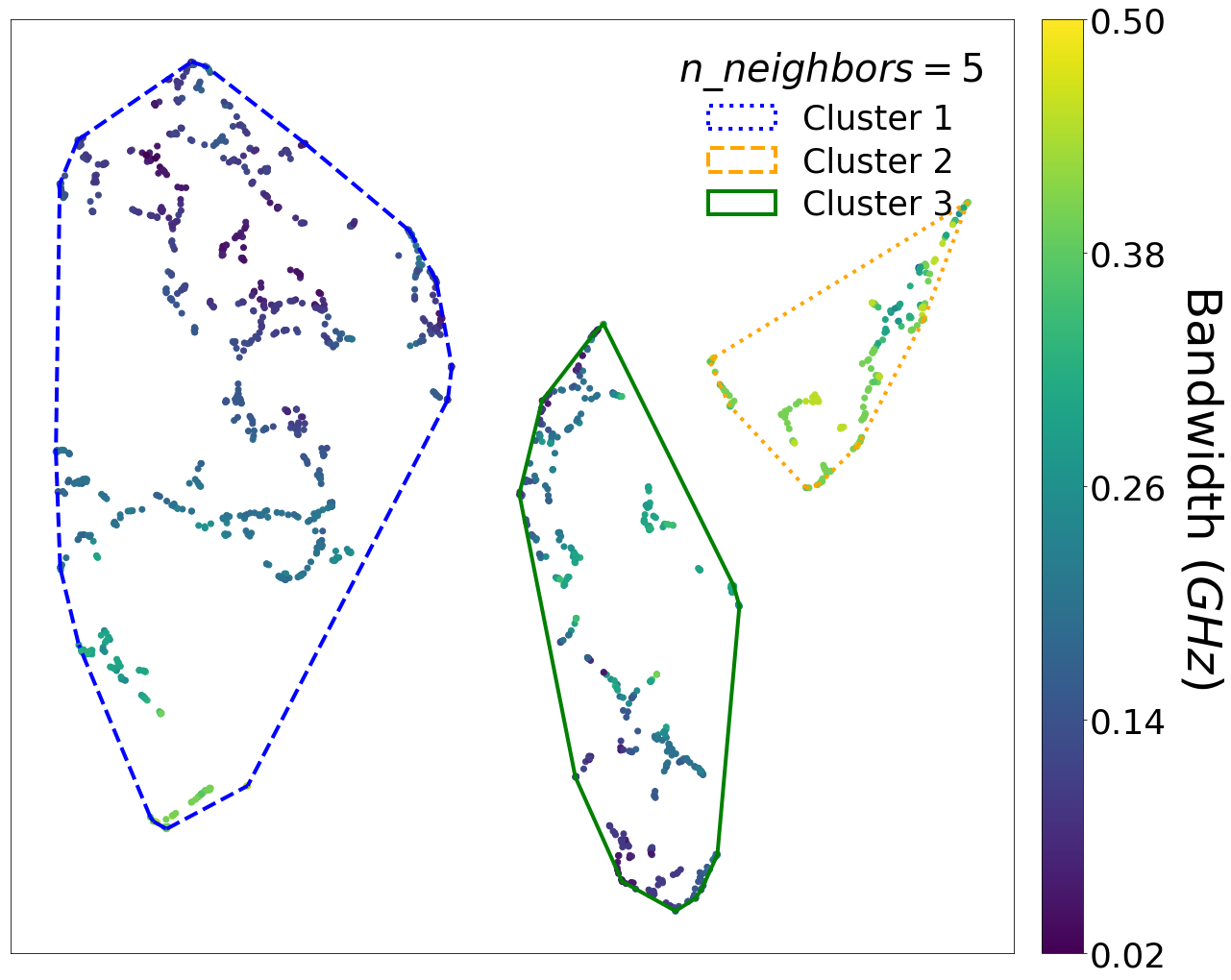}}\par 
                    \subcaptionbox{\label{apcbw6}}{\includegraphics[width=\linewidth, height=0.8\columnwidth]{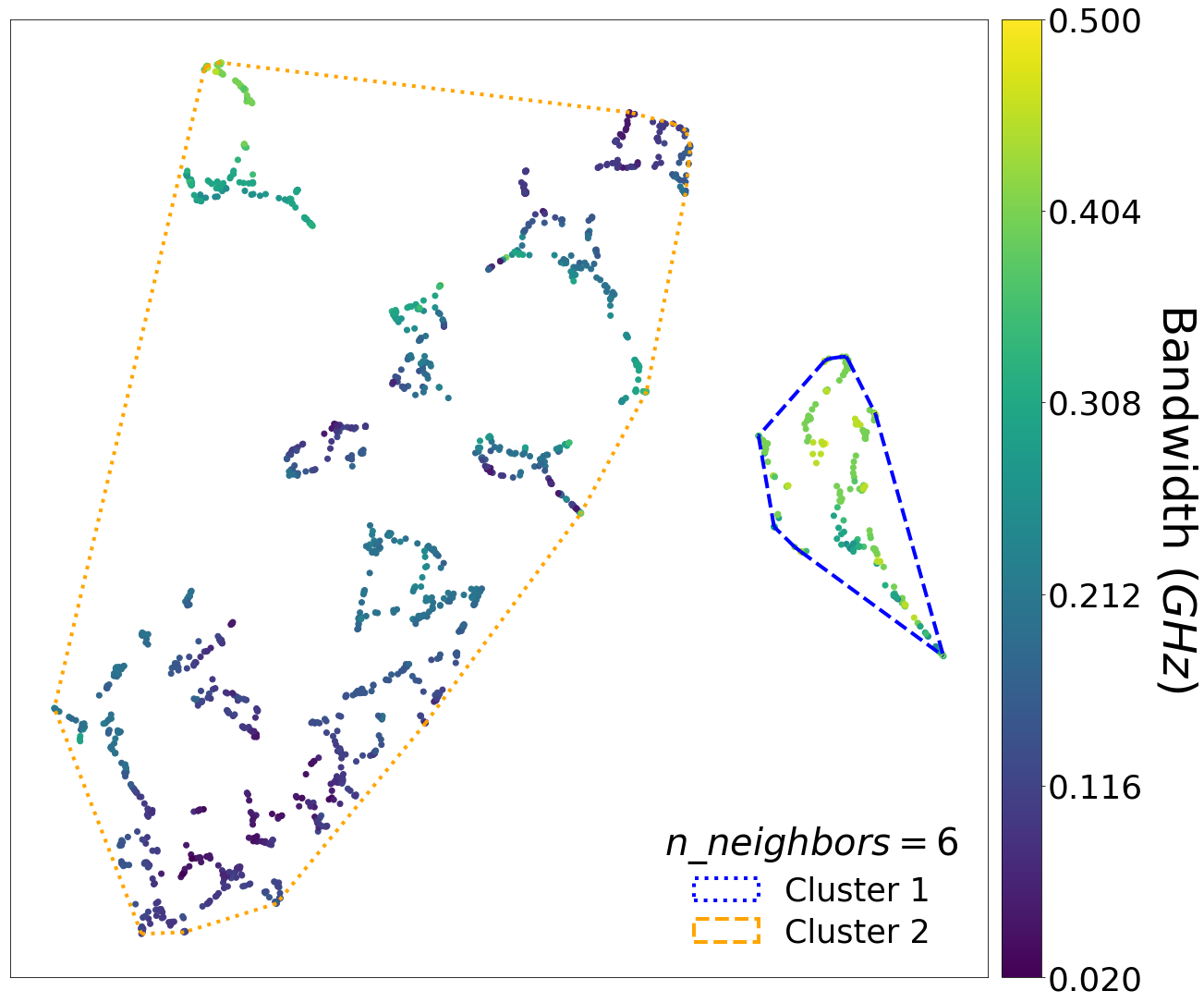}}\par 
                \end{multicols}
                \begin{multicols}{2}
                    \subcaptionbox{\label{apcbw7}}{\includegraphics[width=\linewidth, height=0.8\columnwidth]{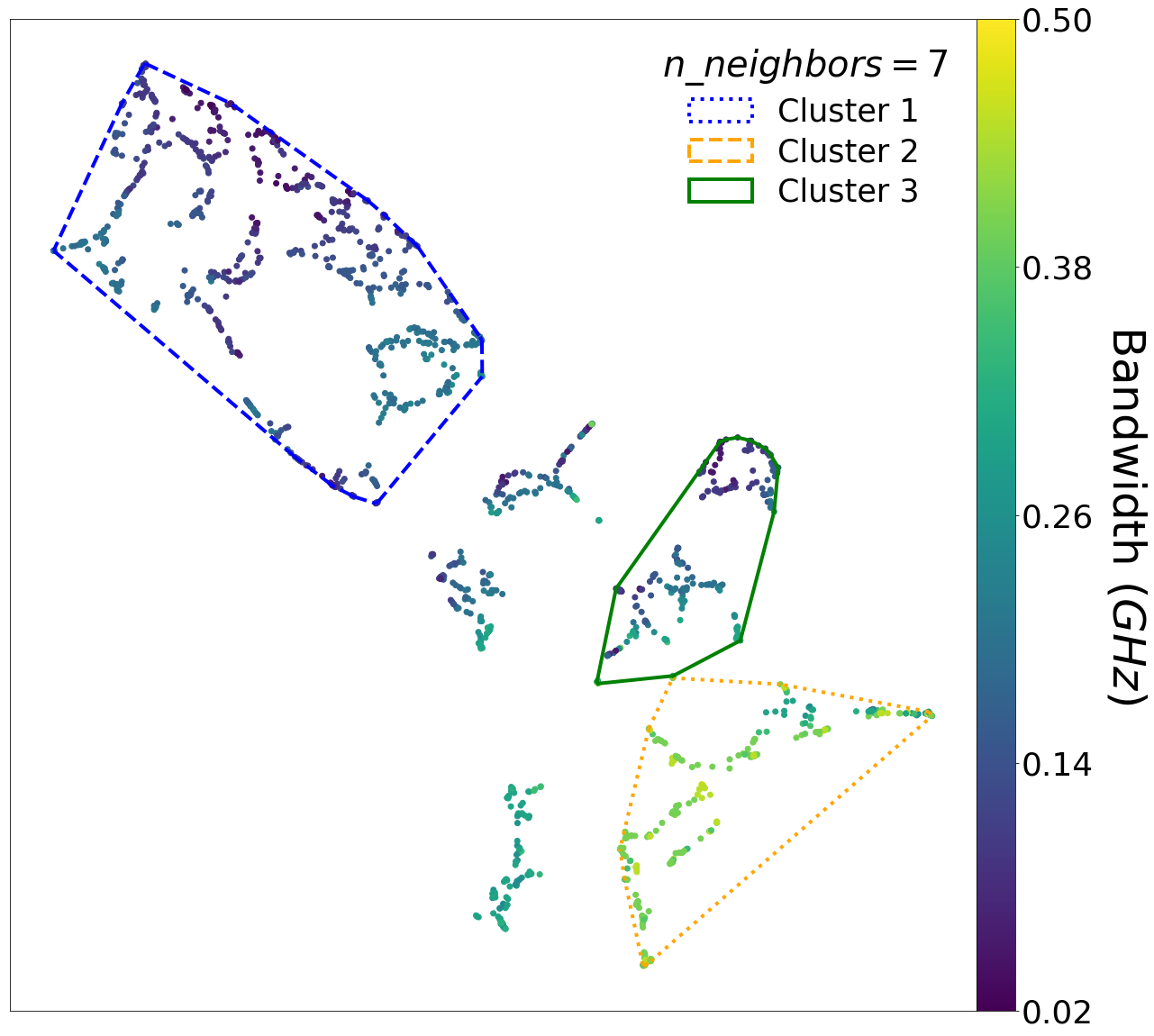}}\par
                    \subcaptionbox{\label{apcbw8}}{\includegraphics[width=\linewidth, height=0.8\columnwidth]{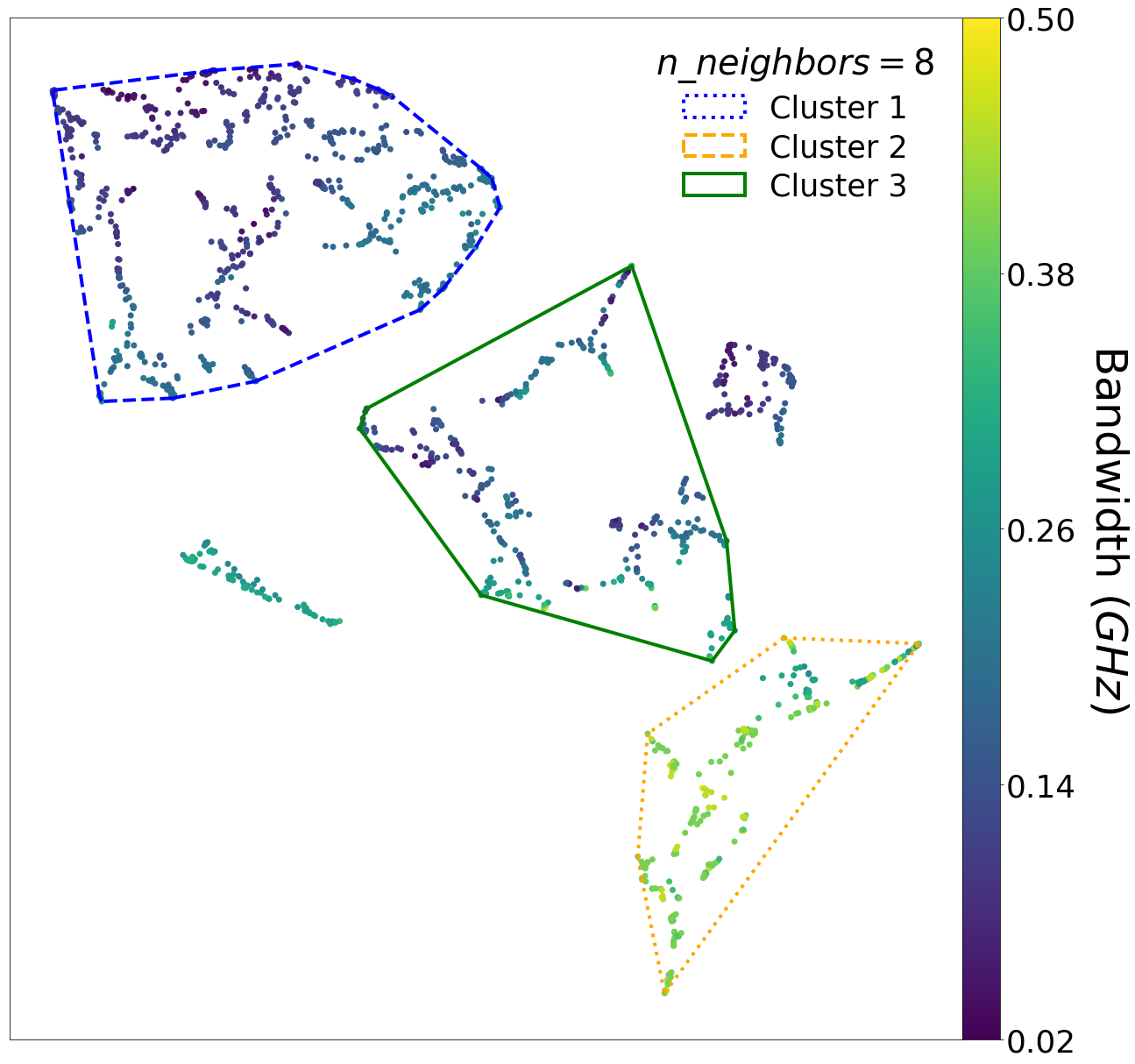}}\par
                \end{multicols}
                \caption{ \large Bandwidth colouring of the clustering results for (\ref{apcbw5}) {\ttfamily n\_neighbors = 5}, (\ref{apcbw6}) {\ttfamily n\_neighbors = 6}, (\ref{apcbw7}) {\ttfamily n\_neighbors = 7}, and (\ref{apcbw8}) {\ttfamily n\_neighbors = 8}. The data which are not surrounded by lines correspond to Noise clusters.}
            \label{apbwparcol}
            \end{figure*}
            
            \begin{figure*} 
                \centering
                \begin{multicols}{2}
                    \subcaptionbox{\label{apce5}}{\includegraphics[width=\linewidth, height=0.8\columnwidth]{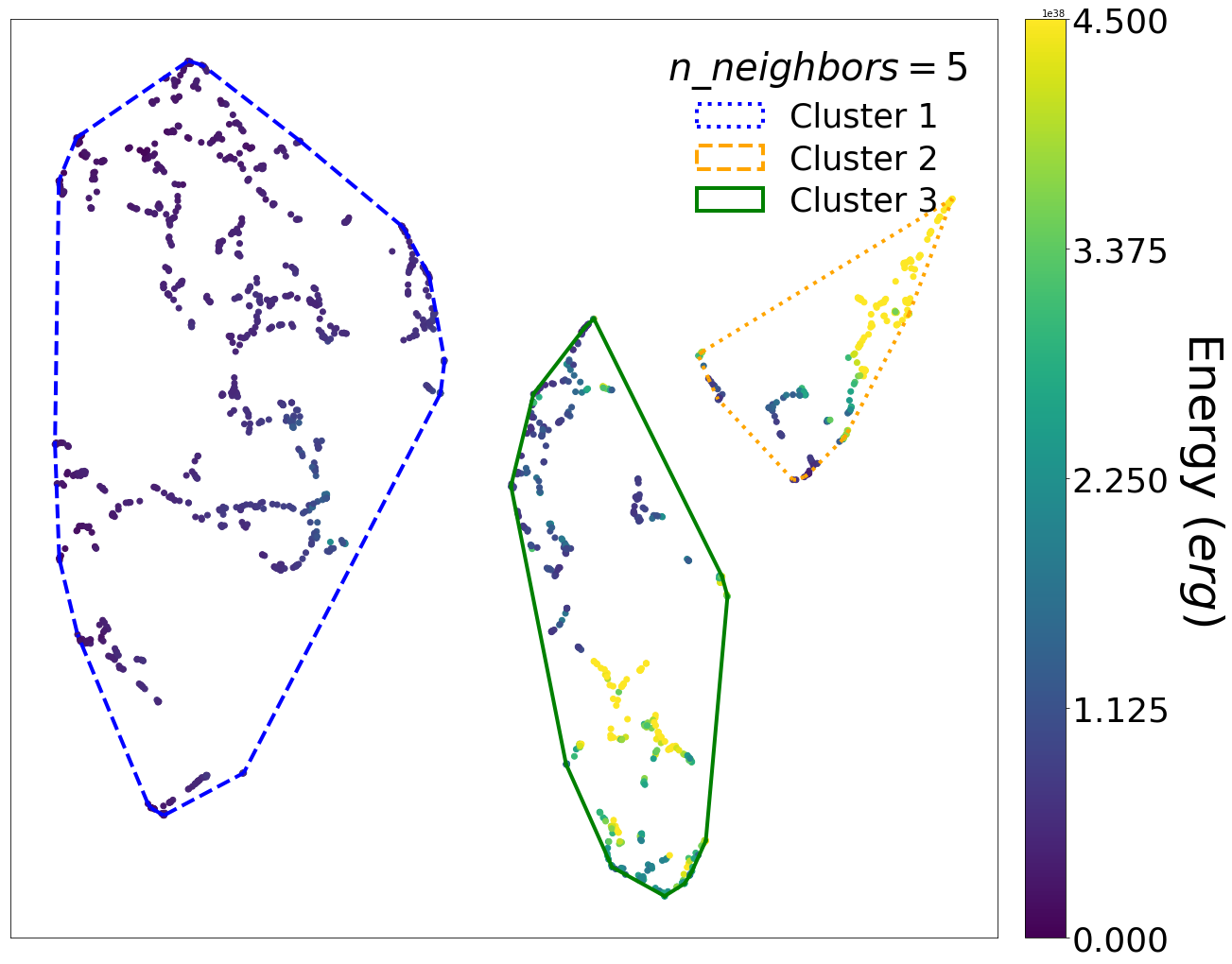}}\par 
                    \subcaptionbox{\label{apce6}}{\includegraphics[width=\linewidth, height=0.8\columnwidth]{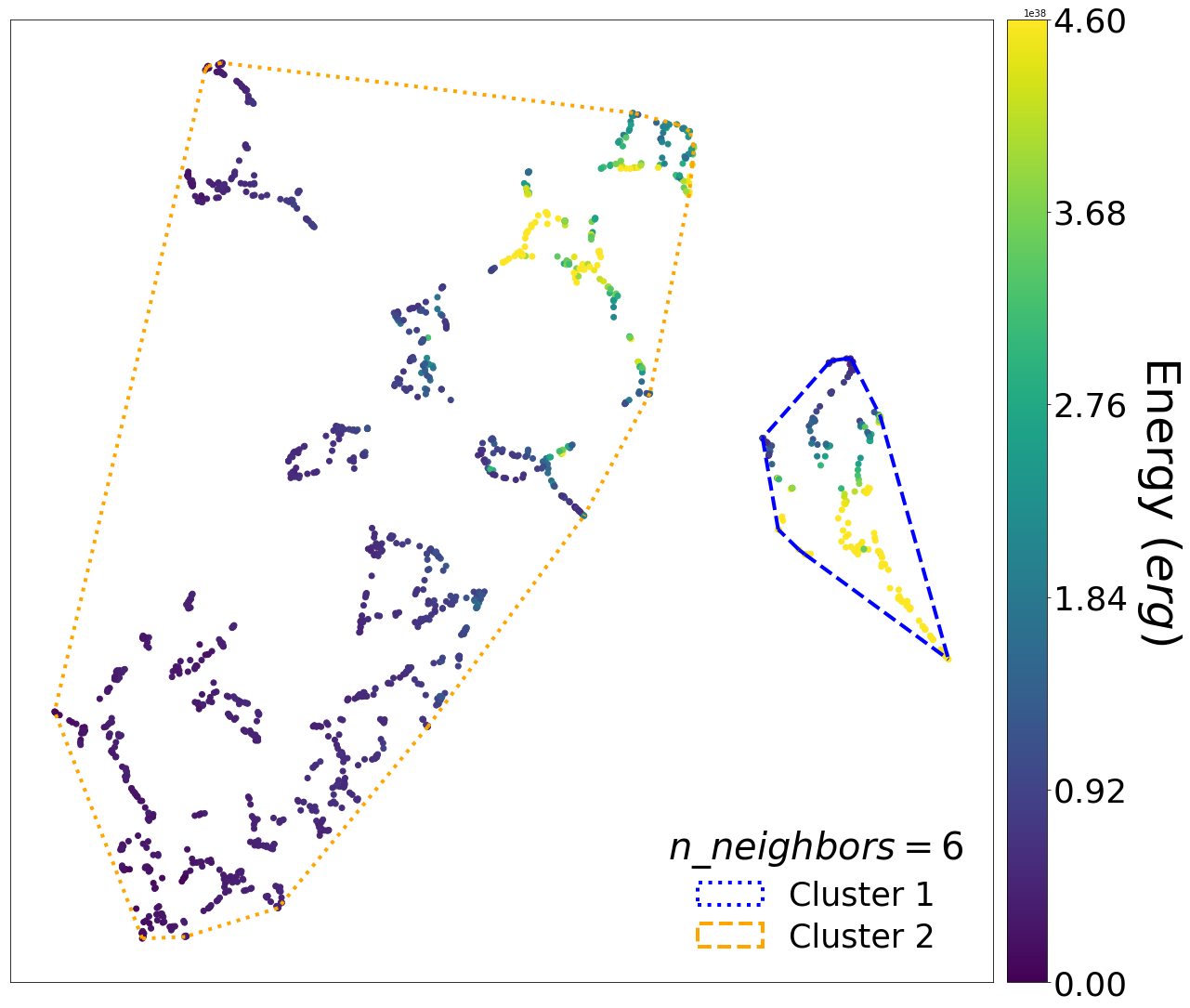}}\par 
                \end{multicols}
                \begin{multicols}{2}
                    \subcaptionbox{\label{apce7}}{\includegraphics[width=\linewidth, height=0.8\columnwidth]{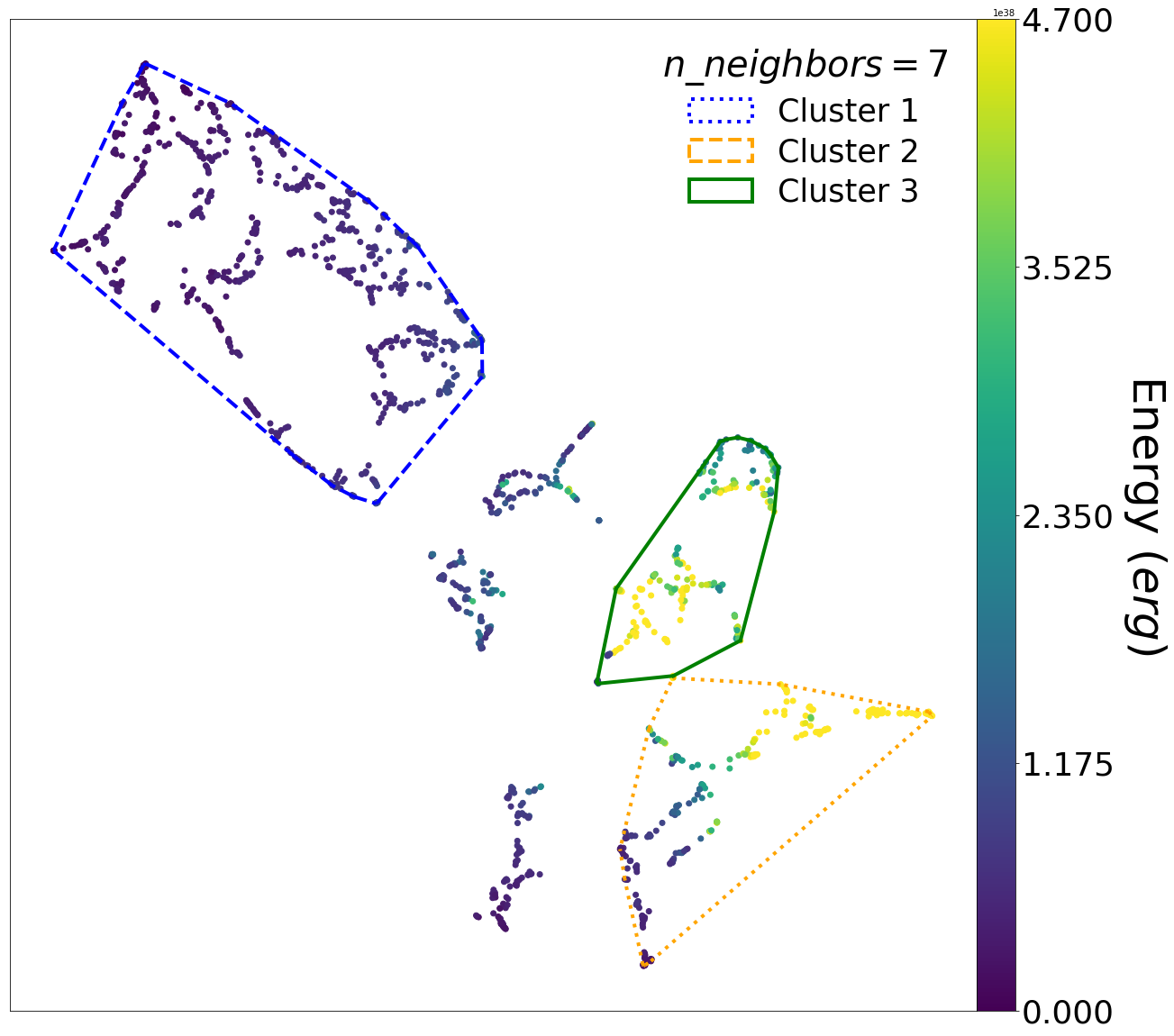}}\par
                    \subcaptionbox{\label{apce8}}{\includegraphics[width=\linewidth, height=0.8\columnwidth]{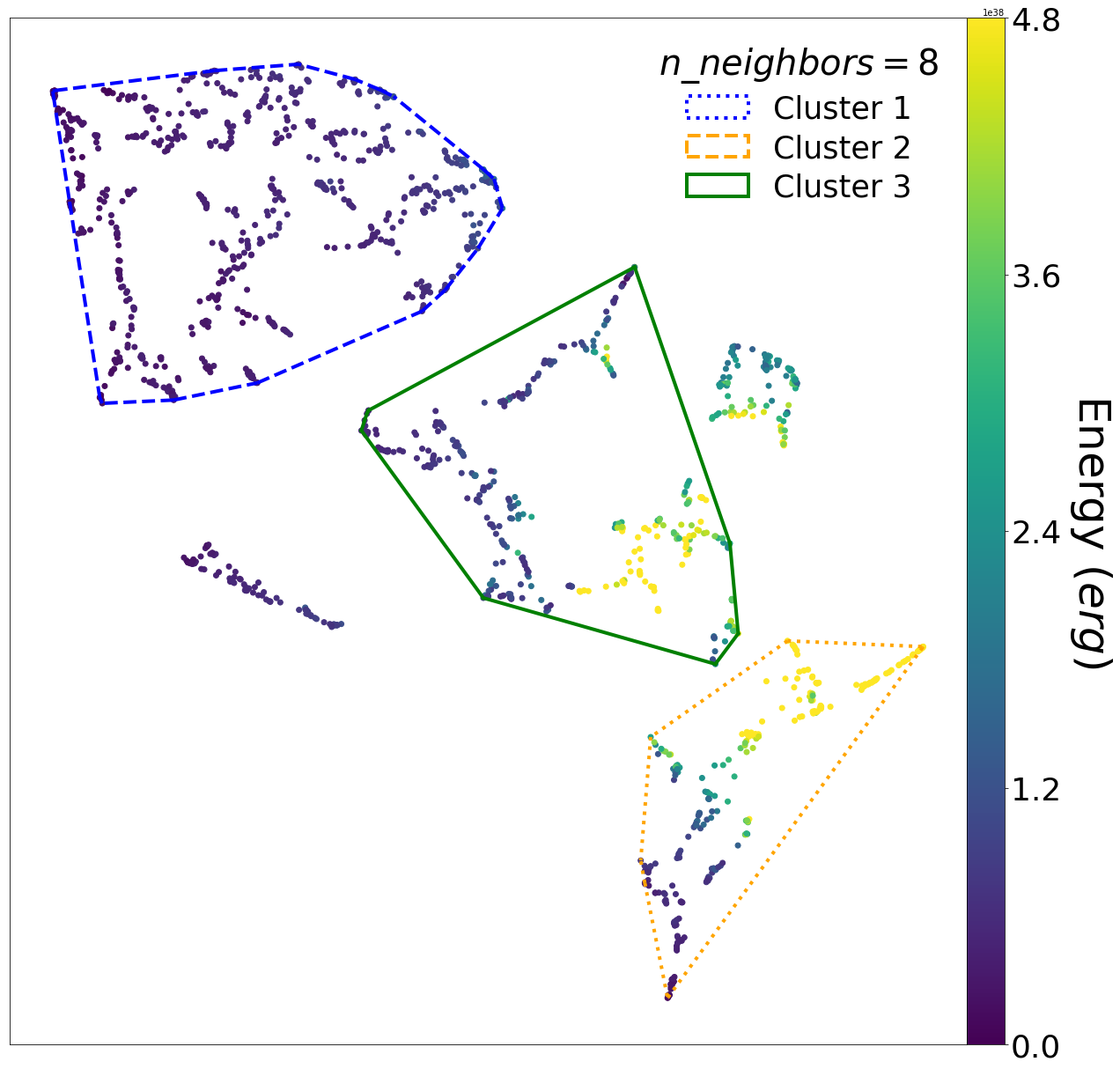}}\par
                \end{multicols}
                \caption{\large Energy colouring of the clustering results for (\ref{apce5}) {\ttfamily n\_neighbors = 5}, (\ref{apce6}) {\ttfamily n\_neighbors = 6}, (\ref{apce7}) {\ttfamily n\_neighbors = 7}, and (\ref{apce8}) {\ttfamily n\_neighbors = 8}. The data which are not surrounded by lines correspond to Noise clusters.}
            \label{apeparcol}
            \end{figure*}

            \begin{figure*} 
                \centering
                \begin{multicols}{2}
                    \subcaptionbox{\label{apcfl5}}{\includegraphics[width=\linewidth, height=0.8\columnwidth]{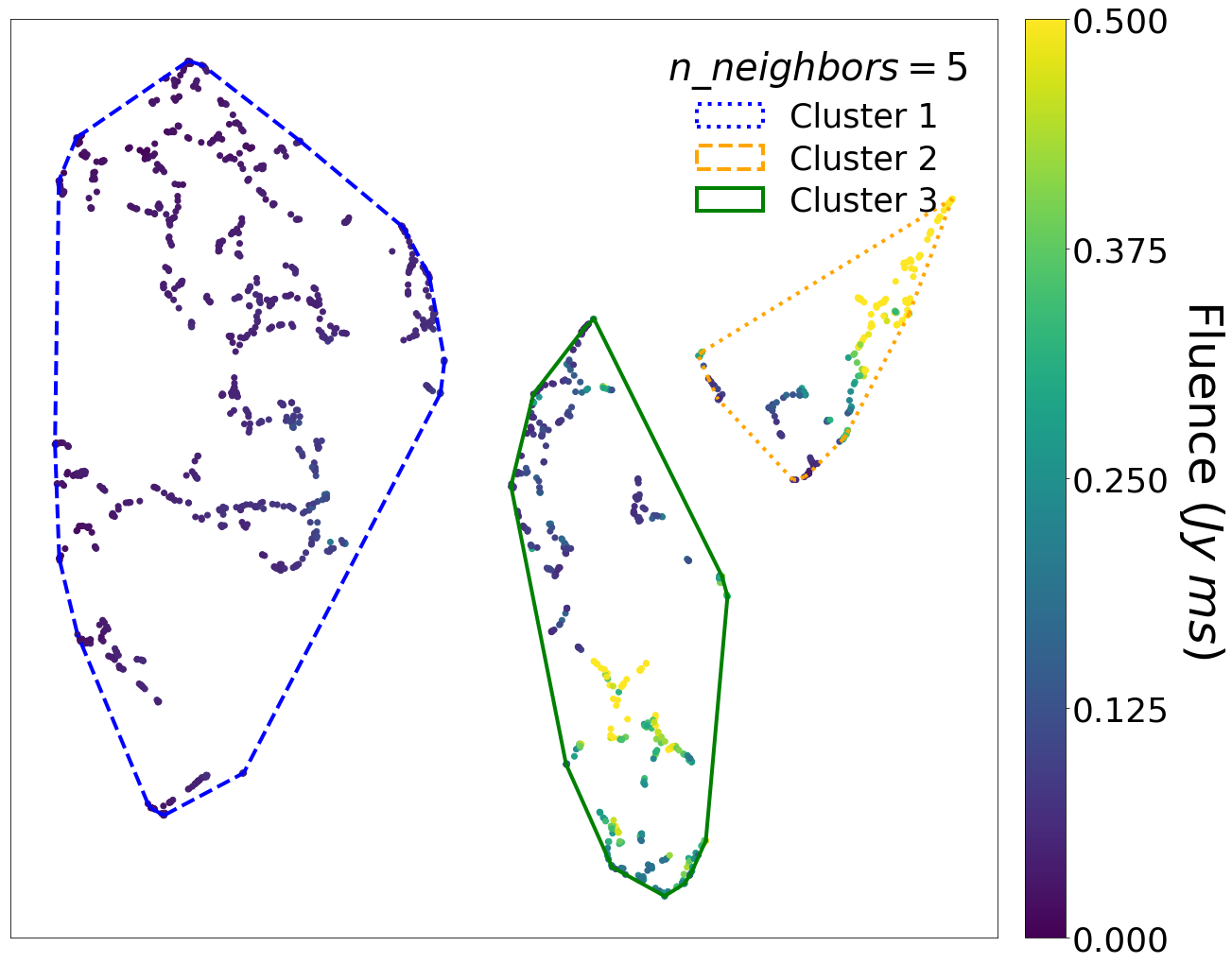}}\par 
                    \subcaptionbox{\label{apcfl6}}{\includegraphics[width=\linewidth, height=0.8\columnwidth]{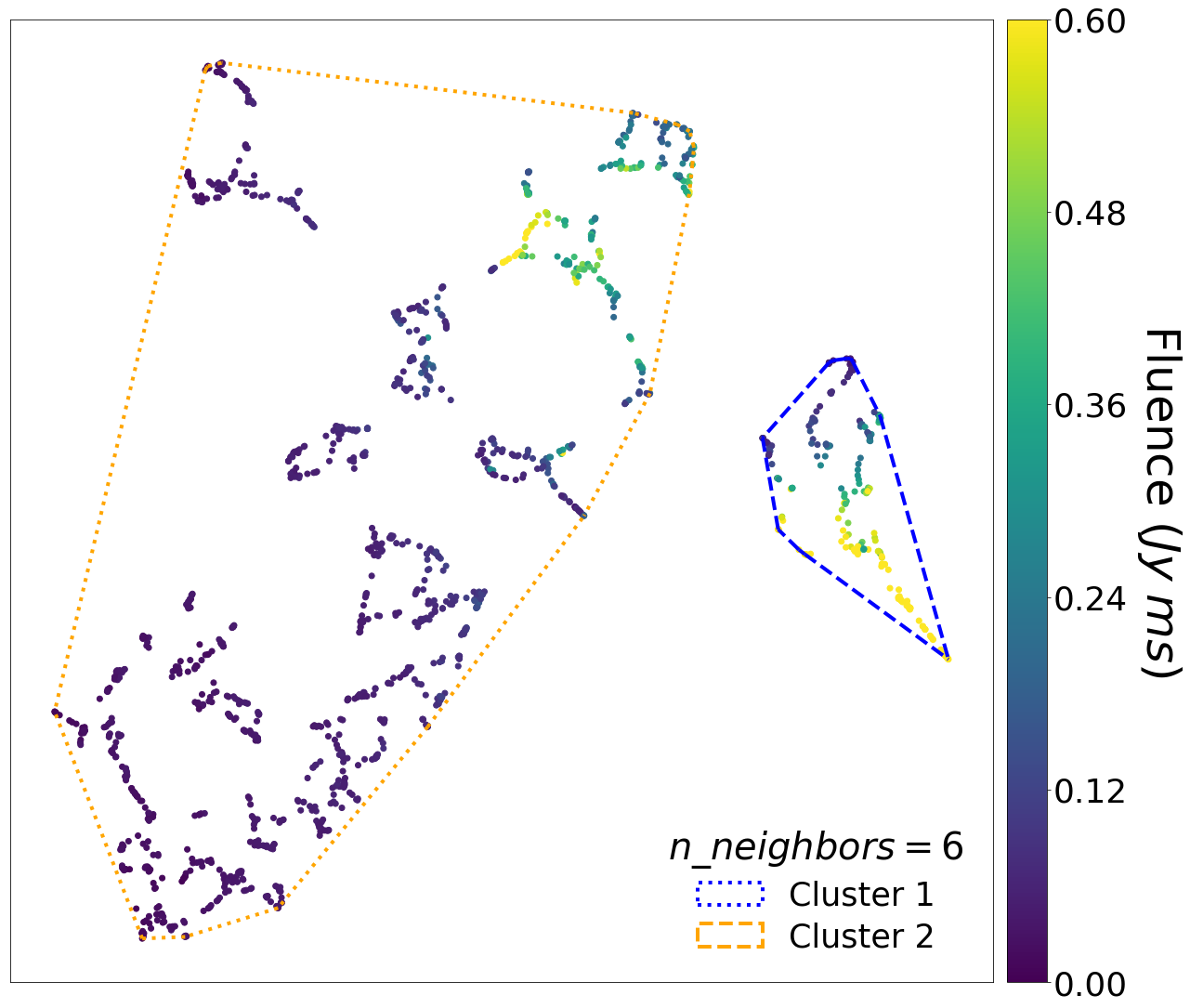}}\par 
                \end{multicols}
                \begin{multicols}{2}
                    \subcaptionbox{\label{apcfl7}}{\includegraphics[width=\linewidth, height=0.8\columnwidth]{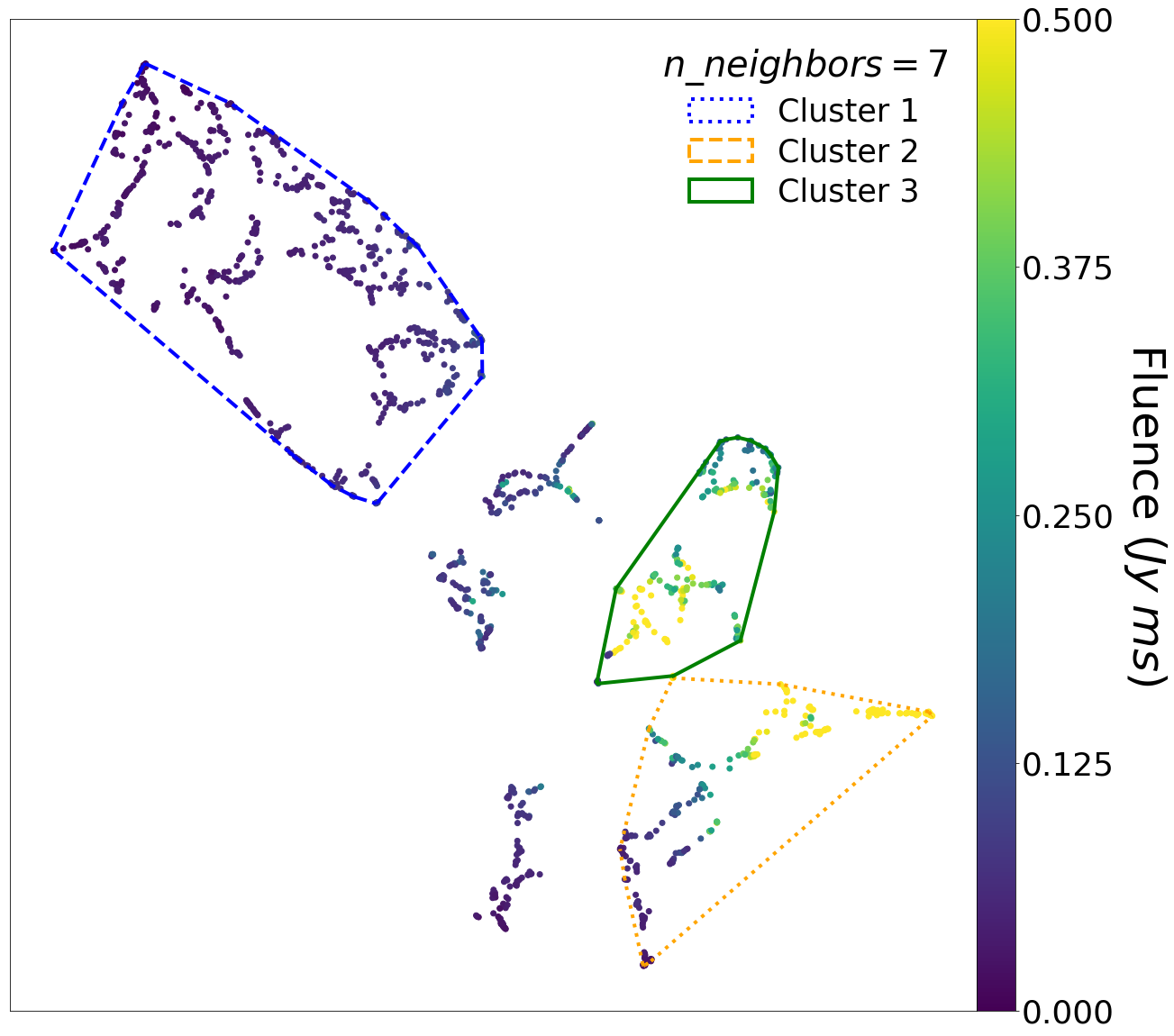}}\par
                    \subcaptionbox{\label{apcfl8}}{\includegraphics[width=\linewidth, height=0.8\columnwidth]{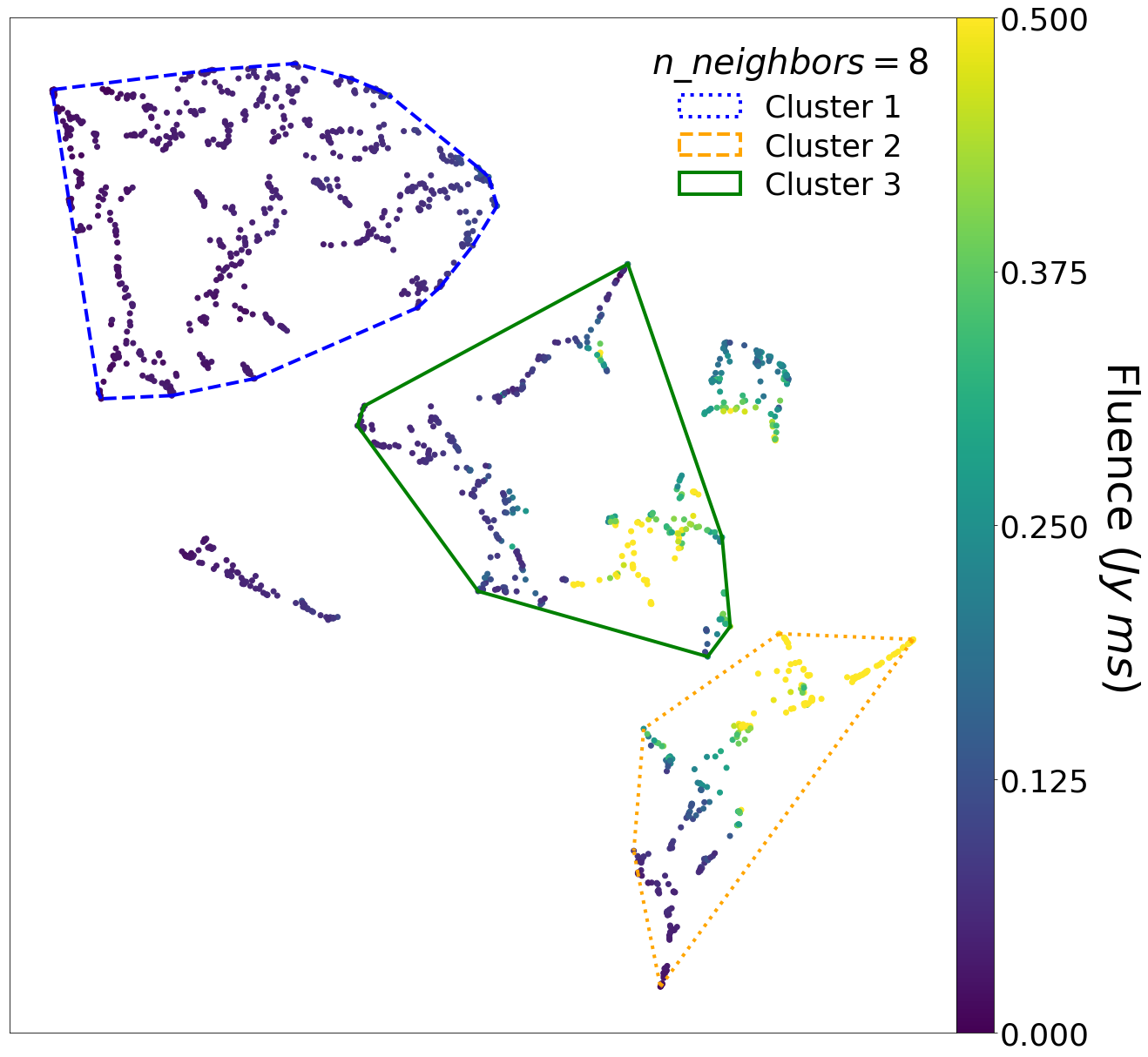}}\par
                \end{multicols}
                \caption{\large Fluence colouring of the clustering results for (\ref{apcfl5}) {\ttfamily n\_neighbors = 5}, (\ref{apcfl6}) {\ttfamily n\_neighbors = 6}, (\ref{apcfl7}) {\ttfamily n\_neighbors = 7}, and (\ref{apcfl8}) {\ttfamily n\_neighbors = 8}. The data which are not surrounded by lines correspond to Noise clusters.}
            \label{apflparcol}
            \end{figure*}
            
            \begin{figure*} 
                \centering
                \begin{multicols}{2}
                    \subcaptionbox{\label{apcpf5}}{\includegraphics[width=\linewidth, height=0.8\columnwidth]{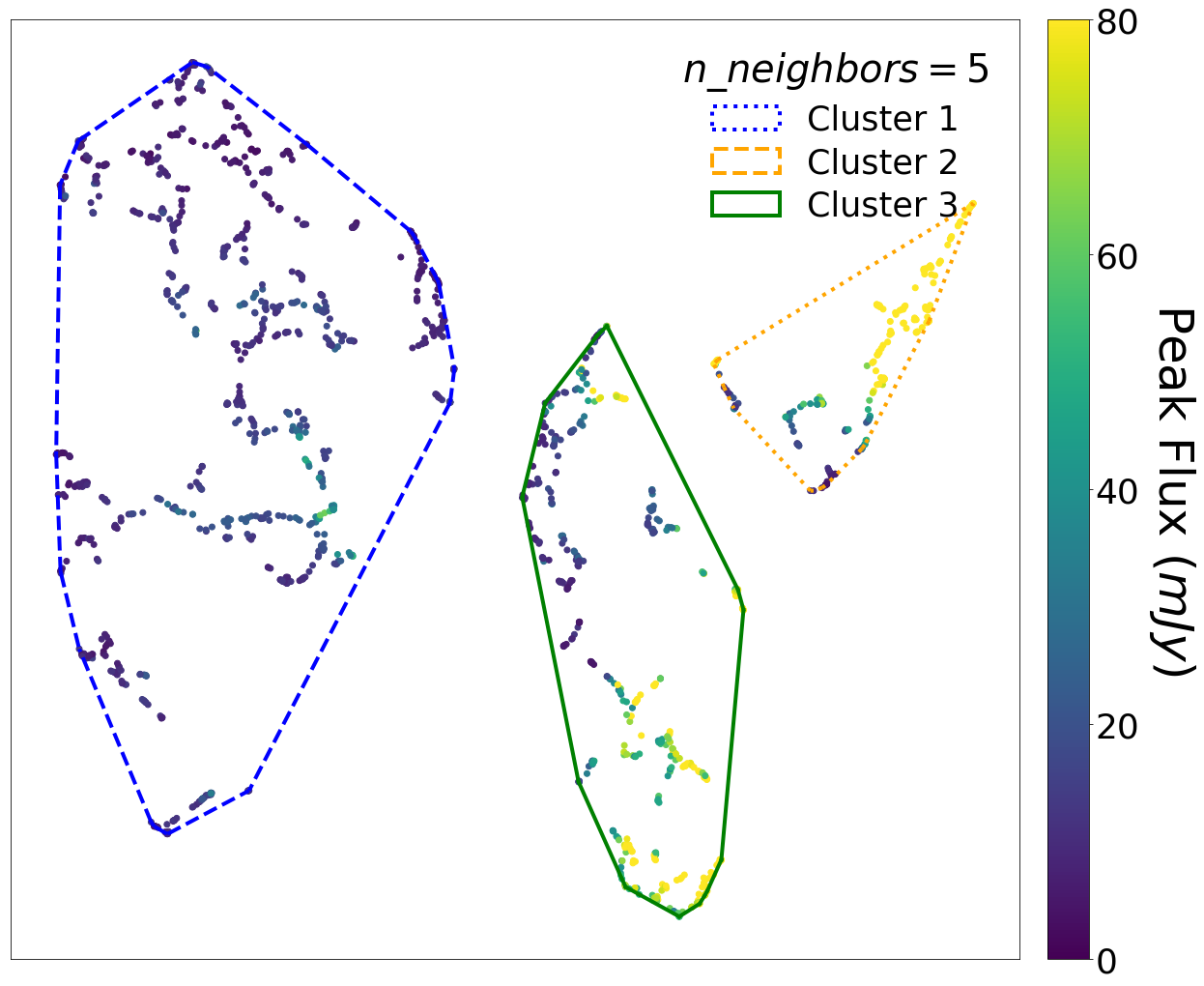}}\par 
                    \subcaptionbox{\label{apcpf6}}{\includegraphics[width=\linewidth, height=0.8\columnwidth]{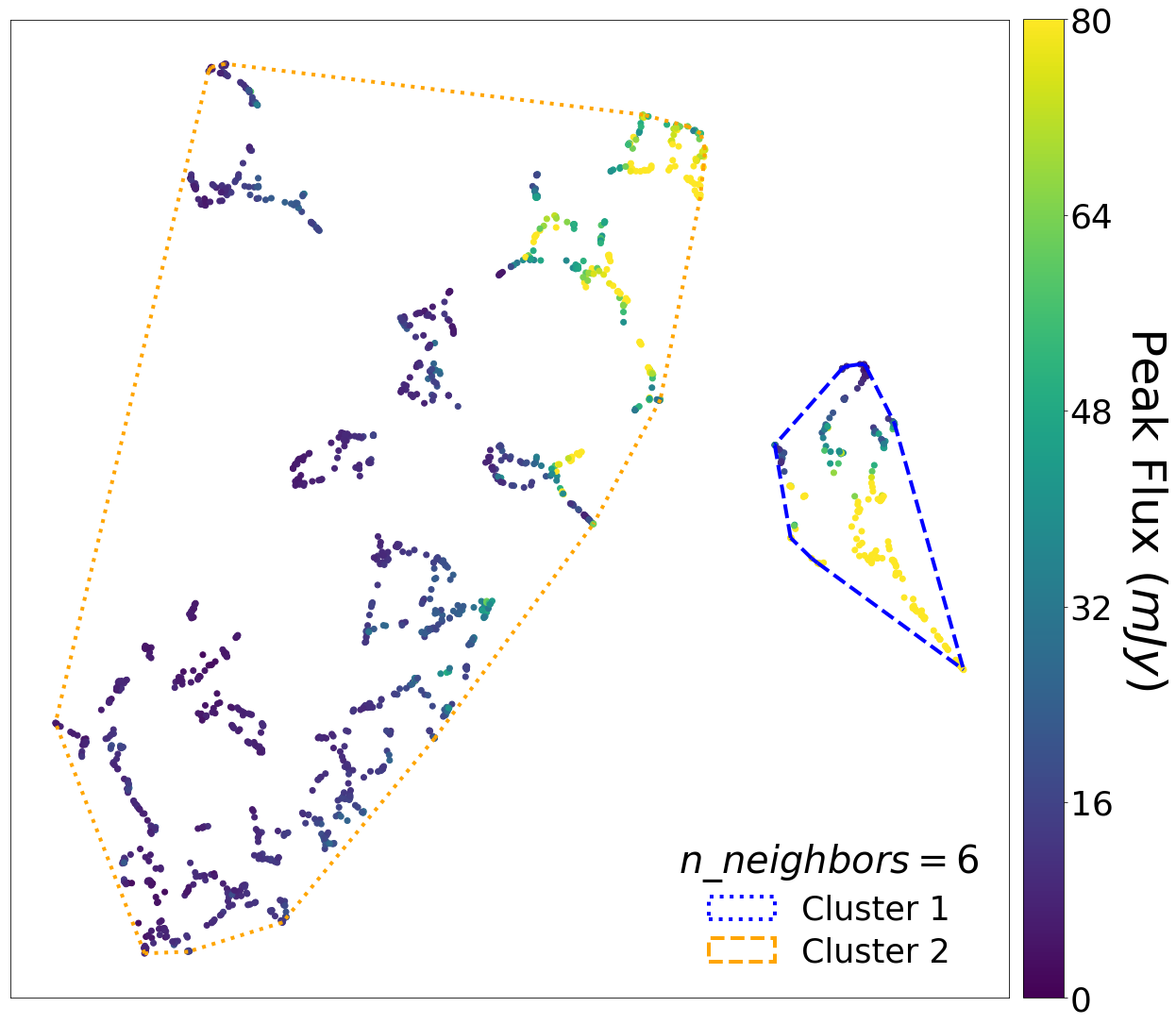}}\par 
                \end{multicols}
                \begin{multicols}{2}
                    \subcaptionbox{\label{apcpf7}}{\includegraphics[width=\linewidth, height=0.8\columnwidth]{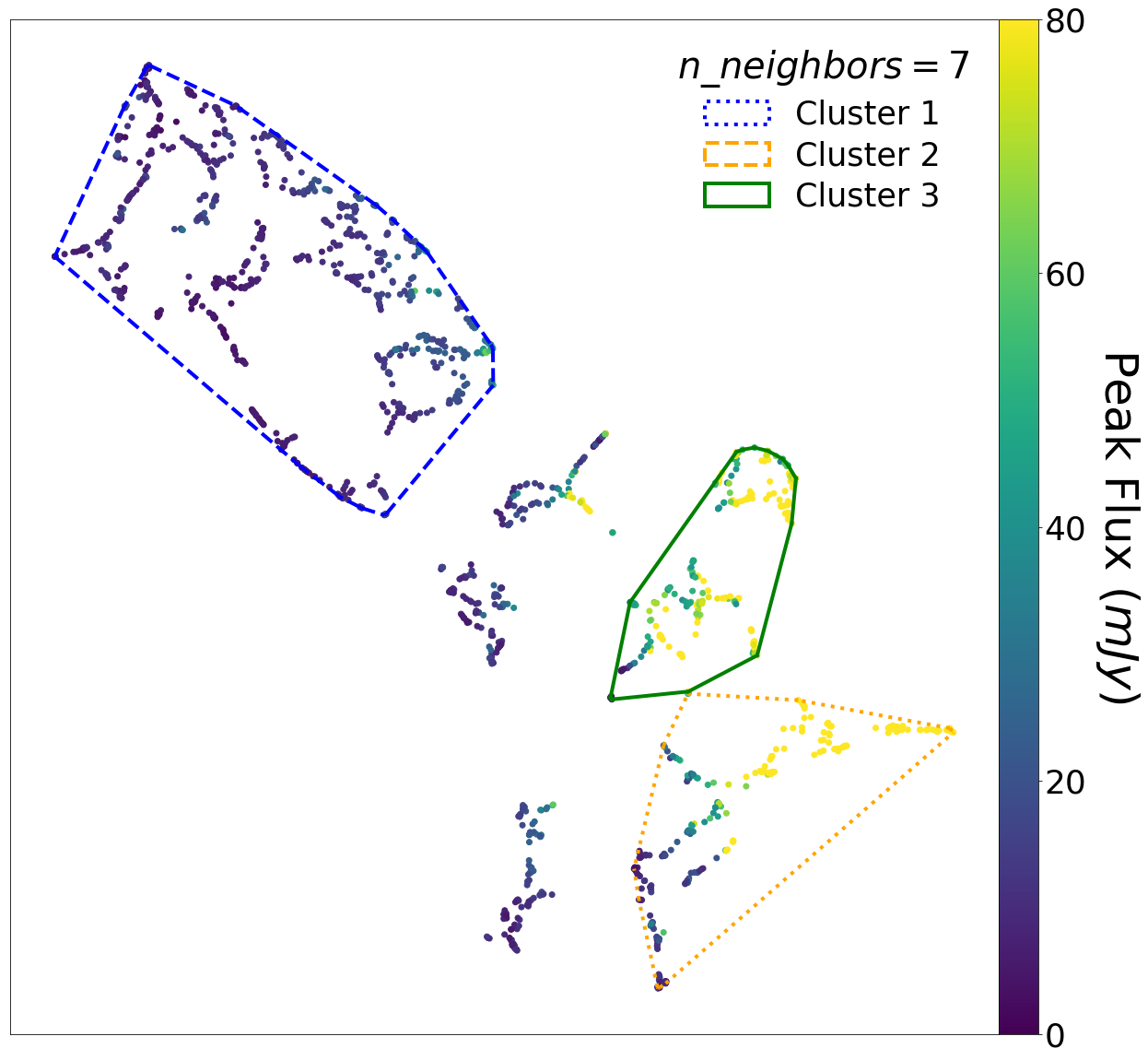}}\par
                    \subcaptionbox{\label{apcpf8}}{\includegraphics[width=\linewidth, height=0.8\columnwidth]{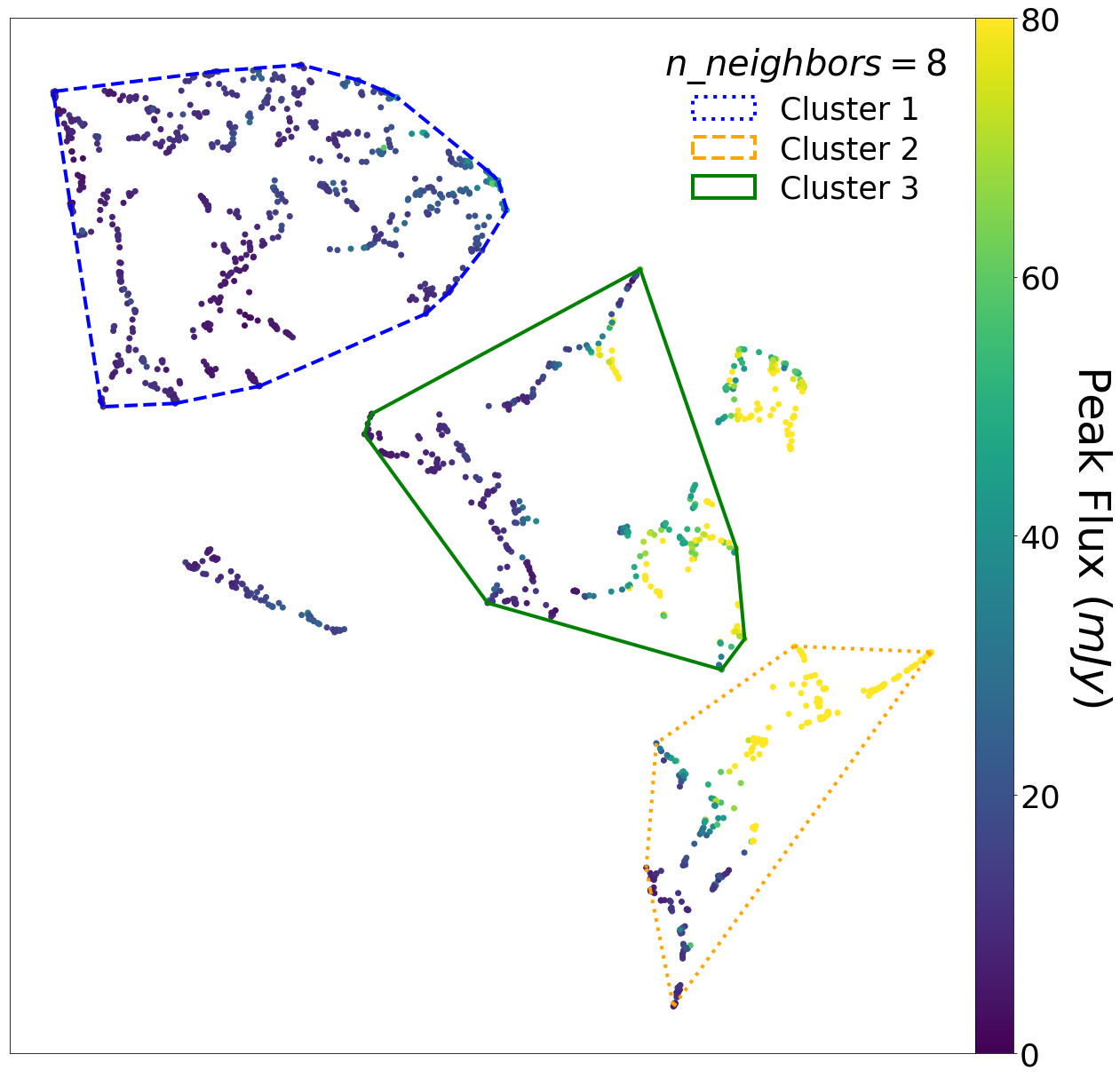}}\par
                \end{multicols}
                \caption{\large Peak Flux colouring of the clustering results for (\ref{apcpf5}) {\ttfamily n\_neighbors = 5}, (\ref{apcpf6}) {\ttfamily n\_neighbors = 6}, (\ref{apcpf7}) {\ttfamily n\_neighbors = 7}, and (\ref{apcpf8}) {\ttfamily n\_neighbors = 8}. The data which are not surrounded by lines correspond to Noise clusters.}
            \label{appfparcol}
            \end{figure*}
            
            \begin{figure*} 
                \centering
                \begin{multicols}{2}
                    \subcaptionbox{\label{apcw5}}{\includegraphics[width=\linewidth, height=0.8\columnwidth]{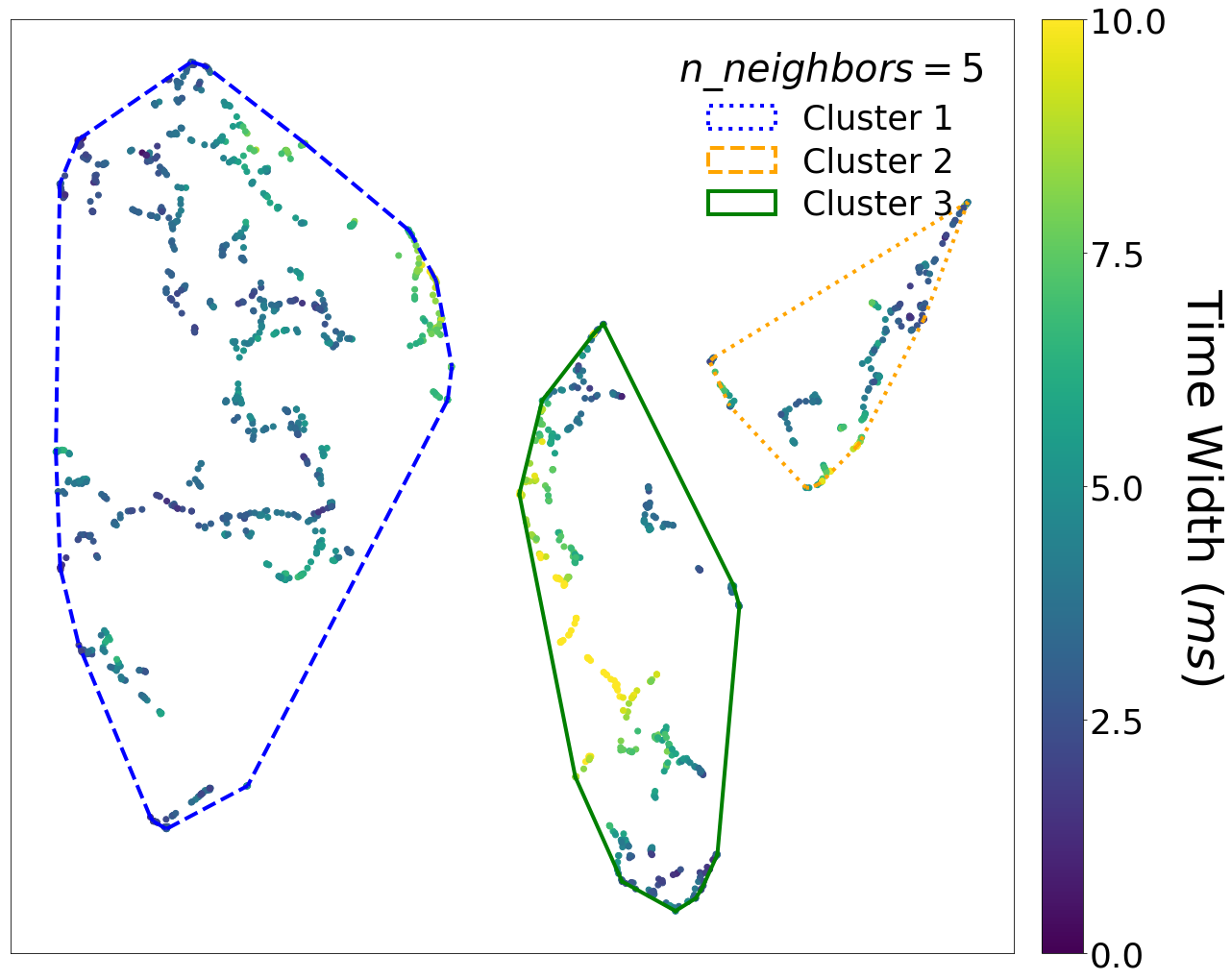}}\par 
                    \subcaptionbox{\label{apcw6}}{\includegraphics[width=\linewidth, height=0.8\columnwidth]{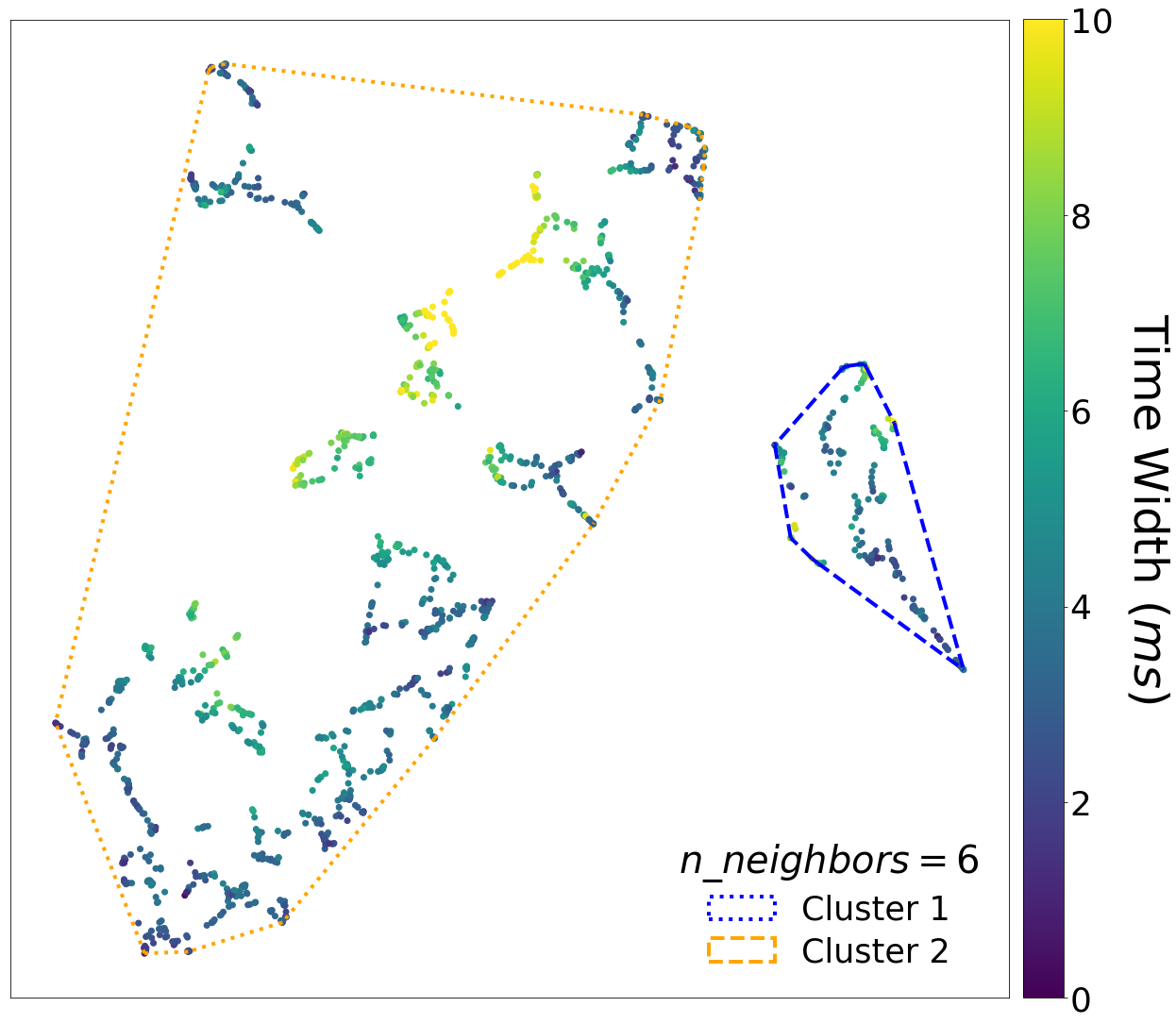}}\par 
                \end{multicols}
                \begin{multicols}{2}
                    \subcaptionbox{\label{apcw7}}{\includegraphics[width=\linewidth, height=0.8\columnwidth]{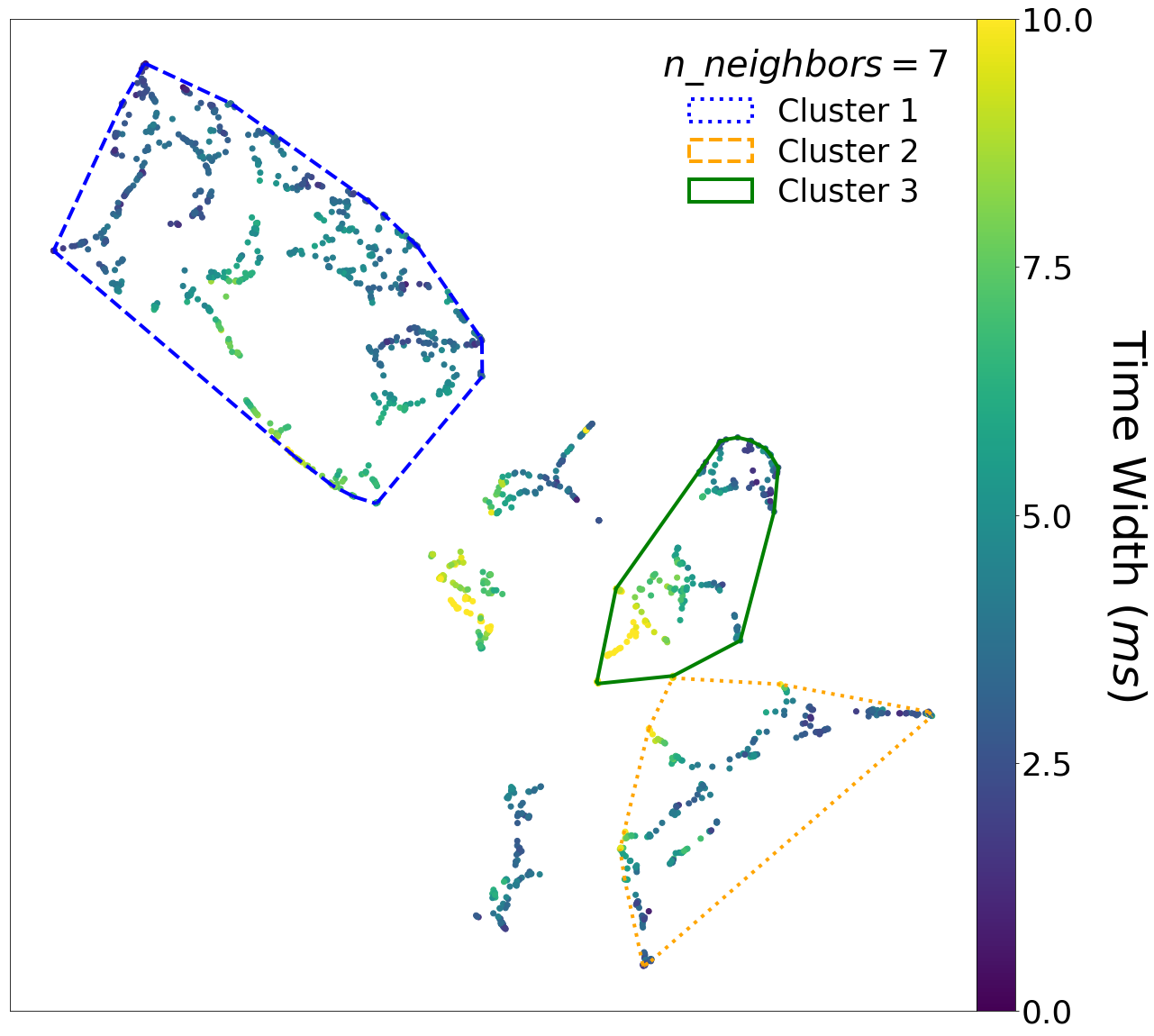}}\par
                    \subcaptionbox{\label{apcw8}}{\includegraphics[width=\linewidth, height=0.8\columnwidth]{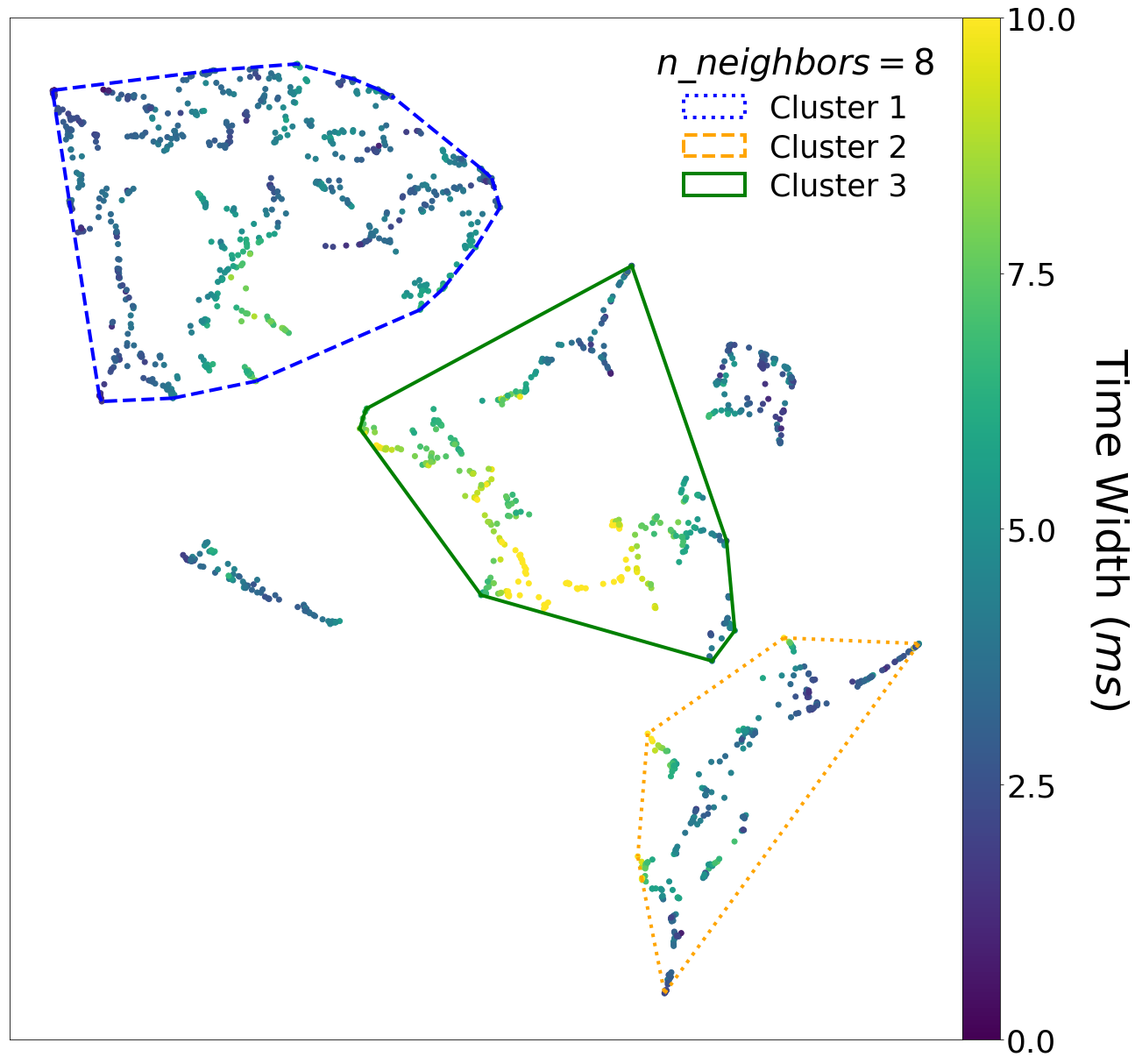}}\par
                \end{multicols}
                \caption{\large Time Width colouring of the clustering results for (\ref{apcw5}) {\ttfamily n\_neighbors = 5}, (\ref{apcw6}) {\ttfamily n\_neighbors = 6}, (\ref{apcw7}) {\ttfamily n\_neighbors = 7}, and (\ref{apcw8}) {\ttfamily n\_neighbors = 8}. The data which are not surrounded by lines correspond to Noise clusters.}
            \label{apwparcol}
            \end{figure*}
            
            \begin{figure*} 
                \centering
                \begin{multicols}{2}
                    \subcaptionbox{\label{apcwt5}}{\includegraphics[width=\linewidth, height=0.8\columnwidth]{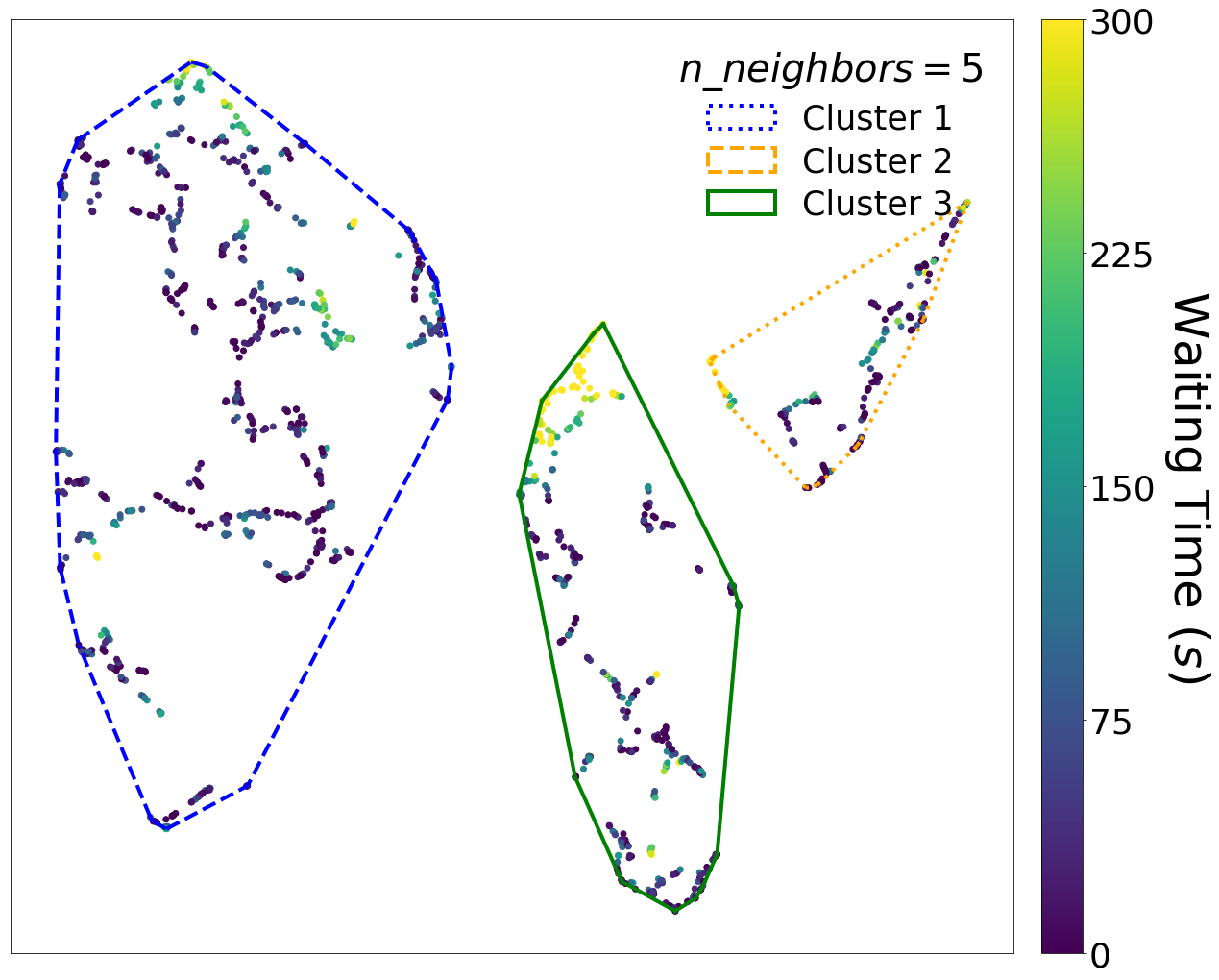}}\par 
                    \subcaptionbox{\label{apcwt6}}{\includegraphics[width=\linewidth, height=0.8\columnwidth]{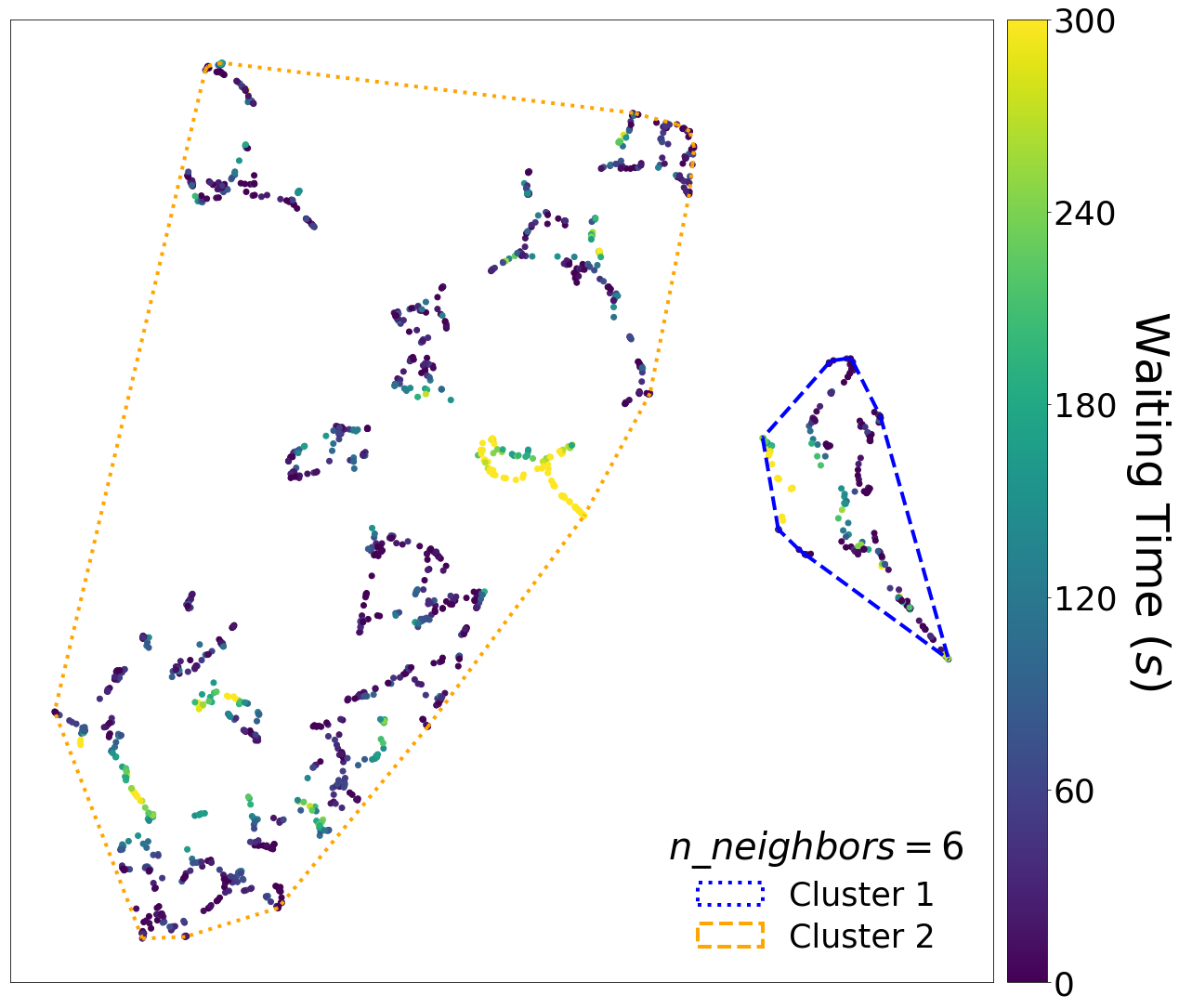}}\par 
                \end{multicols}
                \begin{multicols}{2}
                    \subcaptionbox{\label{apcwt7}}{\includegraphics[width=\linewidth, height=0.8\columnwidth]{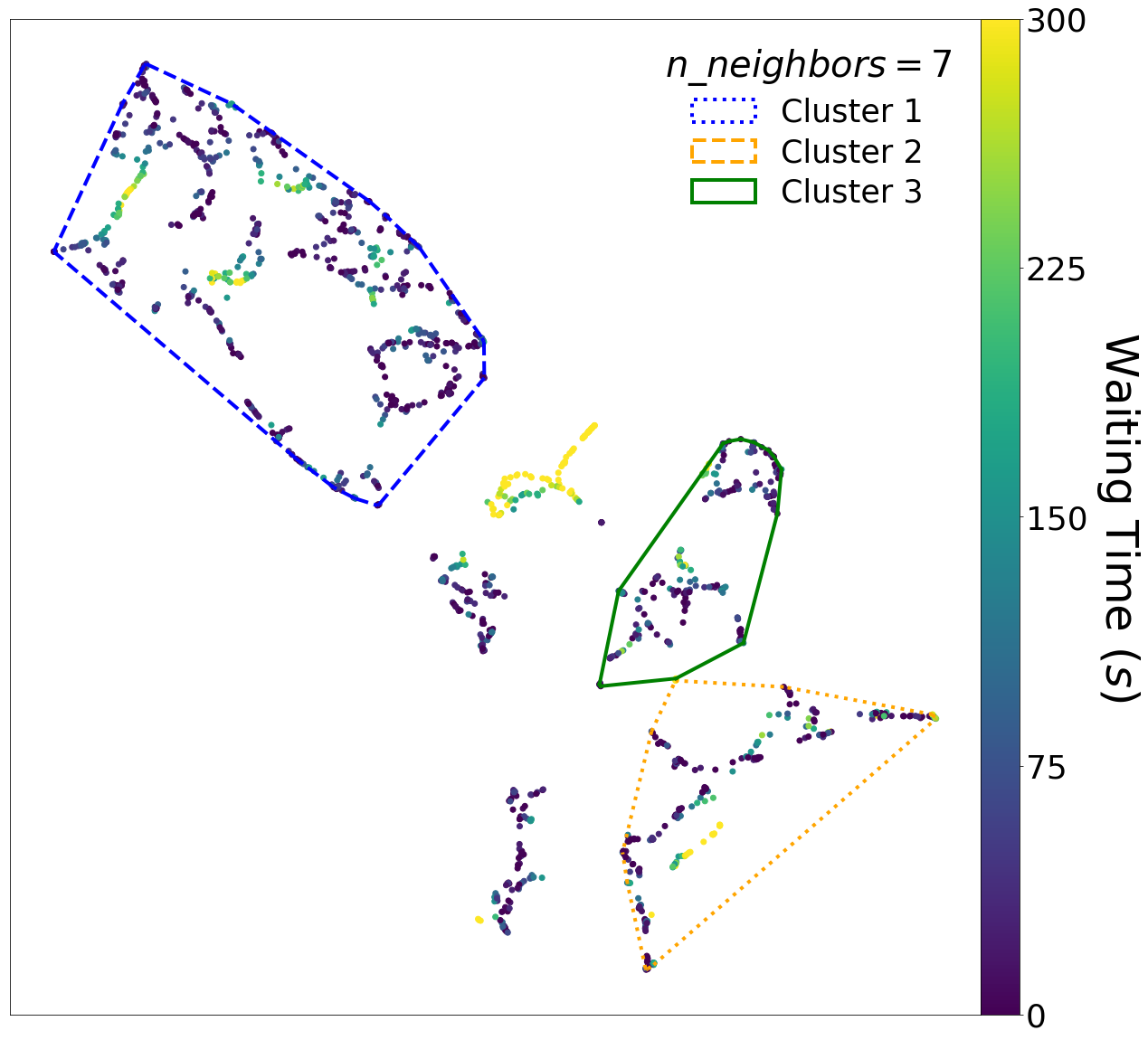}}\par
                    \subcaptionbox{\label{apcwt8}}{\includegraphics[width=\linewidth, height=0.8\columnwidth]{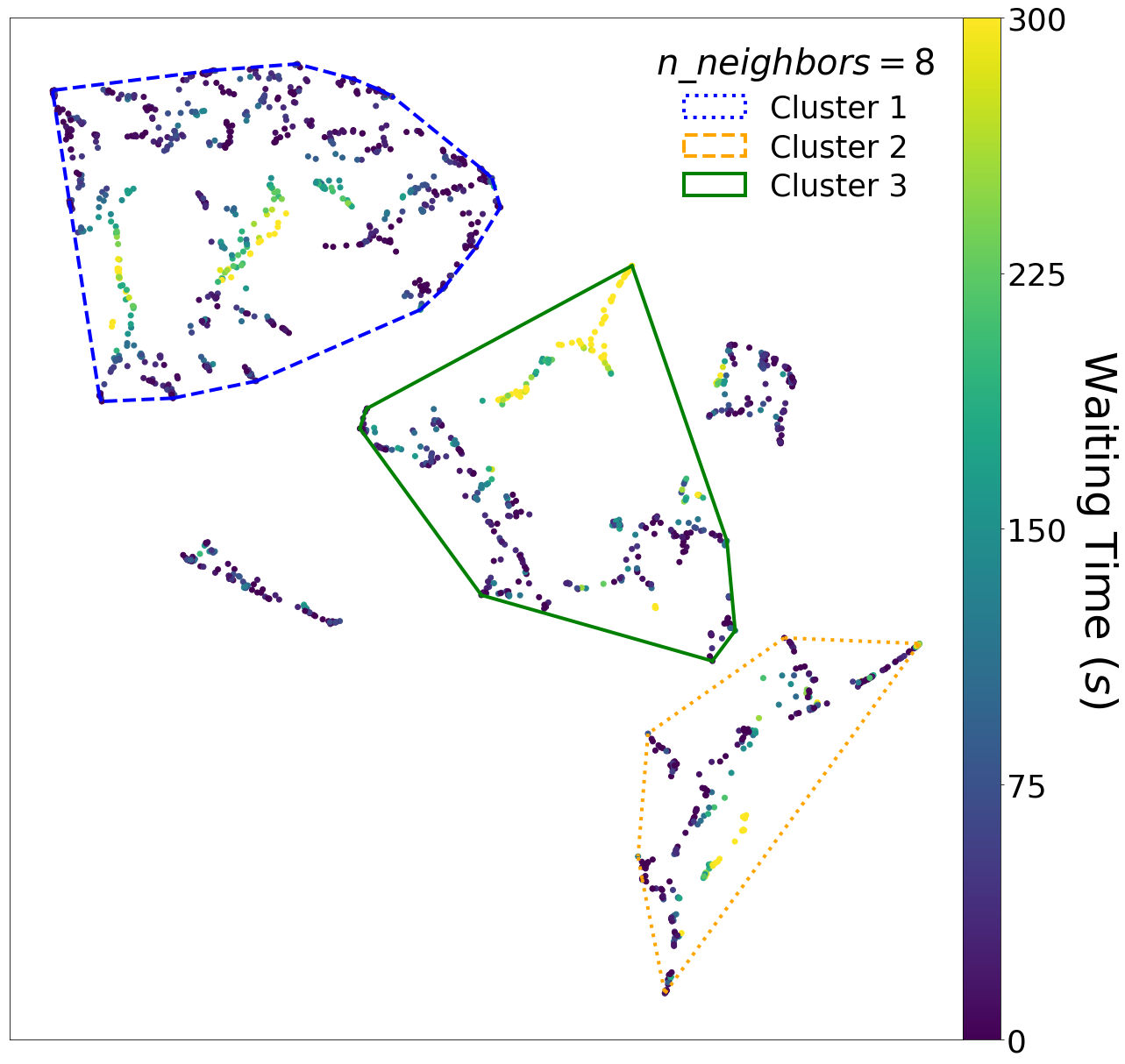}}\par
                \end{multicols}
                \caption{\large Waiting Time colouring of the clustering results for (\ref{apcwt5}) {\ttfamily n\_neighbors = 5}, (\ref{apcwt6}) {\ttfamily n\_neighbors = 6}, (\ref{apcwt7}) {\ttfamily n\_neighbors = 7},  and (\ref{apcwt8}) {\ttfamily n\_neighbors = 8}. The data which are not surrounded by lines correspond to Noise clusters.}
            \label{apwtparcol}
            \end{figure*}

\subsection{Histogram results for {\ttfamily n\_neighbors = 5,6,7,8, and 9}}
\label{histogrammers}
  
                \begin{figure*} 
                \centering
                \begin{multicols}{2}
                    \subcaptionbox{\label{aphbw5}}{\includegraphics[width=\linewidth, height=0.8\columnwidth]{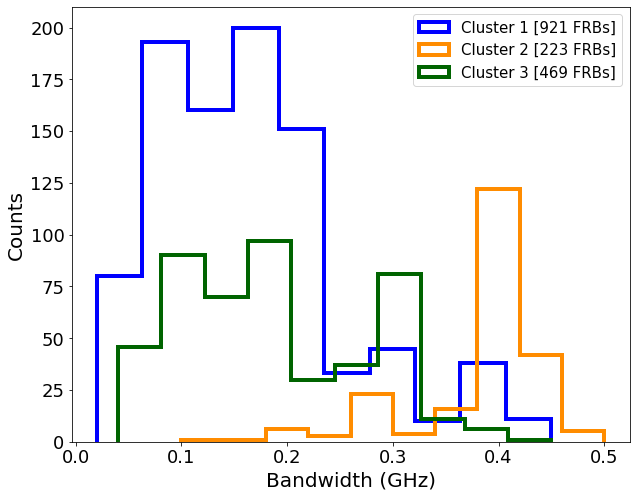}}\par 
                    \subcaptionbox{\label{aphbw6}}{\includegraphics[width=\linewidth, height=0.8\columnwidth]{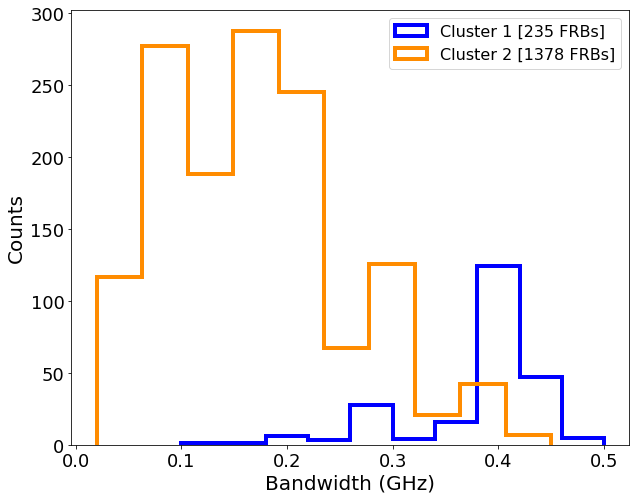}}\par 
                \end{multicols}
                \begin{multicols}{2}
                    \subcaptionbox{\label{aphbw7}}{\includegraphics[width=\linewidth, height=0.8\columnwidth]{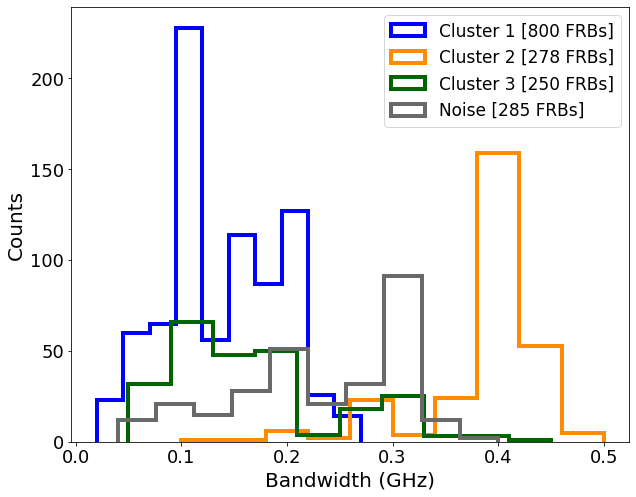}}\par
                    \subcaptionbox{\label{aphbw8}}{\includegraphics[width=\linewidth, height=0.8\columnwidth]{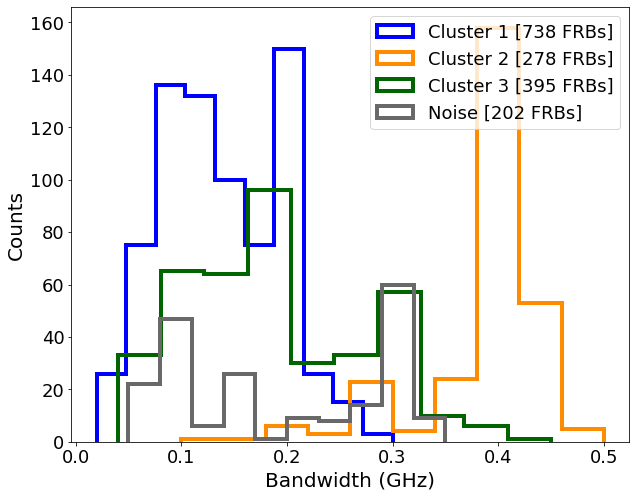}}\par
                \end{multicols}
                \begin{multicols}{2}
                    \subcaptionbox{\label{aphbw9}}{\includegraphics[width=\linewidth, height=0.8\columnwidth]{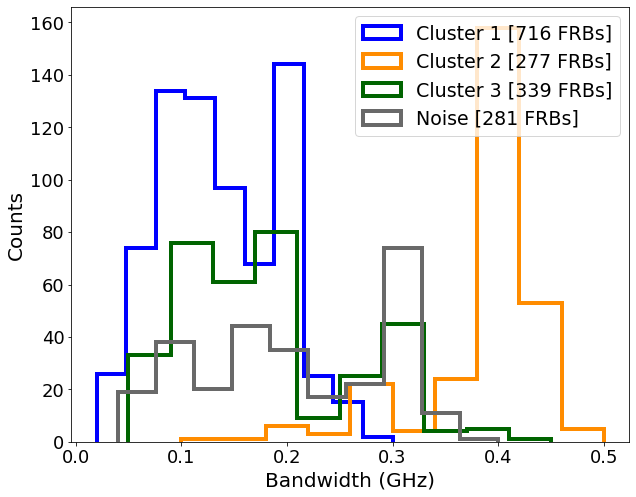}}\par
                \end{multicols}
                \caption{ \large Bandwidth histograms of the clustering results for (\ref{aphbw5}) {\ttfamily n\_neighbors = 5}, (\ref{aphbw6}) {\ttfamily n\_neighbors = 6}, (\ref{aphbw7}) {\ttfamily n\_neighbors = 7}, (\ref{aphbw8}) {\ttfamily n\_neighbors = 8}, and (\ref{aphbw9}) {\ttfamily n\_neighbors = 9}.}
            \label{apbwparcolhisto}
            \end{figure*}
            
            \begin{figure*} 
                \centering
                \begin{multicols}{2}
                    \subcaptionbox{\label{aphe5}}{\includegraphics[width=\linewidth, height=0.8\columnwidth]{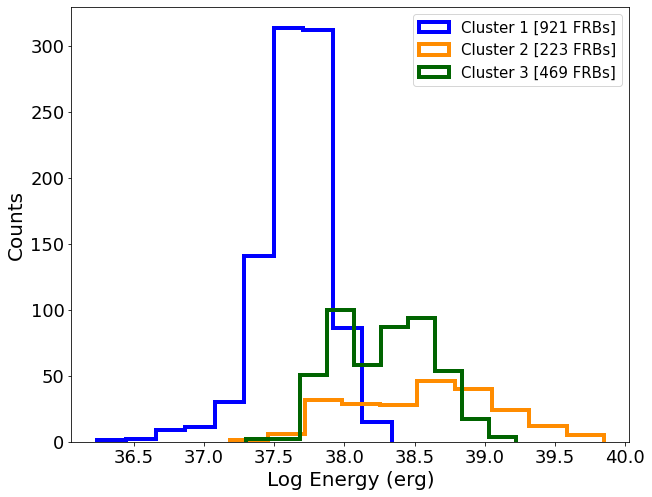}}\par 
                    \subcaptionbox{\label{aphe6}}{\includegraphics[width=\linewidth, height=0.8\columnwidth]{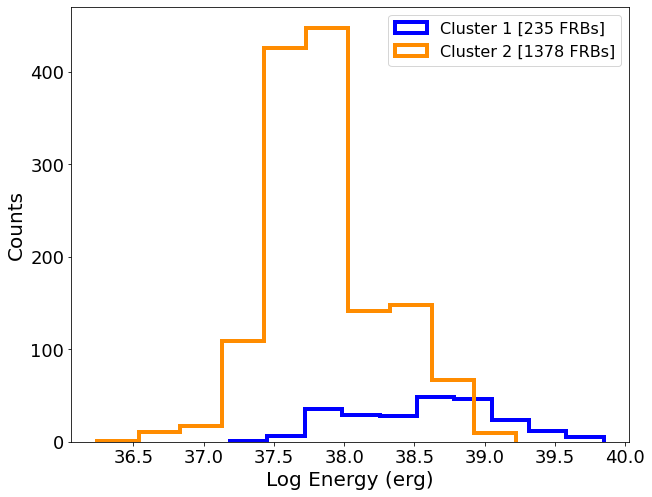}}\par 
                \end{multicols}
                \begin{multicols}{2}
                    \subcaptionbox{\label{aphe7}}{\includegraphics[width=\linewidth, height=0.8\columnwidth]{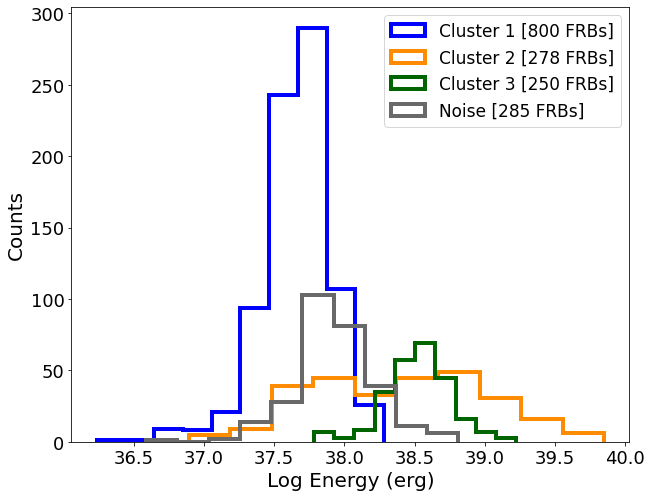}}\par
                    \subcaptionbox{\label{aphe8}}{\includegraphics[width=\linewidth, height=0.8\columnwidth]{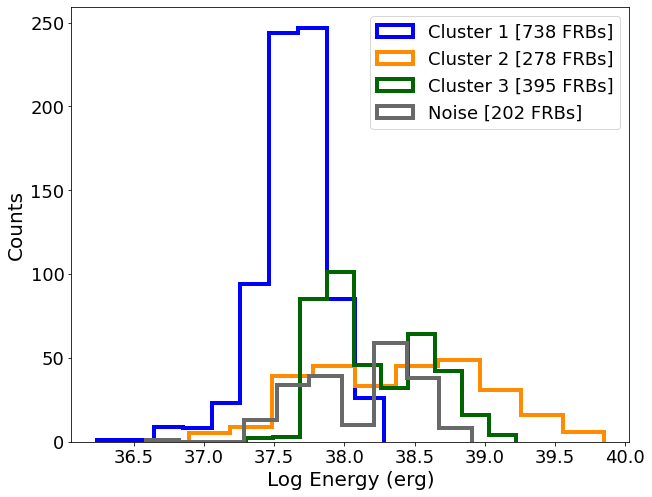}}\par
                \end{multicols}
                \begin{multicols}{2}
                    \subcaptionbox{\label{aphe9}}{\includegraphics[width=\linewidth, height=0.8\columnwidth]{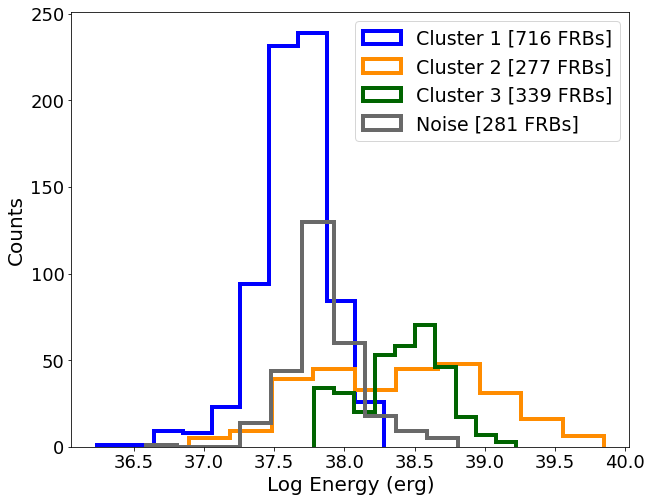}}\par
                \end{multicols}
                \caption{\large Energy histograms of the clustering results for (\ref{aphe5}) {\ttfamily n\_neighbors = 5}, (\ref{aphe6}) {\ttfamily n\_neighbors = 6}, (\ref{aphe7}) {\ttfamily n\_neighbors = 7}, (\ref{aphe8}) {\ttfamily n\_neighbors = 8}, and (\ref{aphe9}) {\ttfamily n\_neighbors = 9}}
            \label{apeparcolhisto}
            \end{figure*}

            \begin{figure*} 
                \centering
                \begin{multicols}{2}
                    \subcaptionbox{\label{aphfl5}}{\includegraphics[width=\linewidth, height=0.8\columnwidth]{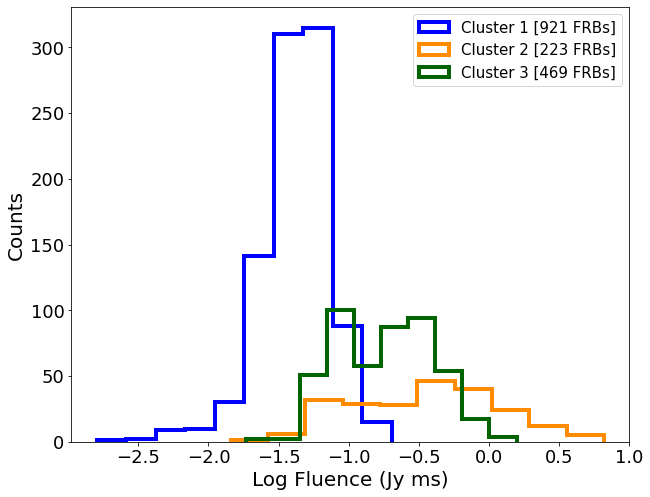}}\par 
                    \subcaptionbox{\label{aphfl6}}{\includegraphics[width=\linewidth, height=0.8\columnwidth]{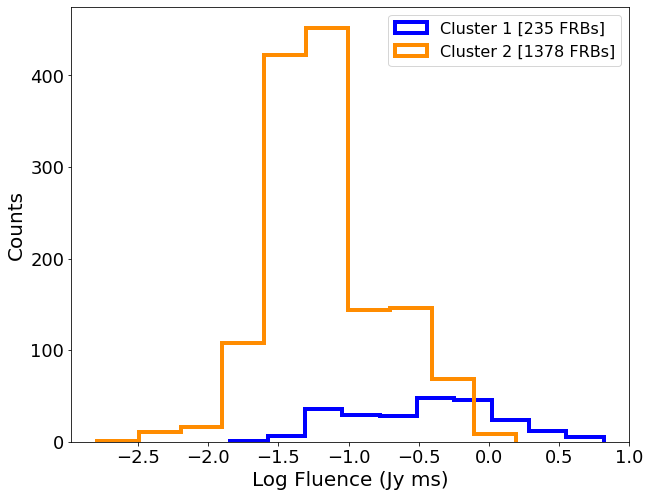}}\par 
                \end{multicols}
                \begin{multicols}{2}
                    \subcaptionbox{\label{aphfl7}}{\includegraphics[width=\linewidth, height=0.8\columnwidth]{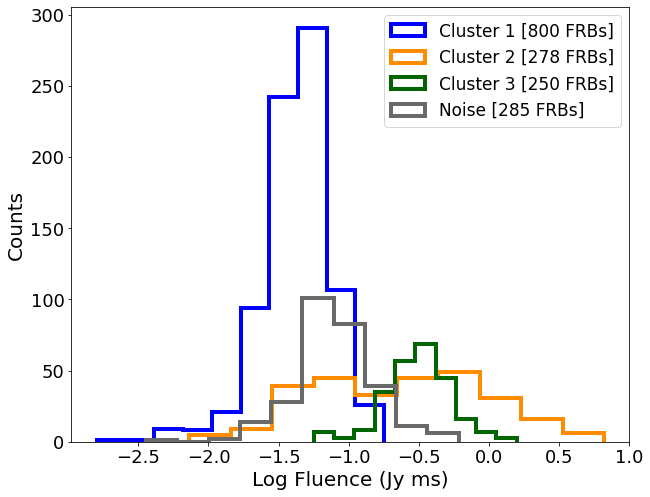}}\par
                    \subcaptionbox{\label{aphfl8}}{\includegraphics[width=\linewidth, height=0.8\columnwidth]{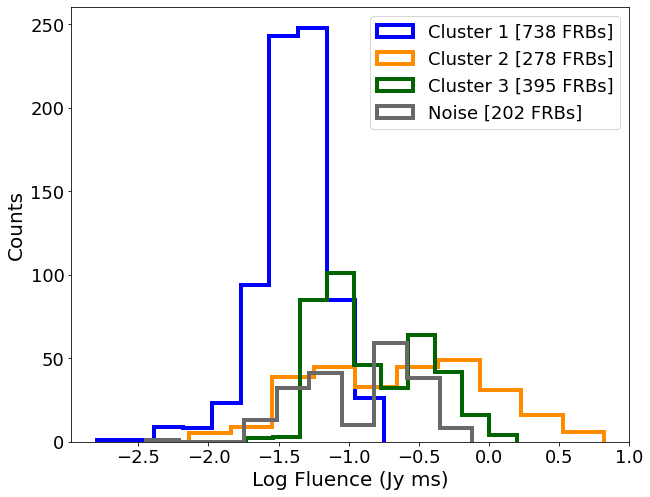}}\par
                \end{multicols}
                \begin{multicols}{2}
                    \subcaptionbox{\label{aphfl9}}{\includegraphics[width=\linewidth, height=0.8\columnwidth]{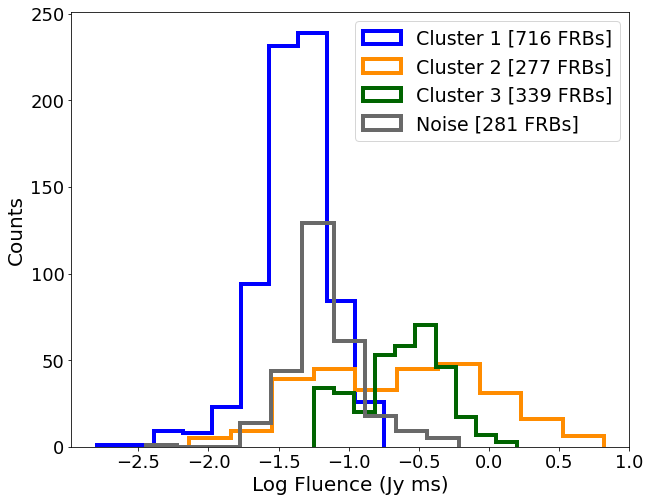}}\par
                \end{multicols}
                \caption{\large Fluence histograms of the clustering results for (\ref{aphfl5}) {\ttfamily n\_neighbors = 5}, (\ref{aphfl6}) {\ttfamily n\_neighbors = 6}, (\ref{aphfl7}) {\ttfamily n\_neighbors = 7}, (\ref{aphe8}) {\ttfamily n\_neighbors = 8}, and (\ref{aphe9}) {\ttfamily n\_neighbors = 9}.}
            \label{apflparcolhisto}
            \end{figure*}
            
            \begin{figure*} 
                \centering
                \begin{multicols}{2}
                    \subcaptionbox{\label{aphpf5}}{\includegraphics[width=\linewidth, height=0.8\columnwidth]{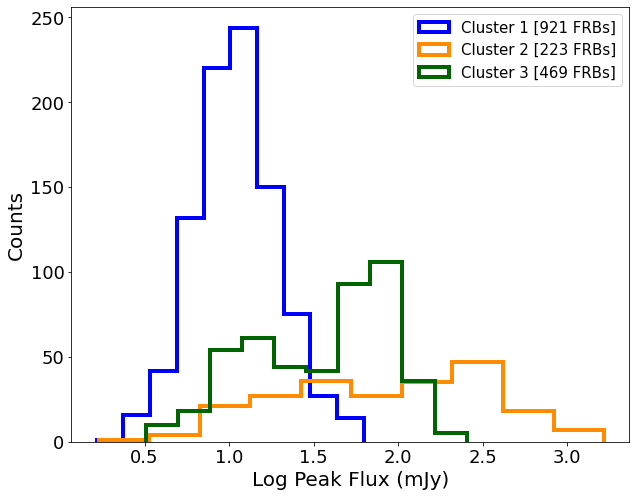}}\par 
                    \subcaptionbox{\label{aphpf6}}{\includegraphics[width=\linewidth, height=0.8\columnwidth]{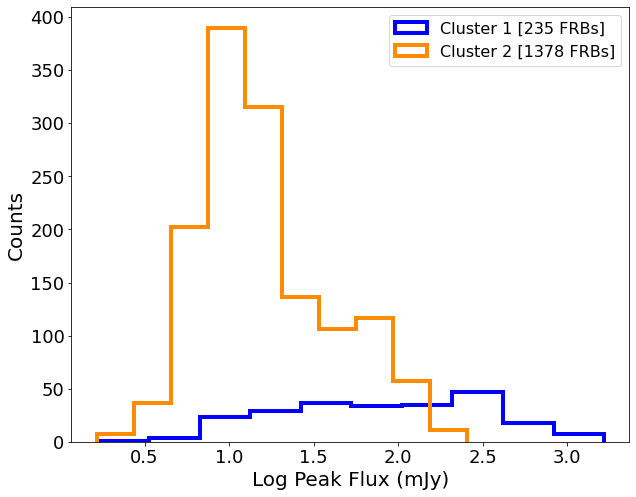}}\par 
                \end{multicols}
                \begin{multicols}{2}
                    \subcaptionbox{\label{aphpf7}}{\includegraphics[width=\linewidth, height=0.8\columnwidth]{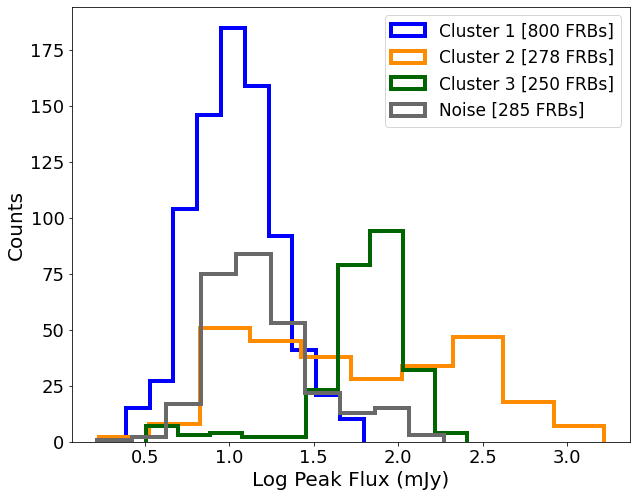}}\par
                    \subcaptionbox{\label{aphpf8}}{\includegraphics[width=\linewidth, height=0.8\columnwidth]{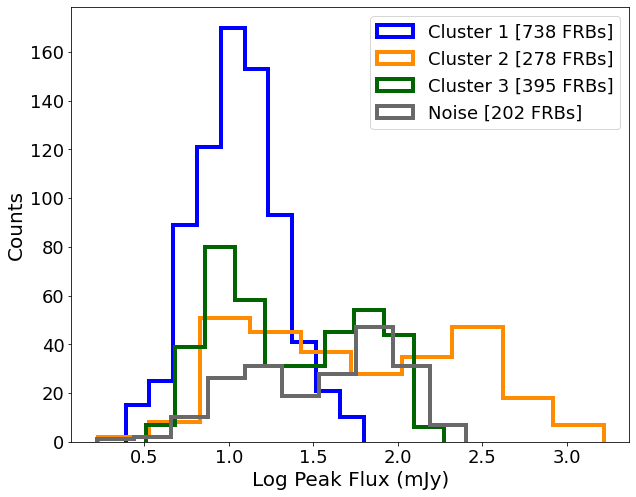}}\par
                \end{multicols}
                \begin{multicols}{2}
                    \subcaptionbox{\label{aphpf9}}{\includegraphics[width=\linewidth, height=0.8\columnwidth]{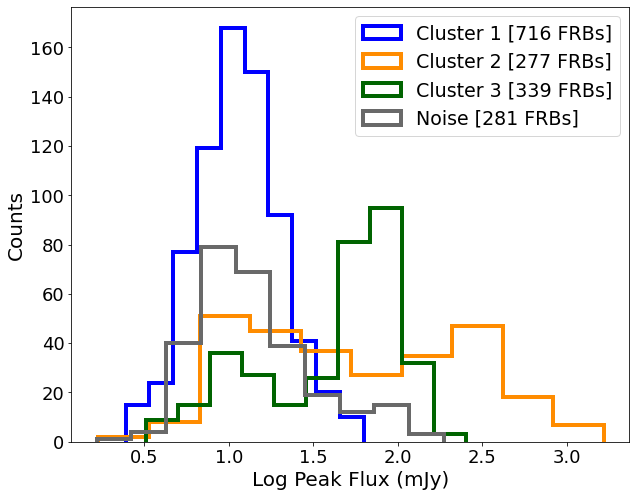}}\par
                \end{multicols}
                \caption{\large Peak Flux histograms of the clustering results for (\ref{aphpf5}) {\ttfamily n\_neighbors = 5}, (\ref{aphpf6}) {\ttfamily n\_neighbors = 6}, (\ref{aphpf7}) {\ttfamily n\_neighbors = 7}, (\ref{aphpf8}) {\ttfamily n\_neighbors = 8}, and (\ref{aphpf9}) {\ttfamily n\_neighbors = 9}.}
            \label{appfparcolhisto}
            \end{figure*}
            
            \begin{figure*} 
                \centering
                \begin{multicols}{2}
                    \subcaptionbox{\label{aphw5}}{\includegraphics[width=\linewidth, height=0.8\columnwidth]{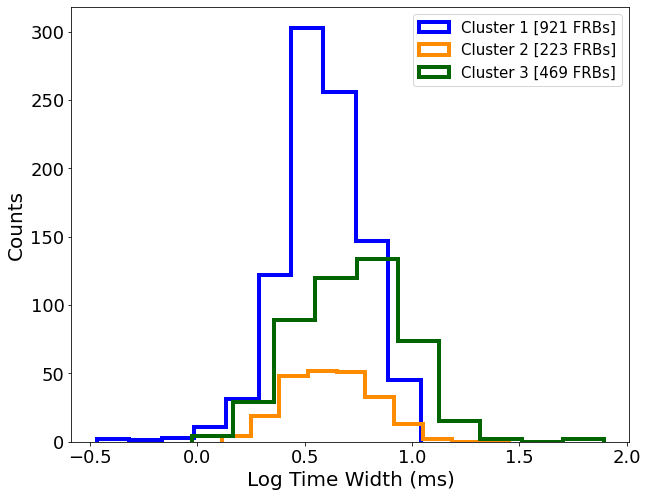}}\par 
                    \subcaptionbox{\label{aphw6}}{\includegraphics[width=\linewidth, height=0.8\columnwidth]{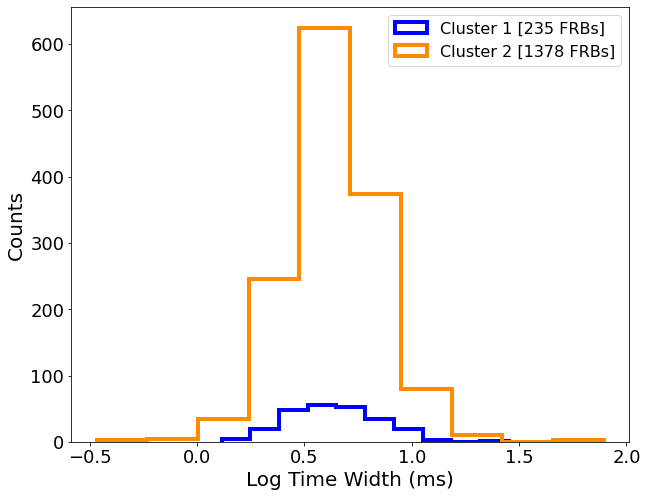}}\par 
                \end{multicols}
                \begin{multicols}{2}
                    \subcaptionbox{\label{aphw7}}{\includegraphics[width=\linewidth, height=0.8\columnwidth]{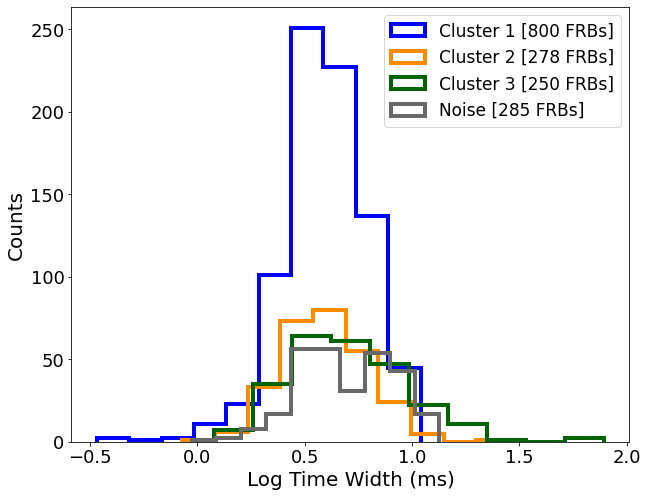}}\par
                    \subcaptionbox{\label{aphw8}}{\includegraphics[width=\linewidth, height=0.8\columnwidth]{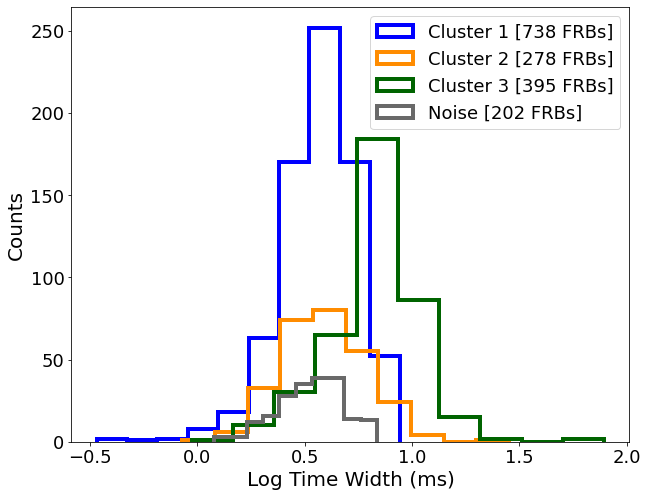}}\par
                \end{multicols}
                \begin{multicols}{2}
                    \subcaptionbox{\label{aphw9}}{\includegraphics[width=\linewidth, height=0.8\columnwidth]{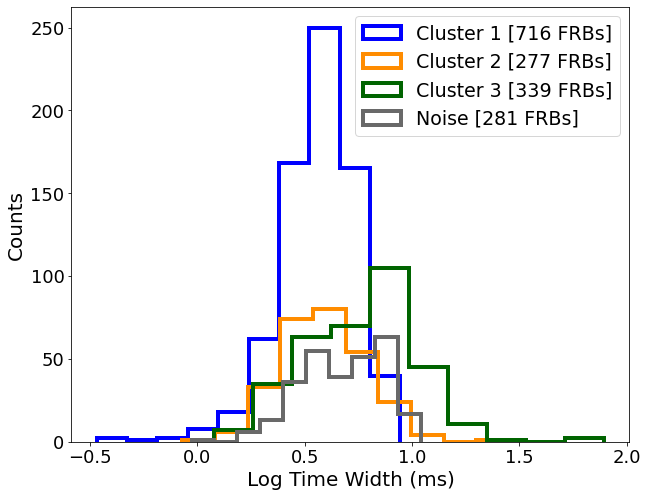}}\par
                \end{multicols}
                \caption{\large Time Width histograms of the clustering results for (\ref{aphw5}) {\ttfamily n\_neighbors = 5}, (\ref{aphw6}) {\ttfamily n\_neighbors = 6}, (\ref{aphw7}) {\ttfamily n\_neighbors = 7}, (\ref{aphw8}) {\ttfamily n\_neighbors = 8}, and (\ref{aphw9}) {\ttfamily n\_neighbors = 9}.}
            \label{apwparcolhisto}
            \end{figure*}
            
            \begin{figure*} 
                \centering
                \begin{multicols}{2}
                    \subcaptionbox{\label{aphwt5}}{\includegraphics[width=\linewidth, height=0.8\columnwidth]{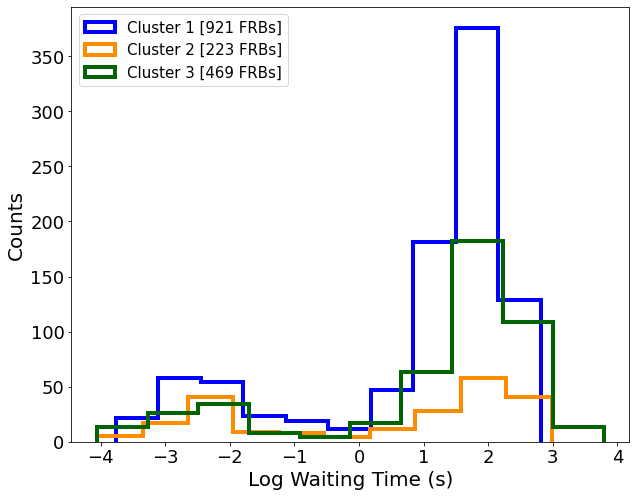}}\par 
                    \subcaptionbox{\label{aphwt6}}{\includegraphics[width=\linewidth, height=0.8\columnwidth]{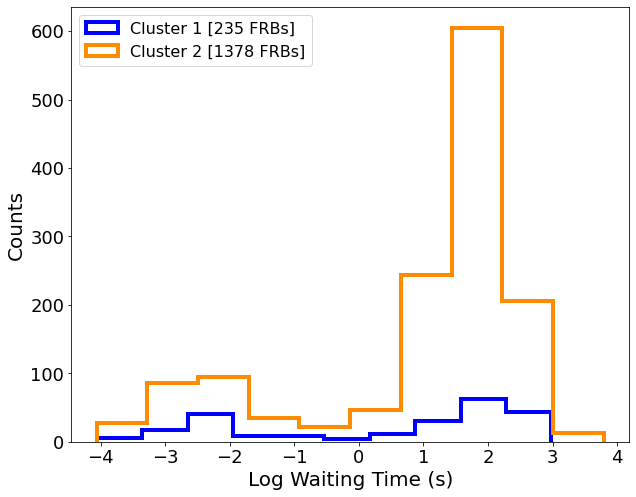}}\par 
                \end{multicols}
                \begin{multicols}{2}
                    \subcaptionbox{\label{aphwt7}}{\includegraphics[width=\linewidth, height=0.8\columnwidth]{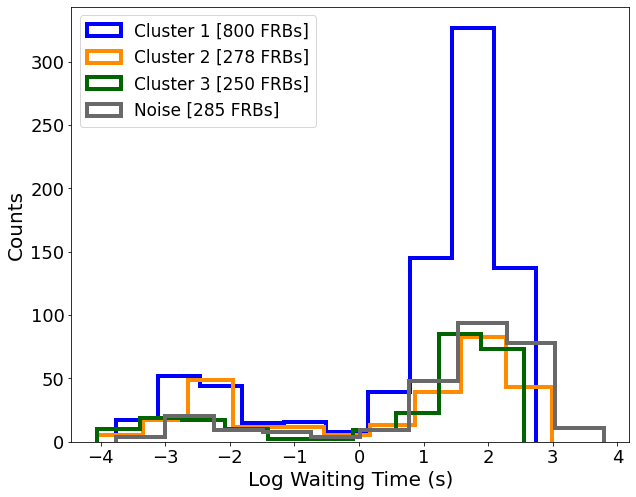}}\par
                    \subcaptionbox{\label{aphwt8}}{\includegraphics[width=\linewidth, height=0.8\columnwidth]{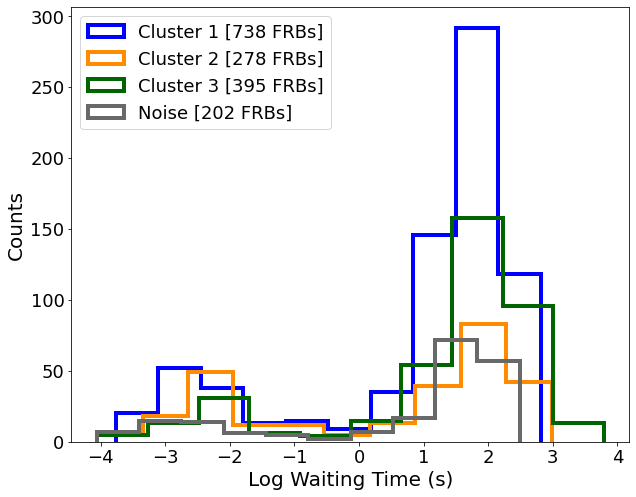}}\par
                \end{multicols}
                \begin{multicols}{2}
                    \subcaptionbox{\label{aphwt9}}{\includegraphics[width=\linewidth, height=0.8\columnwidth]{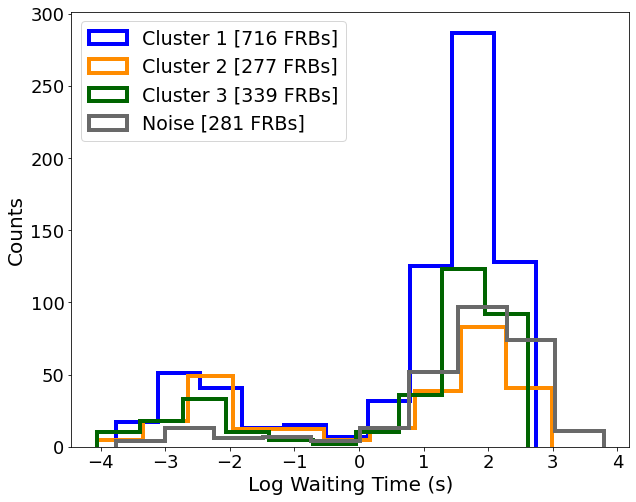}}\par
                \end{multicols}
                \caption{\large Waiting Time histograms of the clustering results for (\ref{aphwt5}) {\ttfamily n\_neighbors = 5}, (\ref{aphwt6}) {\ttfamily n\_neighbors = 6}, (\ref{aphwt7}) {\ttfamily n\_neighbors = 7}, (\ref{aphwt8}) {\ttfamily n\_neighbors = 8}, and (\ref{aphwt9}) {\ttfamily n\_neighbors = 9}.}
            \label{apwtparcolhisto}
            \end{figure*}

%%%%%%%%%%%%%%%%%%%%%%%%%%%%%%%%%%%%%%%%%%%%%%%%%%

% Don't change these lines
\bsp	% typesetting comment
\label{lastpage}
\end{document}